\documentclass[aps]{revtex4}  
\usepackage{graphicx}  
\usepackage{dcolumn}   
\usepackage{amssymb}   
\usepackage{rotating}  
\usepackage{appendix}
\def\MET{{\mbox{$E\kern-0.57em\raise0.19ex\hbox{/}_{T}$}}}
\def\met{{\mbox{$E\kern-0.57em\raise0.19ex\hbox{/}_{T}$}}}
\def\DZ{D0 }
\def\DZero{D0 }
\def\Dzero{D0 }

\def\ifb{~fb$^{-1}$}
\def\pp{$p\bar{p}$}
\def\bb{$b\bar{b}$}
\def\cc{$c\bar{c}$}
\def\ttbar{$t\bar{t}$}
\def\WH{$WH\rightarrow \ell\nu b\bar{b}$}

\def\lmet{$WH\rightarrow \ell\kern-0.45em\raise0.19ex\hbox{/} \nu b\bar{b}$}
\def\ZH{$ZH\rightarrow \nu\bar{\nu} b\bar{b}$}
\def\ZHll{$ZH\rightarrow \ell^+ \ell^- b\bar{b}$}

\def\vww{$VH \rightarrow l^{\pm}l'^{\pm} + X$}

\def\hww{$H\rightarrow W^+ W^-$}
\def\hbb{$H\rightarrow b\bar{b}$}

\def\tevE{$\sqrt{s}=1.96$~TeV}

\begin{document}

\rightline{FERMILAB-CONF-10-257-E}
\rightline{CDF Note 10241}
\rightline{\DZ Note 6096}
\vskip0.5in

\title{Combined CDF and \DZ Upper Limits on Standard Model Higgs-Boson Production with up to 6.7 fb$^{-1}$ of Data\\[2.5cm]}

\author{The TEVNPH Working Group\footnote{The Tevatron
New-Phenomena and Higgs Working Group can be contacted at
TEVNPHWG@fnal.gov. More information can be found at http://tevnphwg.fnal.gov/.}
 }
\affiliation{\vskip0.3cm for the CDF and \DZ Collaborations\\ \vskip0.2cm
\today}
\begin{abstract}
\vskip0.3in
We combine results from CDF and D0 on direct searches for the standard model (SM) 
Higgs boson ($H$) in \pp~collisions at the Fermilab Tevatron at $\sqrt{s}=1.96$~TeV. 
Compared to the previous Tevatron Higgs search combination more data have been added, 
additional new channels have been incorporated, and some previously used channels 
have been reanalyzed to gain sensitivity.  We use the latest parton distribution 
functions and $gg \rightarrow H$ theoretical cross sections when comparing our 
limits to the SM predictions.  With up to 5.9\ifb\ of data analyzed at CDF, and up 
to 6.7\ifb\ at D0, 
the 95\% C.L. upper limits on Higgs boson production are factors of 1.56 and 0.68 
the values of the SM cross section for a Higgs boson mass of $m_{H}=$115~GeV/$c^2$ 
and 165~GeV/$c^2$.
We exclude, at the 95\% C.L., a new and larger region at high mass between 
$158<m_{H}<175$~GeV/$c^{2}$.
\\[2cm]
{\hspace*{5.5cm}\em Preliminary Results}
\end{abstract}

\maketitle

\newpage
\section{Introduction} 

The search for a mechanism for electroweak symmetry breaking, and in
particular for a standard model (SM) Higgs boson has been a major 
goal of particle physics for many years, and is a central part of the
Fermilab Tevatron physics program. Both the CDF and \Dzero collaborations
have performed new combinations~\cite{CDFHiggs,DZHiggs} of multiple
direct searches for the SM Higgs boson.  The new searches include more
data, the inclusion of additional channels, and improved analysis techniques compared to previous analyses.
The sensitivities of these new combinations significantly exceed those 
of previous combinations~\cite{prevhiggs,WWPRLhiggs}.

In this note, we combine the most recent results of all such
searches in \pp~collisions at~\tevE.  The analyses combined
here seek signals of Higgs bosons produced in association with 
vector bosons ($q\bar{q}\rightarrow W/ZH$), through gluon-gluon 
fusion ($gg\rightarrow H$), and through vector boson fusion (VBF) 
($q\bar{q}\rightarrow q^{\prime}\bar{q}^{\prime}H$) corresponding
to integrated luminosities up to 5.9\ifb~at CDF and up to 
6.7\ifb~at D0.  In order to report an integrated luminosity corresponding
to the data sample used to make our results, we must average together the
contributing searches' luminosities in a way that represents their contributions
to the final results.   A search with a low sensitivity contributes less to the
average than searches with higher sensitivity. 
The overall sensitivity-weighted luminosities 
at low and high mass are 5.8\ifb~and 6.0\ifb, respectively. The 
Higgs boson decay modes studied are $H\rightarrow b{\bar{b}}$, 
$H\rightarrow W^+W^-$, $H\rightarrow\tau^+\tau^-$ and $H\rightarrow 
\gamma\gamma$.

To simplify the combination, the searches are separated into 
129 mutually exclusive final states (56 for CDF and 73 for D0; 
see Tables~\ref{tab:cdfacc} and~\ref{tab:dzacc}) referred to 
as ``analysis sub-channels'' in this note.  The selection procedures for each
analysis are detailed in Refs.~\cite{cdfWH2J} through~\cite{dzttH},
and are briefly described below.

\section{Acceptance, Backgrounds, and Luminosity}  

Event selections are similar for the corresponding CDF and D0 analyses.
For the case of \WH, an isolated lepton ($\ell=$ electron or muon) 
and two jets are required, with one or more $b$-tagged jets, i.e., 
identified as containing a weakly-decaying $B$ hadron.  Selected events 
must also display a significant imbalance in transverse momentum
(referred to as missing transverse energy or \met).  Events with more 
than one isolated lepton are vetoed.  

For the D0 \WH\ analyses, the 
data are split by lepton type and jet multiplicity (two or three jet 
sub-channels), and in turn for each of these two non-overlapping 
$b$-tagged samples are defined, one being a single ``tight'' $b$-tag 
(ST) sample, and the other a double ``loose'' $b$-tag (DT) sample. The 
tight and loose $b$-tagging criteria~\cite{Abazov:2010ab} are defined with respect to the 
mis-identification rate that the $b$-tagging algorithm yields for light 
quark or gluon jets (``mistag rate'') typically $\le 0.5\%$ or $\le 
1.5\%$, respectively. Each sub-channel is analyzed separately.  The 
outputs of random forests, trained separately for each sample and for 
each Higgs mass, are used as the final discriminating variables in the 
limit setting procedure.

For the CDF \WH\ analyses, events are analyzed in two and three jet 
sub-channels separately, and in each of these samples the events 
are grouped into various lepton and $b$-tag categories. In addition 
to the selections requiring an identified lepton, events with an
isolated track failing lepton selection requirements in the two jet 
sample, or an identified loose muon in the extended muon coverage 
in the three jet sample, are analyzed separately in their own categories. These 
additional categories provide some acceptance for poorly reconstructed electrons as 
well as single prong tau decays. Within the lepton categories there 
are four $b$-tagging categories considered in the two jet sample: 
two tight $b$-tags (TDT), one tight $b$-tag and one loose $b$-tag (LDT), 
one tight $b$-tag and one looser $b$-tag (LDTX), and a single, tight, 
$b$-tag (ST). The same $b$-tagging categories are used for the three 
jets channel, although the LDTX category is not used (events with one 
tight and one looser $b$-tags propagate into the one $b$-tag category).  
A Bayesian neural network discriminant is trained at each $m_H$ in 
the test range for the two jet sample, separately for each lepton and 
$b$-tagging category, while for the three jet sample a matrix element 
(ME) discriminant is used.

For the \ZH\ analyses, the selection is similar to the $WH$ selection, 
except all events with isolated leptons are vetoed and stronger multijet 
background suppression techniques are applied.  Both CDF and D0 analyses 
use a track-based missing transverse momentum calculation as a discriminant 
against false \met . In addition both CDF and D0 utilize multi-variate 
techniques, a boosted decision tree at D0 and a neural network at CDF, to 
further discriminate against the multi-jet background before $b$-tagging.
There is a sizable fraction of the \WH\ signal in which the lepton is 
undetected that is selected in the \ZH\ samples,  so these analyses are 
also referred to as $VH \rightarrow \met b \bar{b}$.  The CDF analysis 
uses three non-overlapping categories of $b$-tagged events (TDT, LDT and 
ST as for three jet \WH\ channels).  D0 uses orthogonal ST and tight-loose 
double-tag (TLDT) channels.  CDF uses neural-network outputs for the final 
discriminating variables, while D0 uses boosted decision tree outputs.
For this combination D0 has updated the TLDT sample to use 6.4 fb$^{-1}$ 
of data and refined the input variables to the decision trees, including, 
amongst others, event characteristics sensitive to whether the jets 
originated from a color singlet object.

The \ZHll\ analyses require two isolated leptons and at least two jets.   
D0's \ZHll\ analyses separate events into non-overlapping samples of
events with one tight $b$-tag (ST) and two loose $b$-tags (LDT).  
CDF separates events into single tag (ST), double tag (TDT) and loose 
double tag (LDT) samples. To increase signal acceptance D0 has loosened 
the selection criteria for one of the leptons to include an isolated 
track not reconstructed in the muon detector ($\mu\mu_{trk}$) or an 
electron from the inter-cryostat region of the D0 detector ($ee_{ICR}$).
Combined with the dielectron ($ee$) and dimuon ($\mu\mu$) analyses, 
these provide four orthogonal analyses, and each uses 4.2 fb$^{-1}$ of 
data. Most recently the $ee$ and $\mu\mu$ channels have been updated 
to include 6.2 fb$^{-1}$ of data.  CDF has added for this combination 
additional sub-channels for candidate events with two loose muon 
candidates selected using a neural network discriminant.  Events in 
the new category are analyzed separately depending on the trigger path 
(muon or \met) from which they were selected.  D0 applies a kinematic 
to optimize reconstruction.  CDF corrects jet energies for \met\ using 
a neural network approach.  For the D0 analysis random forests of 
decision trees provide the final variables for setting limits, while 
CDF utilizes outputs of two-dimensional neural networks incorporating
likelihoods based on event probabilities obtained from ME calculations 
as additional inputs.

For the \hww~analyses, signal events are characterized by large \met~and 
two opposite-signed, isolated leptons.  The presence of neutrinos in the 
final state prevents the accurate reconstruction of the candidate Higgs boson mass.
D0 selects events containing large \met\ and electrons and/or muons, dividing the data sample 
into three final states: $e^+e^-$, $e^\pm \mu^\mp$, and $\mu^+\mu^-$.  
Final states involving leptonic tau decays and mis-identified hadronic 
tau decays are included. The $e^+e^-$ and $\mu^+\mu^-$ analyses are as in 
Ref.~\cite{dzHWW1}.  The $e^\pm \mu^\mp$ channel has been updated, and now 
uses 6.7 fb$^{-1}$ of data as well as subdividing the dataset according to 
the number of jets in the event: 0, 1, or 2+ jets.  
CDF separates the \hww\ events in five non-overlapping samples, split into 
both ``high $s/b$'' and ``low $s/b$'' categories based on lepton types and 
different categories based on the number of reconstructed jets: 0, 1, or 2+ 
jets.  The sample with two or more jets is not split into low $s/b$ and 
high $s/b$ lepton categories due to low statistics.  A sixth CDF channel 
is the low dilepton mass ($m_{\ell^+\ell^-}$) channel, which accepts events 
with $m_{\ell^+\ell^-}<16$~GeV.  This channel increases the sensitivity of 
the $H\rightarrow W^+W^-$ analyses at low $m_H$, adding 10\% additional 
acceptance at $m_H=120$~GeV.

The division of events into jet categories allows the analysis discriminants 
to separate three different categories of signals from the backgrounds more 
effectively.  The signal production mechanisms considered are $gg\rightarrow 
H\rightarrow W^+W^-$, $WH+ZH\rightarrow jjW^+W^-$, and vector-boson fusion.  
For $gg\rightarrow H$, recent work~\cite{anastasiouwebber} indicates that the 
theoretical uncertainties due to scale and PDF variations are significantly 
different for the different jet categories.  CDF and D0 divide the theoretical 
uncertainty on $gg\rightarrow H$ into PDF and scale pieces, and utilize the 
differential uncertainties of~\cite{anastasiouwebber}.
The D0 $e^+e^-$ and $\mu^+\mu^-$ channels use neural-network discriminants, 
including the number of jets as an input, as the final discriminant while 
the $e^\pm \mu^\mp$ channel relies on boosted decision tree outputs with 
additional input variables now included for the 1 and 2+ jet sub-channels.  
CDF uses neural-network outputs, including likelihoods constructed from 
calculated ME probabilities as additional inputs for the 0-jet bin.

D0 has updated its \vww\ analyses to include additional data and an improved 
discriminant. The associated vector boson and the $W$ boson from the Higgs 
boson decay that has the same charge are required to decay leptonically, 
thereby defining three like-sign dilepton final states ($e^\pm e^\pm$, $e^\pm 
\mu^\pm$, and $\mu^{\pm}\mu^{\pm}$).  The combined output of two decision 
trees, trained against the instrumental and diboson backgrounds respectively, 
is used as the final discriminant.  CDF also includes a separate analysis 
of events with same-sign leptons and large \met\ to incorporate additional 
potential signal from associated production events in which the two leptons 
(one from the associated vector boson and one from a W boson produced in the 
Higgs decay) have the same charge.  CDF for the first time also incorporates 
three tri-lepton channels to include additional associated production 
contributions where leptons result from the associated W boson and the two 
W bosons produced in the Higgs decay or where an associated Z boson decays 
into a dilepton pair and a third lepton is produced in the decay of either 
of the W bosons resulting from the Higgs decay.  In the latter case, CDF 
separates the sample into 1 jet and 2+ jet sub-channels to fully take 
advantage of the Higgs mass constraint available in the 2+ jet case where 
all of the decay products are reconstructed.

For the first time CDF also includes opposite-sign channels in which one of the 
two lepton candidates is a hadronic tau.  Events are separated into $e$-$\tau$
and $\mu$-$\tau$ channels.  The final discriminants are obtained from boosted 
decision trees which incorporate both hadronic tau identification and kinematic 
event variables as inputs.  Also for the first time D0 includes new channels in 
which one of the $W$ bosons in the $H \rightarrow W^+W^-$ process decays
leptonically and the other decays hadronically.  Electron and muon final states 
are studied separately, each with 5.4 fb$^{-1}$ of data, with random forests 
being used for the final discriminants.

CDF contributes an analysis searching for Higgs bosons decaying to a tau lepton 
pair, in three separate production channels: direct $gg \rightarrow H$ production, 
associated $WH$ or $ZH$ production, and vector boson production with $H$ and 
forward jets in the final state.  One or two jets are required in the event 
selection.  In  this analysis, the final variable for setting limits is a 
combination of three boosted signal tree discriminants, each designed to 
discriminate the signal against one of the major backgrounds (QCD multi-jets, 
$Z/\gamma^{*} \rightarrow \tau^+\tau^-$, and $t\overline{t}$).   The theoretical 
systematic uncertainty on the $gg \rightarrow H$ production rate now takes into 
account recent theoretical work~\cite{anastasiouwebber} which provides separate 
uncertainties for each jet category.
D0 also contributes an analysis of the $\tau^+\tau^-+2$~jets  final state, which 
is sensitive to the $VH\rightarrow jj \tau^+\tau^-$, $ZH \rightarrow \tau^-\tau^-b \bar{b}$,
 VBF and gluon gluon fusion (with two additional jets) production 
mechanisms.  A neural network output is used as the discriminant variable for 
RunIIa (the first 1.0 fb$^{-1}$ of data), while a boosted decision tree output 
is used for later data.

CDF also includes an updated all-hadronic analysis, which results in two 
$b$-tagging sub-channels (TDT and LDT) for both $WH/ZH$ and VBF production 
to the $jjb{\bar{b}}$ final state.  Events with either four or five 
reconstructed jets are selected, and at least two must be $b$-tagged.  The 
large QCD multi-jet backgrounds are modeled from the data by applying a 
measured mistag probability to the non $b$-tagged jets in events containing 
a single $b$-tag.  Neural network discriminants based on kinematic event 
variables including ones designed to separate quark and gluon jets are used 
to obtain the final limits.

Both D0 and CDF contribute (CDF for the first time) analyses searching for 
direct Higgs boson production in which the Higgs decays directly into a 
pair of photons.  These analyses look for a signal peak in the diphoton 
invariant mass spectrum above the smooth background originating from 
standard QCD production.  Finally, D0 includes the channel $t \bar{t} 
H \rightarrow t \bar{t} b \bar{b}$.  Here the samples are analyzed 
independently according to the number of $b$-tagged jets and the total 
number of jets.  The scalar sum of the transverse energies of the reconstructed objects 
($H_T$) is used as the discriminant variable.

We normalize our Higgs boson signal predictions to the most recent 
high-order calculations available.  The $gg\rightarrow H$ production 
cross section is calculated at NNLL in QCD and also includes 
two-loop electroweak effects, and handling of the running $b$ quark 
mass~\cite{anastasiou,grazzinideflorian}.  These calculations are 
refinements of the earlier NNLO calculations of the $gg\rightarrow H$ 
production cross section~\cite{harlanderkilgore2002,anastasioumelnikov2002,ravindran2003}.
Electroweak corrections were computed in Refs.~\cite{actis2008,aglietti}. 
Soft gluon resummation was introduced in the prediction of the 
$gg\rightarrow H$ production cross section in Ref.~\cite{catani2003}.

The $gg\rightarrow H$ production cross section depends strongly on 
the gluon parton density function, and the accompanying value of 
$\alpha_s(q^2)$.  The cross sections used here are calculated 
with the MSTW 2008 NNLO PDF set~\cite{mstw2008}.  The Higgs boson 
production cross sections are listed in Table~\ref{tab:higgsxsec}.  
We include the larger theoretical uncertainties 
due to scale variations and PDF variations separately for each 
jet bin for the $gg\rightarrow H$ processes as evaluated in 
Ref.~\cite{anastasiouwebber}.  We treat the scale uncertainties
as 100\% correlated between jet bins and between CDF and D0, and 
also treat the PDF uncertainties in the cross section as correlated 
between jet bins and between CDF and D0.

We include all significant Higgs production modes in the high-mass 
search.   Besides gluon-gluon fusion through virtual quark loops 
(ggH), we include Higgs boson production in association with a $W$ 
or $Z$ vector boson  (VH)~\cite{nnlo2,Brein,Ciccolini}, and vector 
boson  fusion (VBF)~\cite{nnlo2,Berger}.  For the low-mass searches,
we target the $WH$, $ZH$, VBF, and $t{\bar{t}}H$~\cite{tth} production modes with specific searches,
including also those signal components not specifically targeted but which fall
in the acceptance nonetheless.
Our $WH$ and $ZH$ cross sections are from Ref.~\cite{djouadibaglio}.  
We include the $ggH$ production mode in our searches for Higgs bosons decaying to tau pairs and photon pairs.
In order to predict the 
distributions of the kinematics of Higgs boson signal events, CDF 
and D0 use the \textsc{PYTHIA}~\cite{pythia} Monte Carlo program, 
with \textsc{CTEQ5L} and \textsc{CTEQ6L}~\cite{cteq} leading-order 
(LO) parton distribution functions.  The Higgs boson decay branching 
ratio predictions are calculated with \textsc{HDECAY}~\cite{hdecay}, and are
also listed in Table~\ref{tab:higgsxsec}.  We use \textsc{HDECAY} Version 3.53.

For both CDF and D0, events from QCD multijet (instrumental) backgrounds
are measured in independent data samples using several different methods.
For CDF, backgrounds from SM processes with electroweak gauge bosons or 
top quarks were generated using \textsc{PYTHIA}, \textsc{ALPGEN}~\cite{alpgen}, 
\textsc{MC@NLO}~\cite{MC@NLO}, and \textsc{HERWIG}~\cite{herwig} programs. 
For D0, these backgrounds were generated using \textsc{PYTHIA}, 
\textsc{ALPGEN}, and \textsc{COMPHEP}~\cite{comphep}, with \textsc{PYTHIA} 
providing parton-showering and hadronization for all the generators.  These 
background processes were normalized using either experimental data or 
next-to-leading order calculations (including \textsc{MCFM}~\cite{mcfm} for 
the $W+$ heavy flavor process).

\begin{sidewaystable}
\begin{center}
\caption{
The production cross sections and decay branching fractions for the SM
Higgs boson assumed for the combination.}
\vspace{0.2cm}
\label{tab:higgsxsec}
\begin{small}
\begin{tabular}{|c|c|c|c|c|c|c|c|c|c|c|c|}\hline
$m_H$ & $\sigma_{gg\rightarrow H}$ & $\sigma_{WH}$ & $\sigma_{ZH}$ & $\sigma_{VBF}$ & $\sigma_{t{\bar{t}}H}$  &
$B(H\rightarrow b{\bar{b}})$ & $B(H\rightarrow c{\bar{c}})$ & $B(H\rightarrow \tau^+{\tau^-})$ & $B(H\rightarrow W^+W^-)$ & $B(H\rightarrow ZZ)$ & $B(H\rightarrow\gamma\gamma)$ \\ 
(GeV/$c^2$) & (fb) & (fb) & (fb) & (fb) & (fb) & (\%) & (\%) & (\%) & (\%) & (\%) & (\%) \\ \hline
\hline 
   100 &    1861   &  291.9    &  169.8      &   99.5  &  8.000   &  80.33 & 3.542   & 7.920 &  1.052 &  0.1071  & 0.1505  \\ 
   105 &    1618   &  248.4    &  145.9      &   93.3  &  7.062   &  78.57 & 3.463   & 7.821 &  2.307 &  0.2035  & 0.1689  \\
   110 &    1413   &  212.0    &  125.7      &   87.1  &  6.233   &  75.90 & 3.343   & 7.622 &  4.585 &  0.4160  & 0.1870  \\
   115 &    1240   &  181.9    &  108.9      &   79.07 &  5.502   &  71.95 & 3.169   & 7.288 &  8.268 &  0.8298  & 0.2029  \\
   120 &    1093   &  156.4    &  94.4       &   71.65 &  4.857   &  66.49 & 2.927   & 6.789 &  13.64 &  1.527   & 0.2148  \\
   125 &    967    &  135.1    &  82.3       &   67.37 &  4.279   &  59.48 & 2.617   & 6.120 &  20.78 &  2.549   & 0.2204  \\
   130 &    858    &  116.9    &  71.9       &   62.5  &  3.769   &  51.18 & 2.252   & 5.305 &  29.43 &  3.858   & 0.2182  \\
   135 &    764    &  101.5    &  63.0       &   57.65 &  3.320   &  42.15 & 1.854   & 4.400 &  39.10 &  5.319   & 0.2077  \\
   140 &    682    &  88.3     &  55.3       &   52.59 &  2.925   &  33.04 & 1.453   & 3.472 &  49.16 &  6.715   & 0.1897  \\
   145 &    611    &  77.0     &  48.7       &   49.15 &  2.593   &  24.45 & 1.075   & 2.585 &  59.15 &  7.771   & 0.1653  \\
   150 &    548    &  67.3     &  42.9       &   45.67 &  2.298   &  16.71 & 0.7345  & 1.778 &  68.91 &  8.143   & 0.1357  \\
   155 &    492    &  58.9     &  37.9       &   42.19 &  2.037   &  9.88  & 0.4341  & 1.057 &  78.92 &  7.297   & 0.09997  \\
   160 &    439    &  50.8     &  33.1       &   38.59 &  1.806   &  3.74  & 0.1646  & 0.403 &  90.48 &  4.185   & 0.05365  \\
   165 &    389    &  44.6     &  30.0       &   36.09 &  1.607   &  1.29  & 0.05667 & 0.140 &  95.91 &  2.216   & 0.02330  \\
   170 &    349    &  40.2     &  26.6       &   33.58 &  1.430   &  0.854 & 0.03753 & 0.093 &  96.39 &  2.351   & 0.01598  \\
   175 &    314    &  35.6     &  23.7       &   31.11 &  1.272   &  0.663 & 0.02910 & 0.073 &  95.81 &  3.204   & 0.01236  \\
   180 &    283    &  31.4     &  21.1       &   28.57 &  1.132   &  0.535 & 0.02349 & 0.059 &  93.25 &  5.937   & 0.01024  \\
   185 &    255    &  28.2     &  18.9       &   26.81 &  1.004   &  0.415 & 0.01823 & 0.046 &  84.50 &  14.86   & 0.008128  \\
   190 &    231    &  25.1     &  17.0       &   24.88 &  0.890   &  0.340 & 0.01490 & 0.038 &  78.70 &  20.77   & 0.006774  \\
   195 &    210    &  22.4     &  15.3       &   23    &  0.789   &  0.292 & 0.01281 & 0.033 &  75.88 &  23.66   & 0.005919  \\
   200 &    192    &  20.0     &  13.7       &   21.19 &  0.700   &  0.257 & 0.01128 & 0.029 &  74.26 &  25.33   & 0.005285  \\ \hline
\end{tabular}	       
\end{small}	       
\end{center}	       
\end{sidewaystable}    

Tables~\ref{tab:cdfacc} and~\ref{tab:dzacc} summarize, for CDF and D0 respectively,
the integrated luminosities, the Higgs boson mass ranges over which the searches are performed,
and references to further details for each analysis.


\begin{table}[h]
\caption{\label{tab:cdfacc}Luminosity, explored mass range and references
for the different processes
and final states ($\ell=e, \mu$) for the CDF analyses.  The labels ``$2\times$'' and ``$4\times$''
refer to separation into different lepton categories.
}
\begin{ruledtabular}
\begin{tabular}{lccc} \\
Channel & Luminosity (fb$^{-1}$) & $m_H$ range (GeV/$c^2$) & Reference \\ \hline
$WH\rightarrow \ell\nu b\bar{b}$ 2-jet channels\ \ \ 4$\times$(TDT,LDT,ST,LDTX)                                         & 5.7  & 100-150 & \cite{cdfWH2J} \\
$WH\rightarrow \ell\nu b\bar{b}$ 3-jet channels\ \ \ 2$\times$(TDT,LDT,ST)                                              & 5.6  & 100-150 & \cite{cdfWH3J} \\
$ZH\rightarrow \nu\bar{\nu} b\bar{b}$ \ \ \ (TDT,LDT,ST)                                                                & 5.7  & 100-150 & \cite{cdfmetbb} \\
$ZH\rightarrow \ell^+\ell^- b\bar{b}$ \ \ \ 4$\times$(TDT,LDT,ST)                                                       & 5.7  & 100-150 & \cite{cdfZHll1,cdfZHll2} \\
$H\rightarrow W^+ W^-$ \ \ \ 2$\times$(0,1 jets)+(2+ jets)+(low-$m_{\ell\ell}$)+($e$-$\tau_{had}$)+($\mu$-$\tau_{had}$) & 5.9  & 110-200 & \cite{cdfHWW} \\
$WH \rightarrow WW^+ W^-$ \ \ \ (same-sign leptons 1+ jets)+(tri-leptons)                                               & 5.9  & 110-200 & \cite{cdfHWW} \\
$ZH \rightarrow ZW^+ W^-$ \ \ \ (tri-leptons 1 jet)+(tri-leptons 2+ jets)                                               & 5.9  & 110-200 & \cite{cdfHWW} \\
$H$ + $X\rightarrow \tau^+ \tau^-$ \ \ \ (1 jet)+(2 jets)                                                               & 2.3  & 100-150 & \cite{cdfHtt} \\
$WH+ZH\rightarrow jjb{\bar{b}}$ \ \ \  2$\times$(TDT,LDT)                                                               & 4.0  & 100-150 & \cite{cdfjjbb} \\
$H \rightarrow \gamma \gamma$                                                                                           & 5.4  & 100-150 & \cite{cdfHgg} \\
\end{tabular}
\end{ruledtabular}
\end{table}

\vglue 0.5cm

\begin{table}[h]
\caption{\label{tab:dzacc}Luminosity, explored mass range and references
for the different processes
and final states ($\ell=e, \mu$) for the D0 analyses. 
Most analyses are in addition analyzed separately for RunIIa and IIb.
In some cases, not every sub-channel uses the same dataset, and a range of integrated
luminosities is given. 
}
\begin{ruledtabular}
\begin{tabular}{lccc} \\
Channel & Luminosity (fb$^{-1}$) & $m_H$ range (GeV/$c^2$) & Reference \\ \hline
$WH\rightarrow \ell\nu b\bar{b}$ \ \ \ (ST,DT,2,3 jet)             & 5.3  & 100-150 & \cite{dzWHl} \\
$VH\rightarrow \tau^+\tau^- b\bar{b}/q\bar{q} \tau^+\tau^-$ \ \ \      & 4.9  & 105-145 & \cite{dzVHt1,dzVHt2} \\
$ZH\rightarrow \nu\bar{\nu} b\bar{b}$ \ \ \ (ST,TLDT)   & 5.2-6.4  & 100-150 & \cite{dzZHv1,dzZHv2} \\
$ZH\rightarrow \ell^+\ell^- b\bar{b}$ \ \ \ (ST,DT,$ee$,$\mu\mu$,$ee_{ICR}$,$\mu\mu_{trk}$) & 4.2-6.2  & 100-150 & \cite{dzZHll2} \\
$VH \rightarrow \ell^\pm \ell^\pm\ + X $ \ \ \  & 5.3  & 115-200 & \cite{dzWWW} \\
$H\rightarrow W^+ W^- \rightarrow e^\pm\nu e^\mp\nu, \mu^\pm\nu \mu^\mp\nu$ \ \ \    & 5.4  & 115-200 & \cite{dzHWW1}\\
$H\rightarrow W^+ W^- \rightarrow e^\pm\nu \mu^\mp\nu$ \ \ \ (0,1,2+ jet)     & 6.7  & 115-200 & \cite{dzHWW2}\\
$H\rightarrow W^+ W^- \rightarrow \ell\bar{\nu} jj$      & 5.4  & 130-200 & \cite{dzHWWjj}\\
$H \rightarrow \gamma \gamma$                                 & 4.2  & 100-150 & \cite{dzHgg} \\
$t \bar{t} H \rightarrow t \bar{t} b \bar{b}$ \ \ \ (ST,DT,TT,4,5+~jets) & 2.1  & 105-155 & \cite{dzttH} \\
\end{tabular}
\end{ruledtabular}
\end{table}

\section{Distributions of Candidates} 

All analyses provide binned histograms of the final discriminant variables
for the signal and background predictions, itemized separately for each
source, and the observed data.
The number of channels combined is large, and the number of bins
in each channel is large.  Therefore, the task of assembling
histograms and checking whether the expected and observed limits are
consistent with the input predictions and observed data is difficult.
We therefore provide histograms that aggregate all channels' signal,
background, and data together.  In order to preserve most of the
sensitivity gain that is achieved by the analyses by binning the data
instead of collecting them all together and counting, we aggregate the
data and predictions in narrow bins of signal-to-background ratio,
$s/b$.  Data with similar $s/b$ may be added together with no loss in
sensitivity, assuming similar systematic errors on the predictions.
The aggregate histograms do not show the effects of systematic
uncertainties, but instead compare the data with the central
predictions supplied by each analysis.

The range of $s/b$ is quite large in each analysis, and so
$\log_{10}(s/b)$ is chosen as the plotting variable.  Plots of the
distributions of $\log_{10}(s/b)$ are shown for Higgs boson masses 
of 100, 115, 150, and 165~GeV/$c^2$ in Figure~\ref{fig:lnsb}.  These 
distributions can be integrated from the high-$s/b$ side downwards, 
showing the sums of signal, background, and data for the most pure 
portions of the selection of all channels added together.  These 
integrals can be seen in Figure~\ref{fig:integ}.  The most significant 
candidates are found in the bins with the highest $s/b$; an excess 
in these bins relative to the background prediction drives the Higgs 
boson cross section limit upwards, while a deficit drives it downwards.  
The lower-$s/b$ bins show that the modeling of the rates and kinematic 
distributions of the backgrounds is very good.  The integrated plots
show the excess of events in the highest-$s/b$ bins for the analyses 
seeking a Higgs boson mass of 115~GeV/$c^2$, and a deficit of events 
in the highest-$s/b$ bins for the analyses seeking a Higgs boson of 
mass 165~GeV/$c^2$.

We also show the distributions of the data after subtracting the 
expected background, and compare that with the expected signal yield 
for a Standard Model Higgs boson, after collecting all bins in all 
channels sorted by $s/b$.  These background-subtracted distributions
are shown in Figure~\ref{fig:bgsub}.  These graphs also show the 
remaining uncertainty on the background prediction after fitting the 
background model to the data within the systematic uncertainties on 
the rates and shapes in each contributing channel's templates.

 \begin{figure}[t]
 \begin{centering}
 \includegraphics[width=0.4\textwidth]{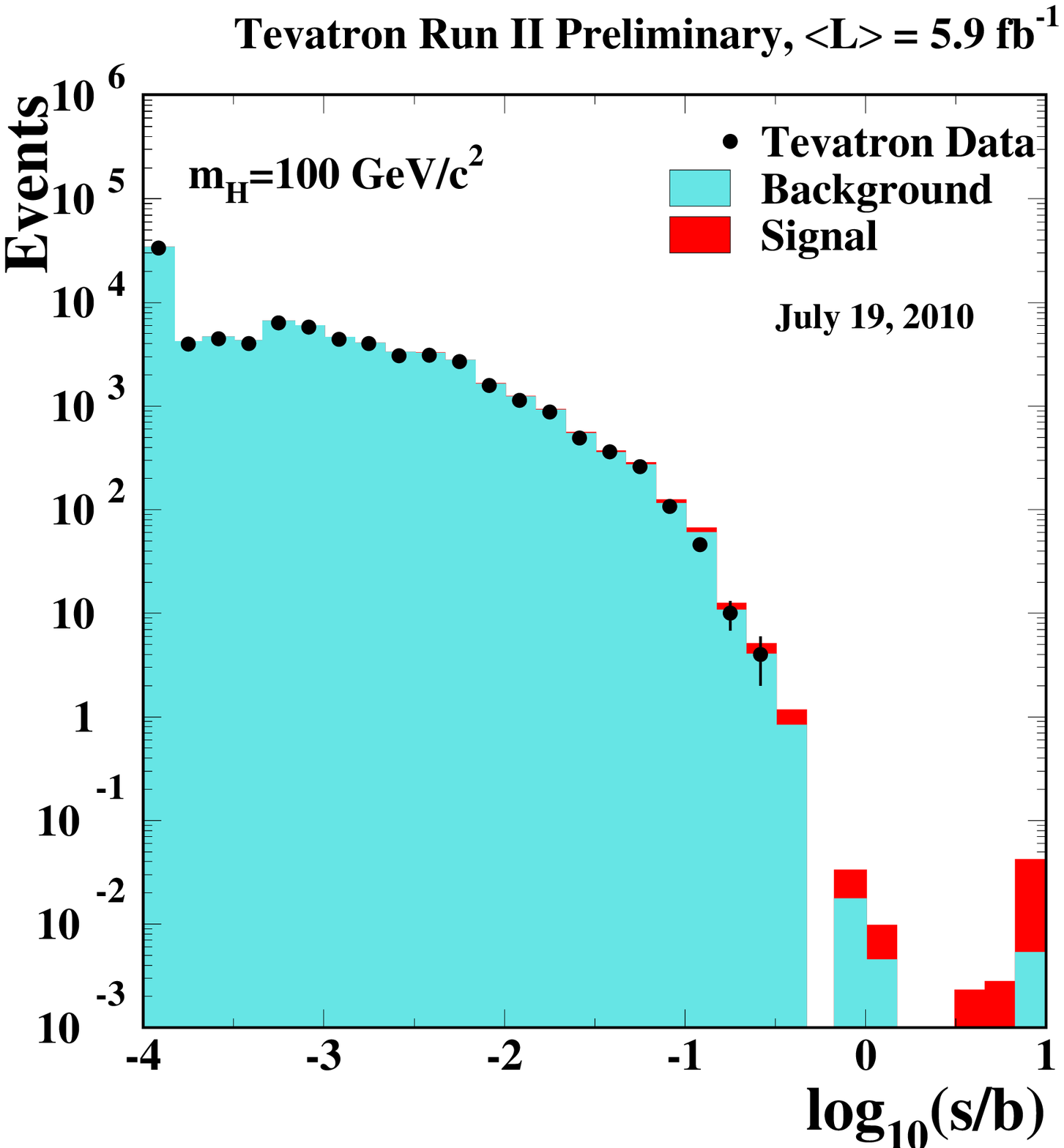}\includegraphics[width=0.4\textwidth]{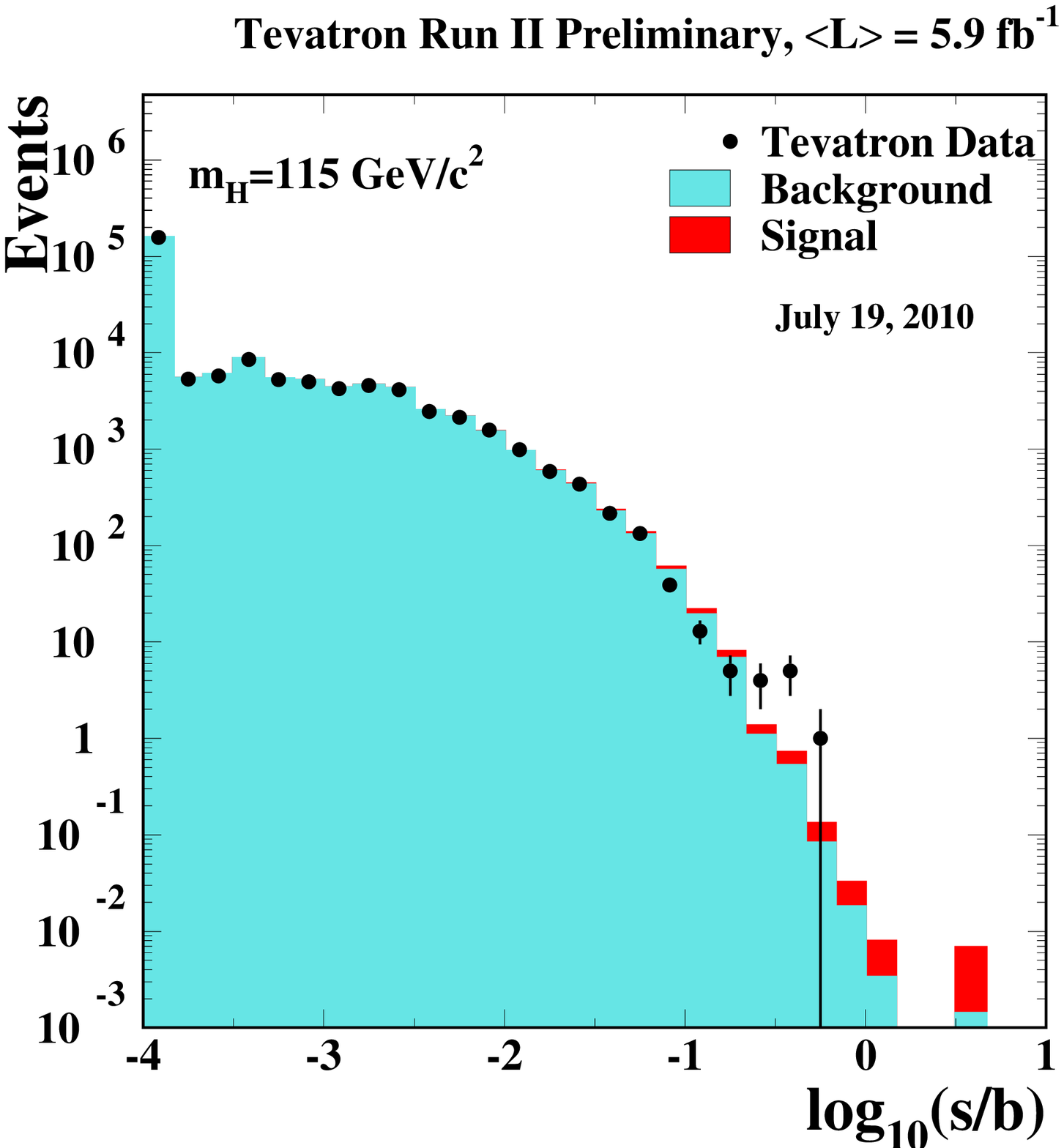} \\
 \includegraphics[width=0.4\textwidth]{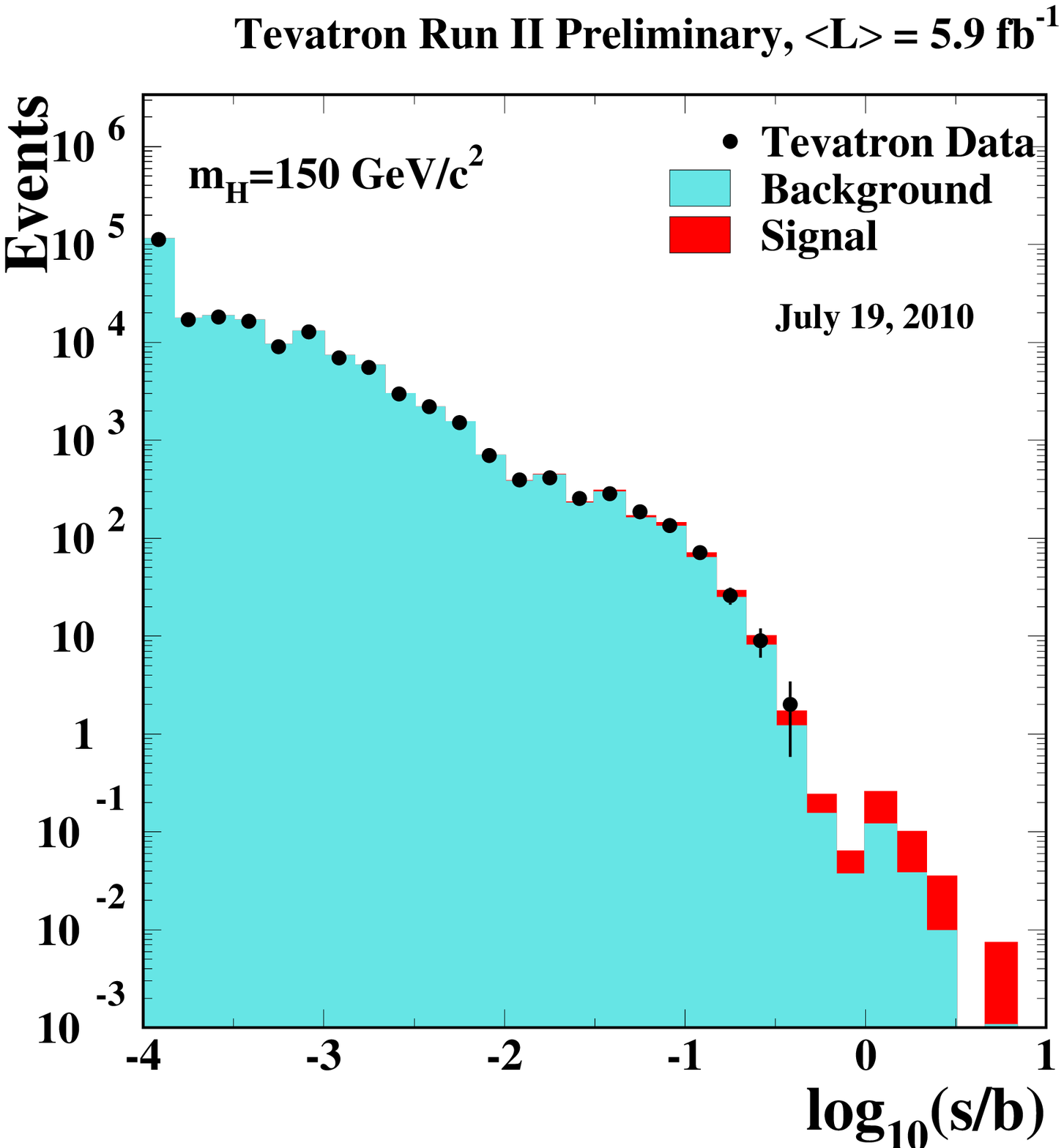}\includegraphics[width=0.4\textwidth]{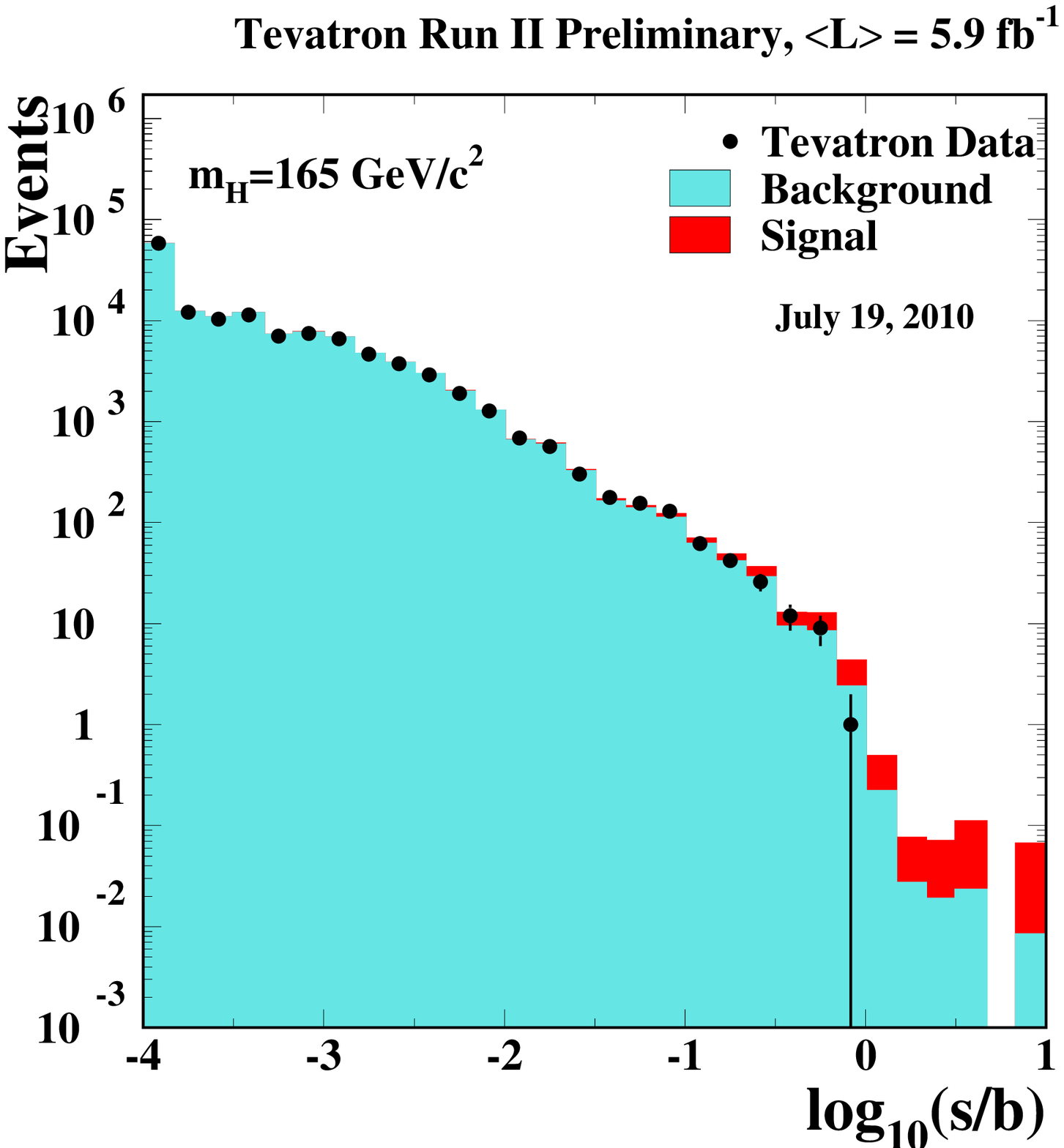}
 \caption{
 \label{fig:lnsb} Distributions of $\log_{10}(s/b)$, for the data from all contributing channels from
CDF and D0, for Higgs boson masses of 100, 115, 150, and 165~GeV/$c^2$.  The
data are shown with points, and the expected signal is shown stacked on top of
the backgrounds.  Underflows and overflows are collected into the
bottom and top bins. }
 \end{centering}
 \end{figure}

 \begin{figure}[t]
 \begin{centering}
 \includegraphics[width=0.4\textwidth]{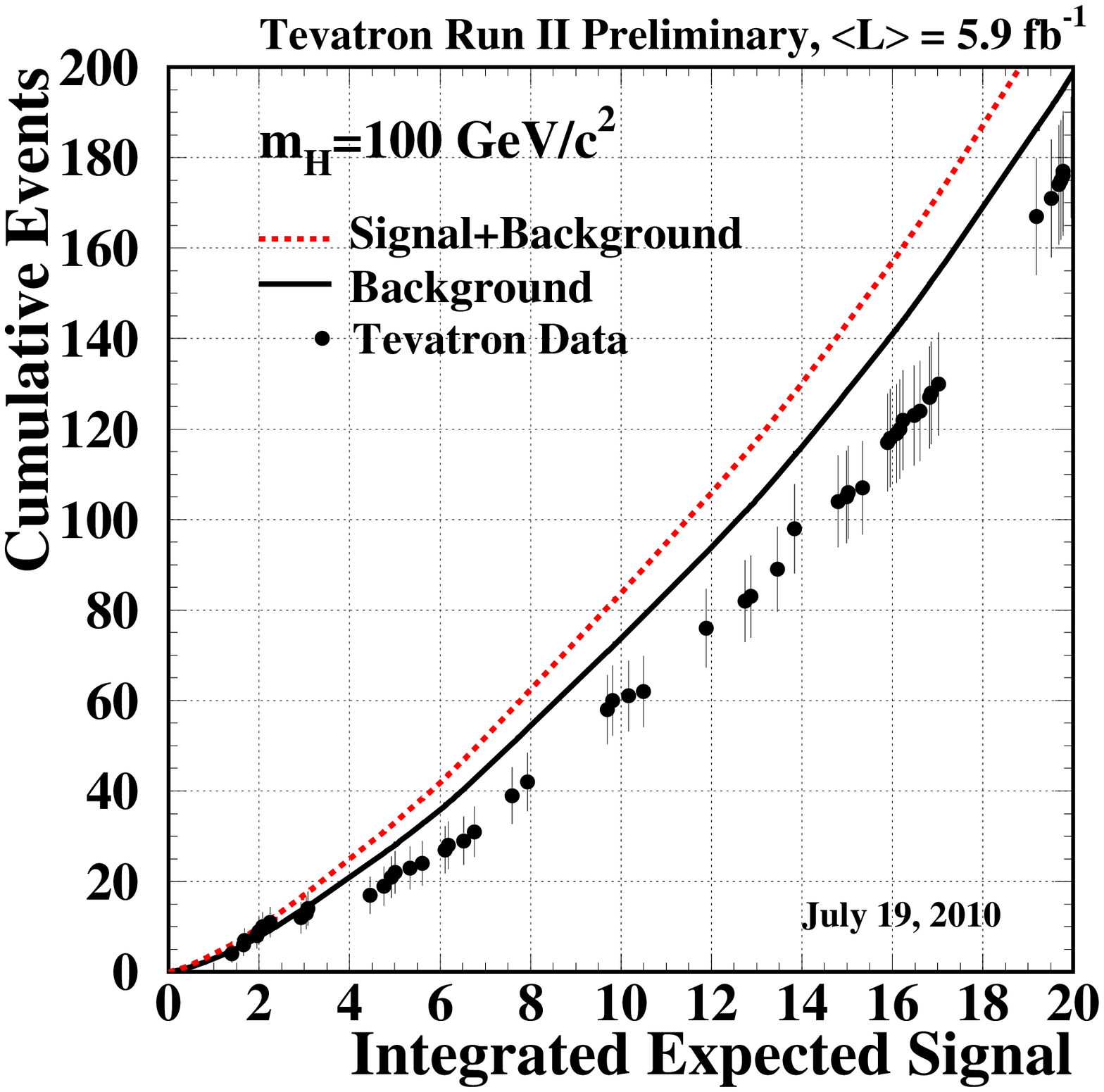}\includegraphics[width=0.4\textwidth]{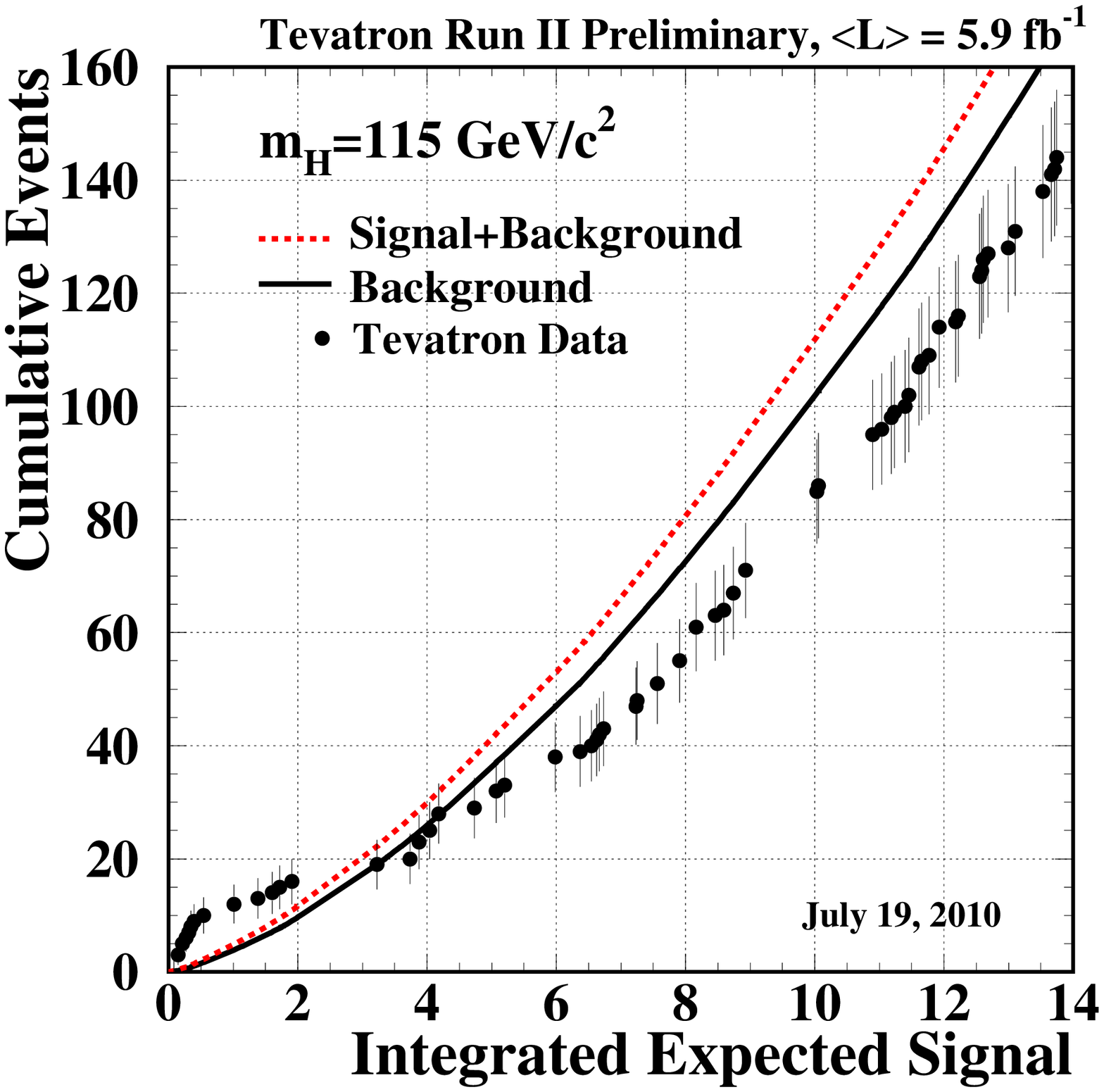} \\
 \includegraphics[width=0.4\textwidth]{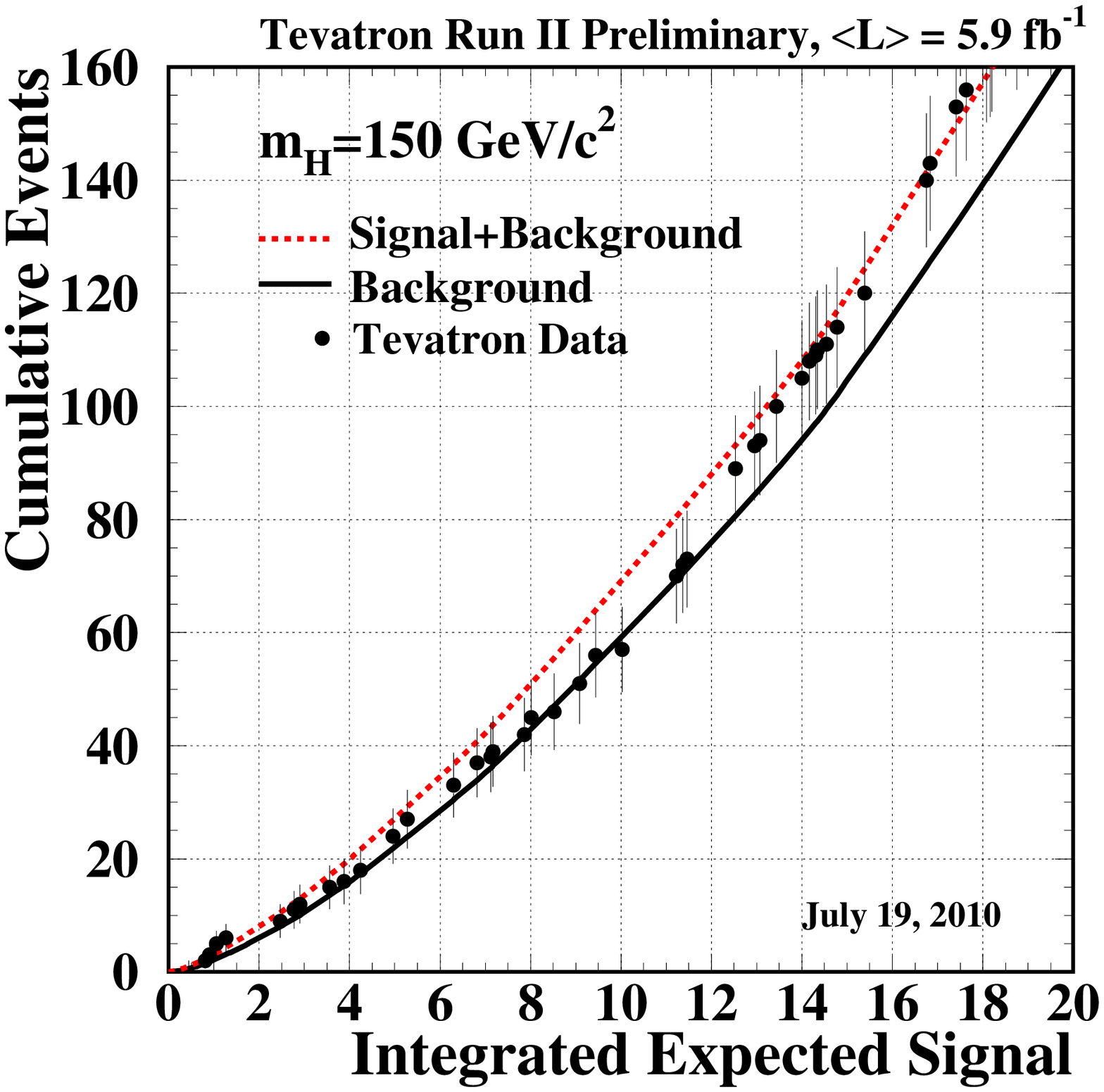}\includegraphics[width=0.4\textwidth]{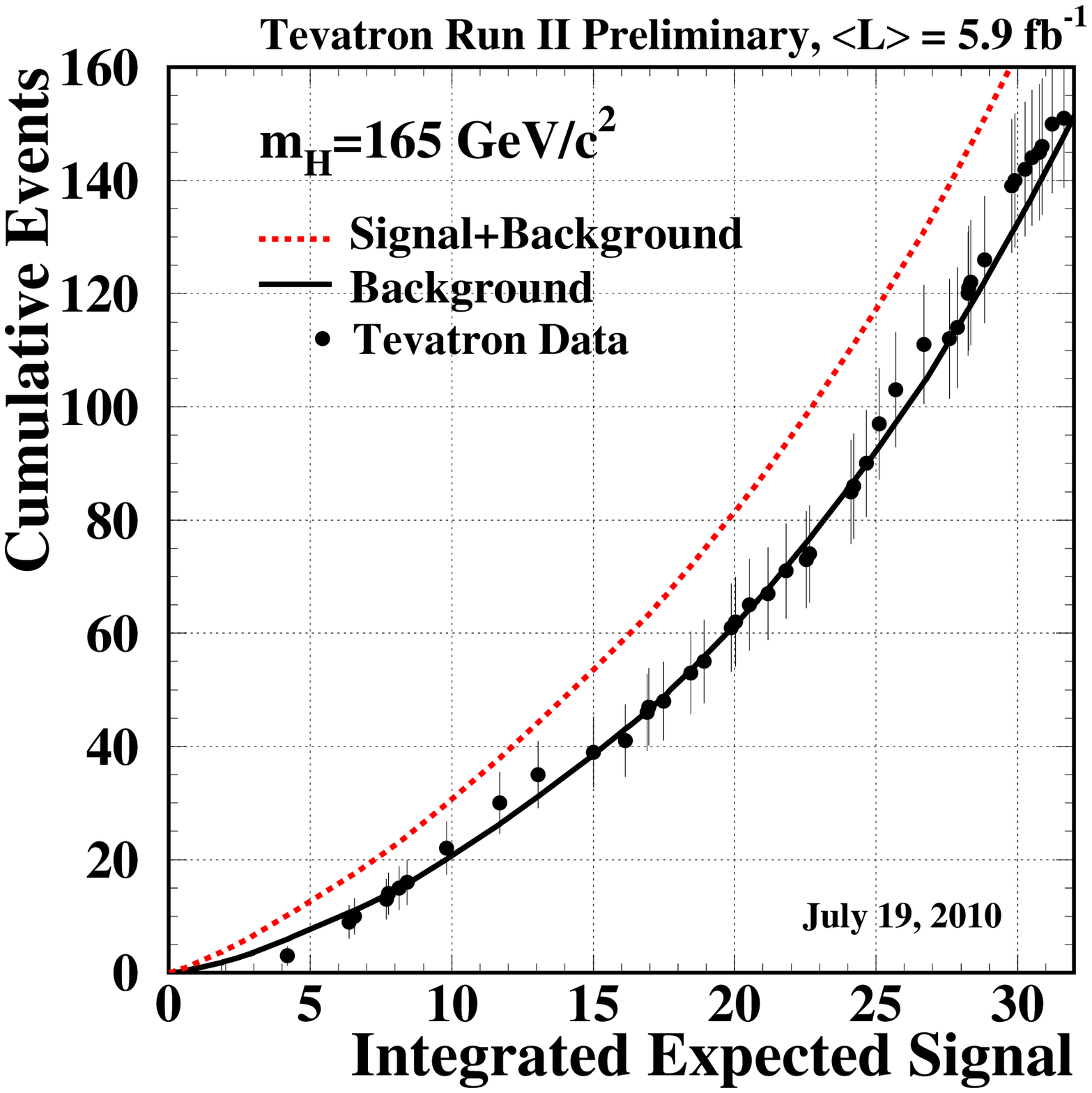}
 \caption{
 \label{fig:integ} Integrated distributions of $s/b$, starting at the high $s/b$ side, for Higgs boson
masses of 100, 115, 150, and 165~GeV/$c^2$.  The total signal+background
and background-only integrals are shown separately, along with the
data sums.  Data are only shown for bins that have
data events in them.}
 \end{centering}
 \end{figure}

 \begin{figure}[t]
 \begin{centering}
 \includegraphics[width=0.45\textwidth]{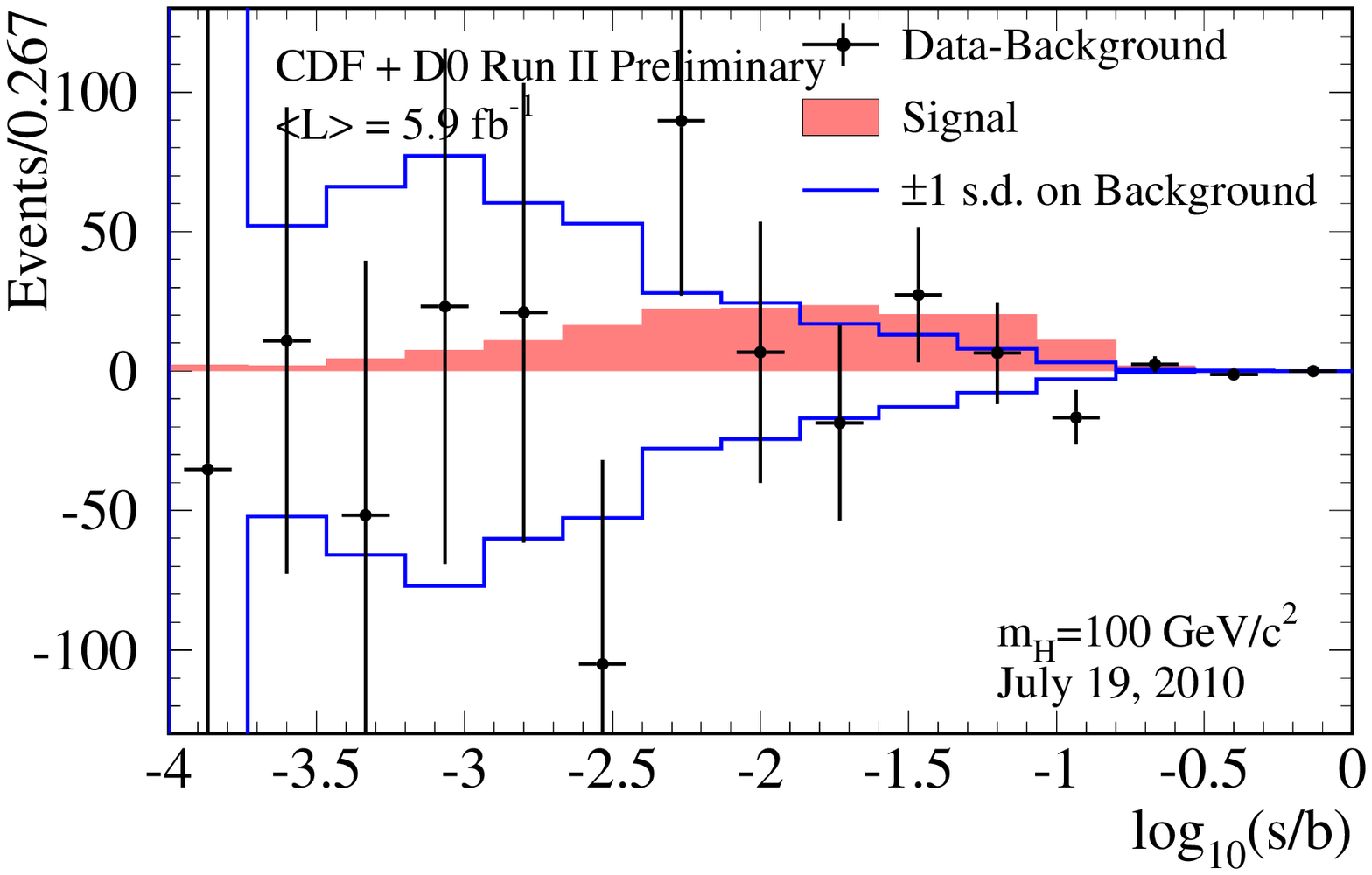}\includegraphics[width=0.45\textwidth]{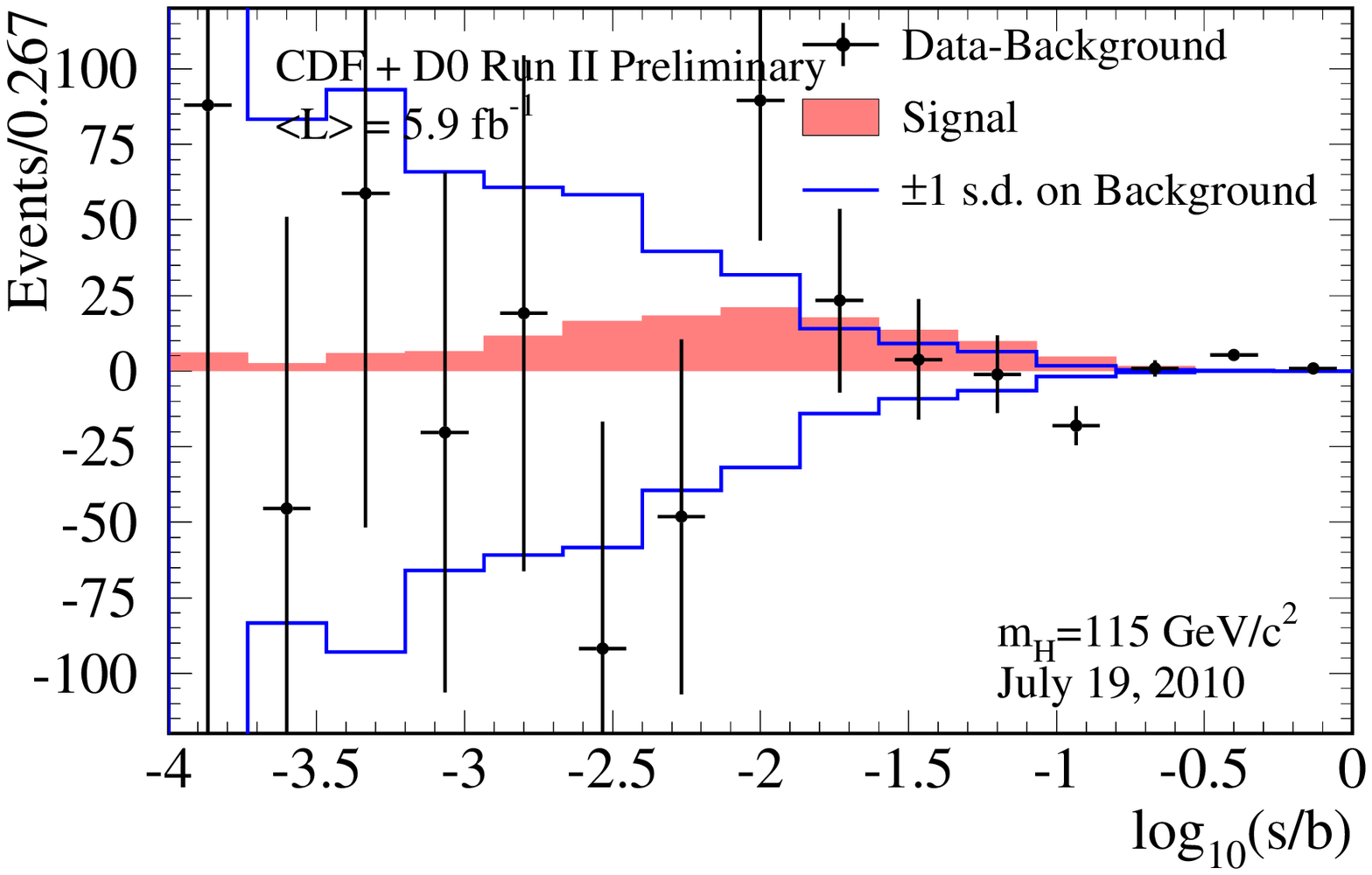} \\
 \includegraphics[width=0.45\textwidth]{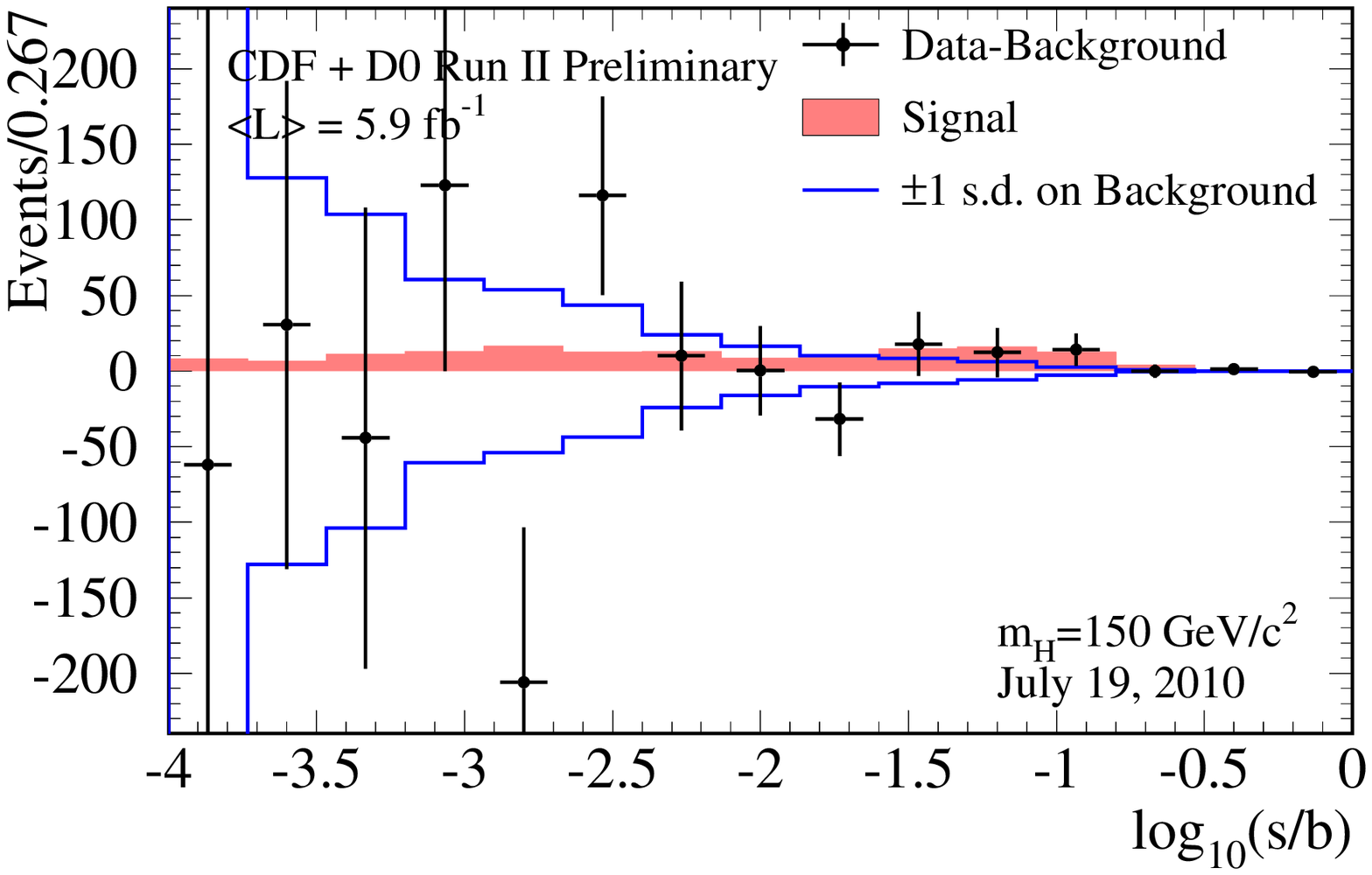}\includegraphics[width=0.45\textwidth]{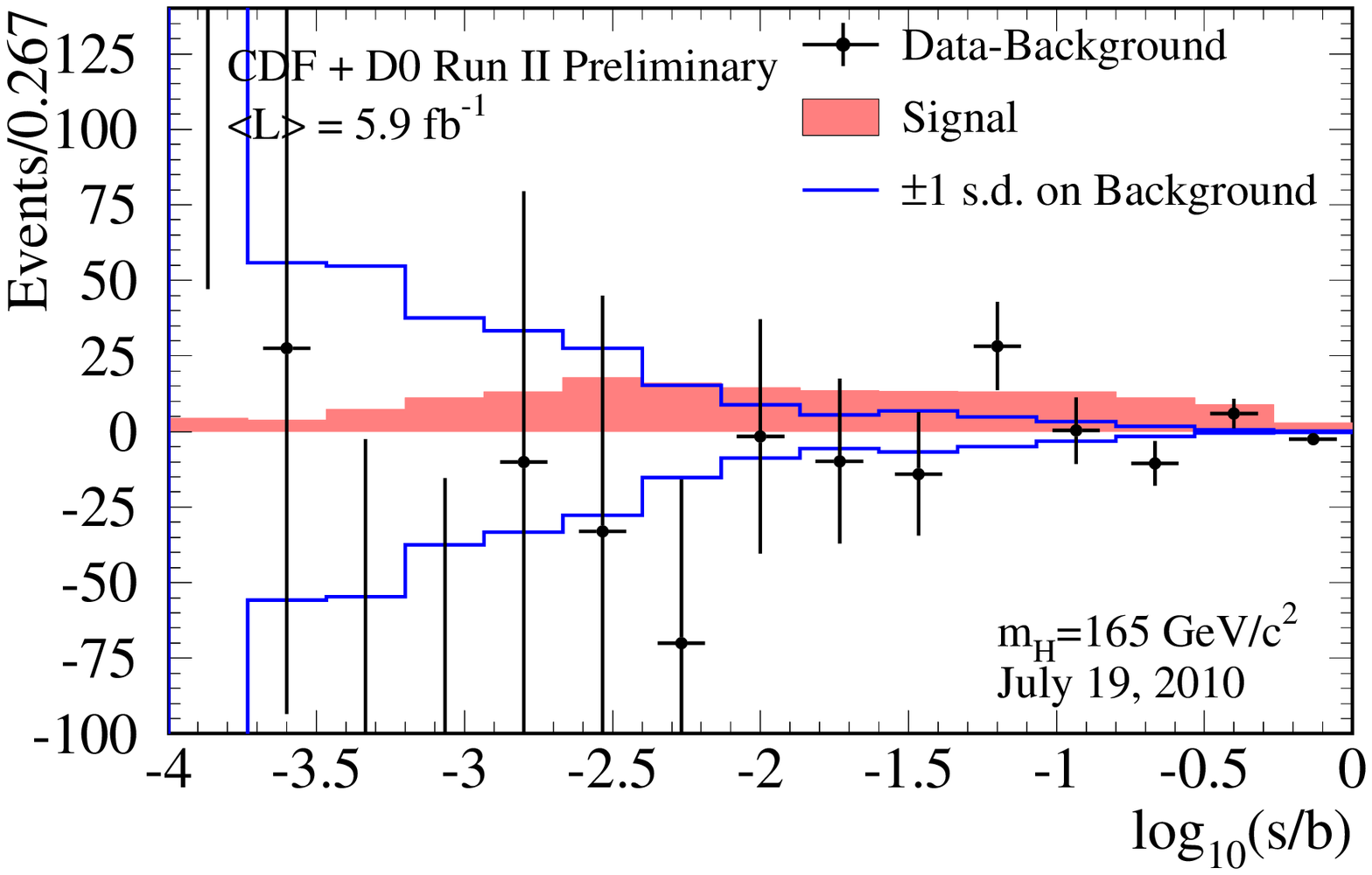}
 \caption{
 \label{fig:bgsub} Background-subtracted data distributions for all channels, summed in bins of $s/b$,
for Higgs boson masses of 100, 115, 150, and 165~GeV/$c^2$.  The background has been fit, within its systematic
uncertainties, to the data.  The points with error bars indicate the background-subtracted data; the sizes of the
error bars are the square roots of the predicted background in each bin.  The unshaded (blue-outline) histogram shows the systematic
uncertainty on the best-fit background model, and the shaded histogram shows the expected signal for a Standard
Model Higgs boson.}
 \end{centering}
 \end{figure}

\section{Combining Channels} 

To gain confidence that the final result does not depend on the
details of the statistical formulation,
we perform two types of combinations, using
Bayesian and  Modified Frequentist approaches, which yield limits on the Higgs boson
production rate that agree within 10\% at each
value of $m_H$, and within 1\% on average.
Both methods rely on distributions in the final discriminants, and not just on
their single integrated values.  Systematic uncertainties enter on the
predicted number of signal and background events as well
as on the distribution of the discriminants in
each analysis (``shape uncertainties'').
Both methods use likelihood calculations based on Poisson
probabilities.

\subsection{Bayesian Method}

Because there is no experimental information on the production cross section for
the Higgs boson, in the Bayesian technique~\cite{CDFHiggs} we assign a flat prior
for the total number of selected Higgs events.  For a given Higgs boson mass, the
combined likelihood is a product of likelihoods for the individual
channels, each of which is a product over histogram bins:

\begin{equation}
{\cal{L}}(R,{\vec{s}},{\vec{b}}|{\vec{n}},{\vec{\theta}})\times\pi({\vec{\theta}})
= \prod_{i=1}^{N_C}\prod_{j=1}^{N_b} \mu_{ij}^{n_{ij}} e^{-\mu_{ij}}/n_{ij}!
\times\prod_{k=1}^{n_{np}}e^{-\theta_k^2/2}
\end{equation}

\noindent where the first product is over the number of channels
($N_C$), and the second product is over $N_b$ histogram bins containing
$n_{ij}$ events, binned in  ranges of the final discriminants used for
individual analyses, such as the dijet mass, neural-network outputs,
or matrix-element likelihoods.
 The parameters that contribute to the
expected bin contents are $\mu_{ij} =R \times s_{ij}({\vec{\theta}}) + b_{ij}({\vec{\theta}})$
for the
channel $i$ and the histogram bin $j$, where $s_{ij}$ and $b_{ij}$
represent the expected background and signal in the bin, and $R$ is a scaling factor
applied to the signal to test the sensitivity level of the experiment.
Truncated Gaussian priors are used for each of the nuisance parameters
$\theta_k$, which define
the
sensitivity of the predicted signal and background estimates to systematic uncertainties.
These
can take the form of uncertainties on overall rates, as well as the shapes of the distributions
used for combination.   These systematic uncertainties can be far larger
than the expected SM Higgs boson signal, and are therefore important in the calculation of limits.
The truncation
is applied so that no prediction of any signal or background in any bin is negative.
The posterior density function is
then integrated over all parameters (including correlations) except for $R$,
and a 95\% credibility level upper limit on $R$ is estimated
by calculating the value of $R$ that corresponds to 95\% of the area
of the resulting distribution.

\subsection{Modified Frequentist Method}

The Modified Frequentist technique relies on the ${\rm CL}_{\rm s}$ method, using
a log-likelihood ratio (LLR) as test statistic~\cite{DZHiggs}:
\begin{equation}
LLR = -2\ln\frac{p({\mathrm{data}}|H_1)}{p({\mathrm{data}}|H_0)},
\end{equation}
where $H_1$ denotes the test hypothesis, which admits the presence of
SM backgrounds and a Higgs boson signal, while $H_0$ is the null
hypothesis, for only SM backgrounds.  The probabilities $p$ are
computed using the best-fit values of the nuisance parameters for each
pseudo-experiment, separately for each of the two hypotheses, and include the
Poisson probabilities of observing the data multiplied by Gaussian
priors for the values of the nuisance parameters.  This technique
extends the LEP procedure~\cite{pdgstats} which does not involve a
fit, in order to yield better sensitivity when expected signals are
small and systematic uncertainties on backgrounds are
large~\cite{pflh}.

The ${\rm CL}_{\rm s}$ technique involves computing two $p$-values, ${\rm CL}_{\rm s+b}$ and ${\rm CL}_{\rm b}$.
The latter is defined by
\begin{equation}
1-{\rm CL}_{\rm b} = p(LLR\le LLR_{\mathrm{obs}} | H_0),
\end{equation}
where $LLR_{\mathrm{obs}}$ is the value of the test statistic computed for the
data. $1-{\rm CL}_{\rm b}$ is the probability of observing a signal-plus-background-like outcome
without the presence of signal, i.e. the probability
that an upward fluctuation of the background provides  a signal-plus-background-like
response as observed in data.
The other $p$-value is defined by
\begin{equation}
{\rm CL}_{\rm s+b} = p(LLR\ge LLR_{\mathrm{obs}} | H_1),
\end{equation}
and this corresponds to the probability of a downward fluctuation of the sum
of signal and background in
the data.  A small value of ${\rm CL}_{\rm s+b}$ reflects inconsistency with  $H_1$.
It is also possible to have a downward fluctuation in data even in the absence of
any signal, and a small value of ${\rm CL}_{\rm s+b}$ is possible even if the expected signal is
so small that it cannot be tested with the experiment.  To minimize the possibility
of  excluding  a signal to which there is insufficient sensitivity
(an outcome  expected 5\% of the time at the 95\% C.L., for full coverage),
we use the quantity ${\rm CL}_{\rm s}={\rm CL}_{\rm s+b}/{\rm CL}_{\rm b}$.  If ${\rm CL}_{\rm s}<0.05$ for a particular choice
of $H_1$, that hypothesis is deemed to be excluded at the 95\% C.L. In an analogous 
way, the expected ${\rm CL}_{\rm b}$, ${\rm CL}_{\rm s+b}$ and ${\rm CL}_{\rm s}$ values are computed from the median of the 
LLR distribution for the background-only hypothesis.

Systematic uncertainties are included  by fluctuating the predictions for
signal and background rates in each bin of each histogram in a correlated way when
generating the pseudo-experiments used to compute ${\rm CL}_{\rm s+b}$ and ${\rm CL}_{\rm b}$.

\subsection{Systematic Uncertainties} 

Systematic uncertainties differ
between experiments and analyses, and they affect the rates and shapes of the predicted
signal and background in correlated ways.  The combined results incorporate
the sensitivity of predictions to  values of nuisance parameters,
and include correlations between rates and shapes, between signals and backgrounds,
and between channels within experiments and between experiments.
More on these issues can be found in the
individual analysis notes~\cite{cdfWH2J} through~\cite{dzttH}.  Here we
consider only the largest contributions and correlations between and
within the two experiments.

\subsubsection{Correlated Systematics between CDF and D0}

The uncertainties on the measurements of the integrated luminosities are 6\%
(CDF) and 6.1\% (D0).
Of these values, 4\% arises from the uncertainty
on the inelastic \pp~scattering cross section, which is correlated
between CDF and D0.
CDF and D0 also share the assumed values and uncertainties on the production cross sections
for top-quark processes (\ttbar~and single top) and for electroweak processes
($WW$, $WZ$, and $ZZ$).  In order to provide a consistent combination, the values of these
cross sections assumed in each analysis are brought into agreement.  We use
$\sigma_{t\bar{t}}=7.04^{+0.24}_{-0.36}~{\rm (scale)}\pm 0.14{\rm (PDF)}\pm 0.30{\rm (mass)}$, 
following the calculation of Moch and Uwer~\cite{mochuwer}, assuming
a top quark mass $m_t=173.0\pm 1.2$~GeV/$c^2$~\cite{tevtop09},
and using the MSTW2008nnlo PDF set~\cite{mstw2008}.  Other
calculations of $\sigma_{t\bar{t}}$ are similar~\cite{otherttbar}.  

For single top, we use the NLL $t$-channel calculation of Kidonakis~\cite{kid1}, 
which has been updated using the MSTW2008nnlo PDF set~\cite{mstw2008}~\cite{kidprivcomm}.
For the $s$-channel process we use~\cite{kid2}, again based on the MSTW2008nnlo PDF set.
Both of the cross section values below are the sum of the single $t$ and single ${\bar{t}}$
cross sections, and both assume $m_t=173\pm 1.2$ GeV.
\begin{equation}
\sigma_{t-{\rm{chan}}} = 2.10\pm 0.027 {\rm{(scale)}} \pm 0.18 {\rm{(PDF)}}  \pm 0.045 {\rm{(mass)}}  {\rm {pb}}. 
\end{equation}
\begin{equation}
\sigma_{s-{\rm{chan}}} = 1.046\pm 0.006 {\rm{(scale)}} \pm 0.059~{\rm{(PDF)}}  \pm 0.030~{\rm{(mass)}}~{\rm {pb}}. 
\end{equation}
Other calculations of $\sigma_{\rm{SingleTop}}$ are
similar for our purposes~\cite{harris}.  

MCFM~\cite{mcfm} has been used to compute the NLO cross sections for $WW$, $WZ$, 
and $ZZ$ production~\cite{dibo}.  Using a scale choice $\mu_0=M_V^2+p_T^2(V)$ and 
the MSTW2008 PDF set~\cite{mstw2008}, the cross section for inclusive $W^+W^-$ 
production is
\begin{equation}
\sigma_{W^+W^-} = 11.34^{+0.56}_{-0.49}~{\rm{(scale)}}~^{+0.35}_{-0.28} {\rm(PDF)} {\rm{pb}} 
\end{equation}
and the cross section for inclusive $W^\pm Z$ production is
\begin{equation}
\sigma_{W^\pm Z} = 3.22^{+0.20}_{-0.17}~{\rm{(scale)}}~^{+0.11}_{-0.08}~{\rm(PDF)}~{\rm{pb}} 
\end{equation}
For the $Z$, leptonic decays are used in the definition, with both $\gamma$ 
and $Z$ exchange.  The cross section quoted above involves the requirement 
$75\leq m_{\ell^+\ell^-}\leq 105$~GeV for the leptons from the neutral current 
exchange.  The same dilepton invariant mass requirement is applied to both 
sets of leptons in determining the $ZZ$ cross section which is
\begin{equation}
\sigma_{ZZ} = 1.20^{+0.05}_{-0.04}~{\rm{(scale)}}~^{+0.04}_{-0.03}~{\rm(PDF)}~{\rm{pb}} 
\end{equation}
For the diboson cross section calculations, $|\eta_{\ell}|<5$ for all calculations.  
Loosening this requirement to include all leptons leads to $\sim$+0.4\% change in 
the predictions.  Lowering the factorization and renormalization scales by a factor 
of two increases the cross section, and raising the scales by a factor of two 
decreases the cross section.  The PDF uncertainty has the same fractional impact on 
the predicted cross section independent of the scale choice.  All PDF uncertainties 
are computed as the quadrature sum of the twenty 68\% C.L. eigenvectors provided with 
MSTW2008 (MSTW2008nlo68cl).

In many analyses, the dominant background yields are calibrated with data control 
samples.  Since the methods of measuring the multijet (``QCD'') backgrounds differ 
between CDF and D0, and even between analyses within the collaborations, there is 
no correlation assumed between these rates.  Similarly, the large uncertainties on 
the background rates for $W$+heavy flavor (HF) and $Z$+heavy flavor are considered 
at this time to be uncorrelated, as both CDF and D0 estimate these rates using data 
control samples, but employ different techniques.  The calibrations of fake leptons, 
unvetoed $\gamma\rightarrow e^+e^-$ conversions, $b$-tag efficiencies and mistag 
rates are performed by each collaboration using independent data samples and 
methods, and are therefore also treated as uncorrelated.

\subsubsection{Correlated Systematic Uncertainties for CDF}
The dominant systematic uncertainties for the CDF analyses are shown
in the Appendix in Tables~\ref{tab:cdfsystwh2jet} and~\ref{tab:cdfsystwh3jet} for the 
\WH\ channels, in Table~\ref{tab:cdfvvbb1} for the $WH,ZH\rightarrow\MET 
b{\bar{b}}$ channels, in Tables~\ref{tab:cdfllbb1} and~\ref{tab:cdfllbb2} 
for the $ZH\rightarrow\ell^+\ell^-b{\bar{b}}$ channels, in 
Tables~\ref{tab:cdfsystww0}, \ref{tab:cdfsystww4}, and~\ref{tab:cdfsystww5} 
for the $H \rightarrow W^+W^-\rightarrow \ell^{\prime \pm}\nu \ell^{\prime 
\mp}\nu$ channels, in Table~\ref{tab:cdfsystwww} for the $WH \rightarrow 
WWW \rightarrow\ell^{\prime \pm}\ell^{\prime \pm}$ and $WH\rightarrow 
WWW \rightarrow \ell^{\pm}\ell^{\prime \pm} \ell^{\prime \prime \mp}$ 
channels, in Table~\ref{tab:cdfsystzww} for the $ZH \rightarrow ZWW 
\rightarrow \ell^{\pm}\ell^{\mp} \ell^{\prime \pm}$ channels, in 
Table~\ref{tab:cdfsysttautau} for the $H \rightarrow \tau^+\tau^-$ 
channels, in Table~\ref{tab:cdfallhadsyst} for the $WH/ZH$ and VBF 
$\rightarrow jjb{\bar{b}}$ channels, and in Table~\ref{tab:cdfsystgg} 
for the $H \rightarrow \gamma \gamma$ channel.  Each source induces a 
correlated uncertainty across all CDF channels' signal and background 
contributions which are sensitive to that source.  For \hbb, the largest 
uncertainties on signal arise from measured $b$-tagging efficiencies,
jet energy scale, and other Monte Carlo modeling.  Shape dependencies of 
templates on jet energy scale, $b$-tagging, and gluon radiation (``ISR'' 
and ``FSR'') are taken into account for some analyses (see tables).  
For \hww, the largest uncertainties on signal acceptance originate from 
Monte Carlo modeling.  Uncertainties on background event rates vary 
significantly for the different processes.  The backgrounds with the 
largest systematic uncertainties are in general quite small. Such 
uncertainties are constrained by fits to the nuisance parameters, and 
they do not affect the result significantly.  Because the largest 
background contributions are measured using data, these uncertainties 
are treated as uncorrelated for the \hbb~channels.  The differences in 
the resulting limits when treating the remaining uncertainties as either 
correlated or uncorrelated, is less than $5\%$.

\subsubsection{Correlated Systematic Uncertainties for D0 }
The dominant systematic uncertainties for the D0 analyses are shown in the Appendix, in Tables~\ref{tab:d0systwh1}, 
\ref{tab:d0vvbb}, \ref{tab:d0llbb1}, \ref{tab:d0systww}, \ref{tab:d0systwww}, \ref{tab:d0lvjj}, 
\ref{tab:d0systtth}, and \ref{tab:d0systgg}.  Each source induces a correlated uncertainty 
across all D0 channels sensitive to that source. Wherever appropriate the impact of systematic 
effects on both the rate and shape of the predicted signal and background is included.  For 
the low mass, \hbb~analyses, the largest sources of uncertainty originate from the measured 
$b$-tagging rate, the determination of the jet energy scale, simulated acceptances, jet 
resolution,  normalization of the W and Z plus heavy flavor backgrounds, and determination of 
the multijet background contribution.  For the \hww and \vww analyses, a significant source 
of uncertainty is the measured efficiencies for selecting leptons.  Significant sources for 
all analyses are the uncertainties on the luminosity and the cross sections for the simulated 
backgrounds.  All systematic uncertainties arising from the same source are taken to be 
correlated among the different backgrounds and between signal and background.

 \begin{figure}[t]
 \begin{centering}
 \includegraphics[width=14.0cm]{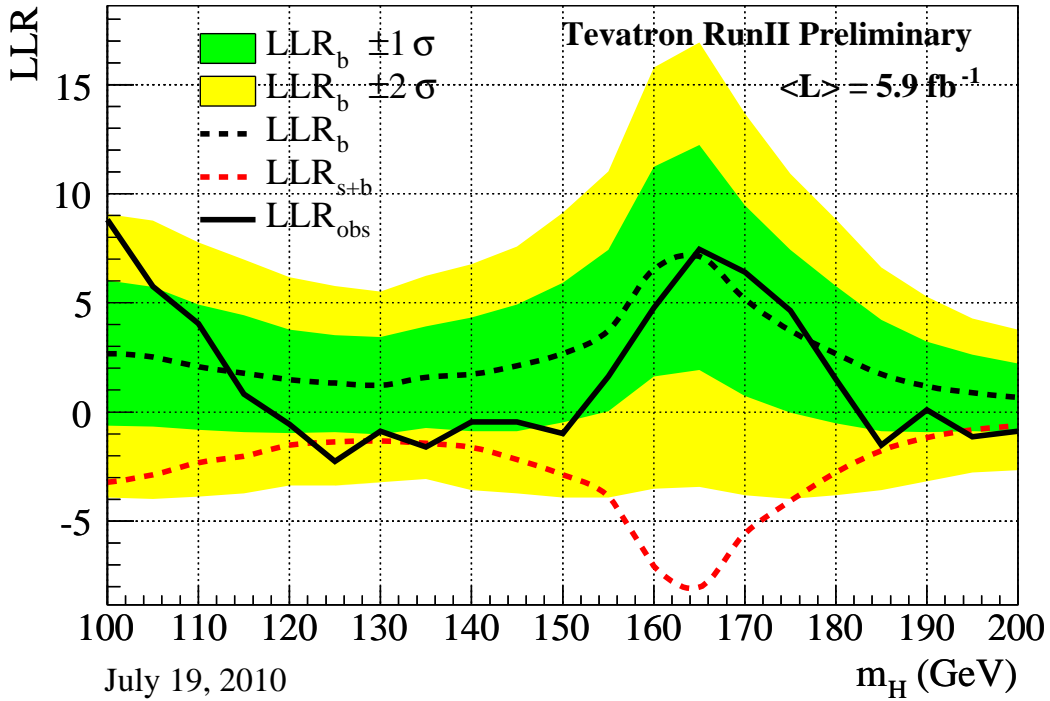}
 \caption{
 \label{fig:comboLLR} {
Distributions of the log-likelihood ratio (LLR) as a function of Higgs mass obtained with the ${\rm CL}_{\rm s}$ method for the combination of all CDF and D0 analyses.
}}
 \end{centering}
 \end{figure}

\vspace*{1cm}
\section{Combined Results} 

Before extracting the combined limits we study the distributions of the
log-likelihood ratio (LLR) for different hypotheses, to quantify the expected
sensitivity across the mass range tested.
Figure~\ref{fig:comboLLR} displays the LLR distributions for the combined
analyses as functions of $m_{H}$. Included are the median of the LLR distributions for the
background-only hypothesis (LLR$_{b}$), the signal-plus-background
hypothesis (LLR$_{s+b}$), and the observed value for the data (LLR$_{\rm{obs}}$).  The
shaded bands represent the one and two standard deviation ($\sigma$)
departures for LLR$_{b}$ centered on the median. Table~\ref{tab:llrVals} lists the observed
and expected LLR values shown in Figure~\ref{fig:comboLLR}.

\begin{table}[htpb]
\caption{\label{tab:llrVals} Log-likelihood ratio (LLR) values for the combined CDF + \Dzero Higgs boson search obtained using the {\rm CL}$_{S}$ method.}
\begin{ruledtabular}
\begin{tabular}{lccccccc}
$m_{H}$ (GeV/$c^2$ &  LLR$_{\rm{obs}}$ & LLR$_{S+B}^{\rm{med}}$ &
LLR$_{B}^{-2\sigma}$ & LLR$_{B}^{-1\sigma}$ & LLR$_{B}^{\rm{med}}$ &  LLR$_{B}^{+1\sigma}$ & LLR$_{B}^{+2\sigma}$ \\
\hline
100 & 8.81 & -3.23 & 9.07 & 6.03 & 2.67 & -0.62 & -3.92 \\ 
105 & 5.75 & -2.88 & 8.78 & 5.72 & 2.52 & -0.68 & -3.98 \\ 
110 & 4.06 & -2.33 & 7.78 & 4.92 & 2.08 & -0.82 & -3.88 \\ 
115 & 0.83 & -2.02 & 6.97 & 4.42 & 1.77 & -0.93 & -3.73 \\ 
120 & -0.56 & -1.52 & 6.17 & 3.77 & 1.48 & -0.97 & -3.38 \\ 
125 & -2.26 & -1.38 & 5.78 & 3.52 & 1.32 & -0.93 & -3.38 \\ 
130 & -0.87 & -1.32 & 5.53 & 3.42 & 1.23 & -1.02 & -3.23 \\ 
135 & -1.61 & -1.44 & 6.24 & 3.91 & 1.59 & -0.74 & -3.06 \\ 
140 & -0.45 & -1.62 & 6.78 & 4.33 & 1.73 & -0.88 & -3.58 \\ 
145 & -0.45 & -2.17 & 7.58 & 4.92 & 2.12 & -0.88 & -3.73 \\ 
150 & -0.99 & -2.88 & 9.12 & 5.92 & 2.67 & -0.47 & -3.92 \\ 
155 & 1.63 & -3.88 & 11.03 & 7.42 & 3.73 & 0.03 & -3.92 \\ 
160 & 4.78 & -7.08 & 15.78 & 11.22 & 6.53 & 1.62 & -3.52 \\ 
165 & 7.46 & -8.03 & 16.93 & 12.22 & 7.12 & 1.93 & -3.42 \\ 
170 & 6.41 & -5.58 & 13.68 & 9.47 & 5.17 & 0.72 & -3.83 \\ 
175 & 4.64 & -4.08 & 10.93 & 7.42 & 3.73 & -0.03 & -3.98 \\ 
180 & 1.50 & -2.77 & 8.82 & 5.78 & 2.67 & -0.53 & -3.83 \\ 
185 & -1.52 & -1.77 & 6.62 & 4.22 & 1.73 & -0.88 & -3.58 \\ 
190 & 0.09 & -1.18 & 5.28 & 3.23 & 1.18 & -0.93 & -3.17 \\ 
195 & -1.13 & -0.82 & 4.28 & 2.62 & 0.88 & -0.88 & -2.77 \\ 
200 & -0.89 & -0.62 & 3.77 & 2.23 & 0.68 & -0.93 & -2.67 \\
\hline
\end{tabular}
\end{ruledtabular}
\end{table}

These
distributions can be interpreted as follows:
The separation between the medians of the LLR$_{b}$ and LLR$_{s+b}$ distributions provides a
measure of the discriminating power of the search.  The sizes
of the one- and two-$\sigma$ LLR$_{b}$ bands indicate the width of the LLR$_{b}$ distribution,
assuming no signal is truly present and only statistical fluctuations and systematic effects are
present.  The value of LLR$_{\rm{obs}}$ relative to LLR$_{s+b}$ and LLR$_{b}$
indicates whether the data distribution appears to resemble what we expect if a signal is present
(i.e. closer to the LLR$_{s+b}$ distribution, which is negative by
construction)
or whether it resembles the background expectation more closely; the significance of any departures
of LLR$_{\rm{obs}}$ from LLR$_{b}$ can be evaluated by the width of the
LLR$_{b}$ bands.

Using the combination procedures outlined in Section III, we extract 
limits on SM Higgs boson production $\sigma \times B(H\rightarrow X)$ 
in \pp~collisions at $\sqrt{s}=1.96$~TeV for $100\leq m_H \leq 200$ GeV/$c^2$.
To facilitate comparisons with the standard model and to accommodate 
analyses with different degrees of sensitivity, we present our results 
in terms of the ratio of obtained limits to the SM Higgs boson production 
cross section, as a function of Higgs boson mass, for test masses for 
which both experiments have performed dedicated searches in different 
channels.  A value of the combined limit ratio which is less than or 
equal to one indicates that that particular Higgs boson mass is excluded 
at the 95\% C.L.

The combinations of results~\cite{CDFHiggs,DZHiggs} of each single experiment, 
as used in this Tevatron combination, yield the following ratios of 95\% C.L. 
observed (expected) limits to the SM cross section:
1.79~(1.90) for CDF and 2.52~(2.36) for D0 at $m_{H}=115$~GeV/$c^2$, and
1.13~(1.00) for CDF and 1.02~(1.14) for D0 at $m_{H}=165$~GeV/$c^2$.

The ratios of the 95\% C.L. expected and observed limit to the SM cross 
section are shown in Figure~\ref{fig:comboRatio} for the combined CDF 
and D0 analyses.  The observed and median expected ratios are listed 
for the tested Higgs boson masses in Table~\ref{tab:ratios} for $m_{H} 
\leq 150$~GeV/$c^2$, and in Table~\ref{tab:ratios-3} for $m_{H} \geq 
155$~GeV/$c^2$, as obtained by the Bayesian and the ${\rm CL}_{\rm s}$ methods.  In 
the following summary we quote only the limits obtained with the Bayesian 
method, which was decided upon {\it a priori}.  It turns out that the 
Bayesian limits are slightly less stringent.  The corresponding limits 
and expected limits obtained using the ${\rm CL}_{\rm s}$ method are shown alongside 
the Bayesian limits in the tables.  We obtain the observed (expected) 
values of 
0.87~(1.24) at $m_{H}=105$~GeV/$c^2$, 1.56~(1.45) at $m_{H}=115$~GeV/$c^2$, 1.28~(1.07) at
$m_{H}=155$~GeV/$c^2$, 0.68~(0.76) at $m_{H}=165$~GeV/$c^2$, 0.95~(1.04) at $m_{H}=175$~GeV/$c^2$ 
and 2.55~(1.61) at $m_{H}=185$~GeV/$c^2$.

\begin{figure}[hb]
\begin{centering}
\includegraphics[width=16.5cm]{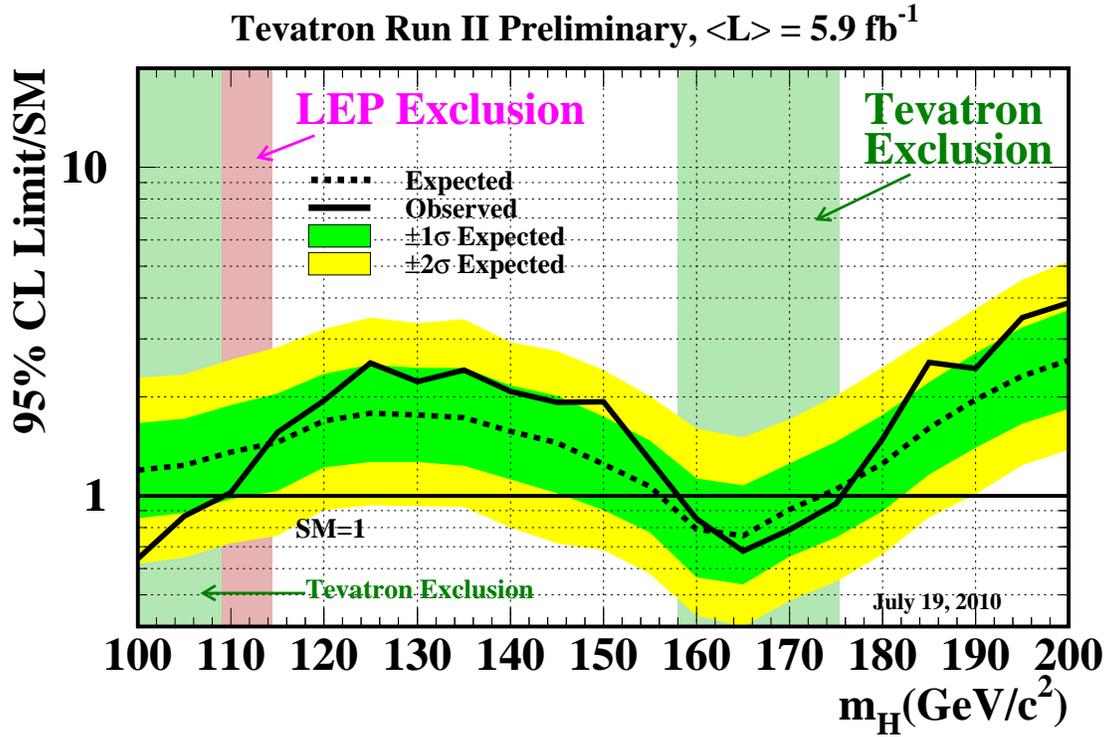}
\caption{
\label{fig:comboRatio}
Observed and expected (median, for the background-only hypothesis)
95\% C.L. upper limits on the ratios to the SM cross section, as
functions of the Higgs boson mass
for the combined CDF and D0 analyses.
The limits are expressed as a multiple of the SM prediction
for test masses (every 5 GeV/$c^2$)
for which both experiments have performed dedicated
searches in different channels.
The points are joined by straight lines
for better readability.
  The bands indicate the
68\% and 95\% probability regions where the limits can
fluctuate, in the absence of signal.
The limits displayed in this figure
are obtained with the Bayesian calculation.
}
\end{centering}
\end{figure}

\begin{table}[ht]
\caption{\label{tab:ratios} Ratios of median expected and observed 95\% C.L.
limit to the SM cross section for the combined CDF and D0 analyses as a function
of the Higgs boson mass in GeV/$c^2$, obtained with the Bayesian and with the ${\rm CL}_{\rm s}$ method.}
\begin{ruledtabular}
\begin{tabular}{lccccccccccc}\\
Bayesian       &  100 &  105 &  110 &  115 &  120 &  125 &  130 &  135 &  140 &  145 &  150 \\ \hline
Expected       & 1.20 & 1.24 & 1.36 & 1.45 & 1.69 & 1.78 & 1.76 & 1.73 & 1.57 & 1.45 & 1.25 \\
Observed       & 0.64 & 0.87 & 1.02 & 1.56 & 1.95 & 2.54 & 2.23 & 2.41 & 2.07 & 1.92 & 1.93 \\

\hline
\hline\\
${\rm CL}_{\rm s}$         &  100 &  105 &  110 &  115 &  120 &  125 &  130 &  135 &  140 &  145 &  150 \\ \hline
Expected       &1.17 &1.24 &1.36 &1.50 &1.66 &1.73 &1.78 &1.69 &1.56 &1.39 &1.20 \\
Observed       &0.61 &0.86 &1.06 &1.64 &2.05 &2.72 &2.38 &2.53 &2.07 &1.90 &1.79 \\
\end{tabular}
\end{ruledtabular}
\end{table}

\begin{table}[ht]
\caption{\label{tab:ratios-3}
Ratios of median expected and observed 95\% C.L.
limit to the SM cross section for the combined CDF and D0 analyses as a function
of the Higgs boson mass in GeV/$c^2$, obtained with the Bayesian and with the ${\rm CL}_{\rm s}$ method.}
\begin{ruledtabular}
\begin{tabular}{lccccccccccc}
Bayesian             &  155 &  160 &  165 &  170 &  175 &  180 &  185 &  190 &  195 &  200 \\ \hline
Expected             & 1.07 & 0.79 & 0.76 & 0.91 & 1.04 & 1.25 & 1.61 & 1.96 & 2.31 & 2.58 \\
Observed             & 1.28 & 0.85 & 0.68 & 0.79 & 0.95 & 1.49 & 2.55 & 2.44 & 3.49 & 3.87 \\
\hline
\hline\\
${\rm CL}_{\rm s}$               &  155 &  160 &  165 &  170 &  175 &  180 &  185 &  190 &  195 &  200 \\ \hline
Expected             &1.05 &0.77 &0.73 &0.89 &1.04 &1.25 &1.59 &1.96 &2.32 &2.66 \\
Observed              &1.26 &0.84 &0.69 &0.77 &0.93 &1.43 &2.46 &2.29 &3.44 &3.80 \\
\end{tabular}
\end{ruledtabular}
\end{table}

We also show in Figure~\ref{fig:comboLLR-2} and list in Table~\ref{tab:clsVals} the observed 1-${\rm CL}_{\rm s}$
and its expected distribution for the background-only hypothesis as a function of the Higgs boson mass.
This is directly interpreted as the level of exclusion of our search.  This
figure is obtained using the ${\rm CL}_{\rm s}$ method.

In summary, we combine all available CDF and D0 results on SM Higgs boson searches,
based on luminosities ranging from 2.1 to 6.7 fb$^{-1}$.
Compared to our previous combination, more data have been added to the existing
channels, additional channels have been included, and analyses have been further 
optimized to gain sensitivity. We use the latest parton distribution functions and 
$gg \rightarrow H$ theoretical cross sections when comparing our limits to the SM 
predictions at high mass.

The 95\% C.L. upper limits on Higgs boson production are a factor of 1.56 and 0.68 
times the SM cross section for a Higgs boson mass of $m_{H}=$115 and 165~GeV/$c^2$, respectively.
Based on simulation, the corresponding median expected upper limits are 1.45 and 0.76, respectively.
Standard Model branching ratios, calculated as functions of the Higgs boson mass, are assumed.

We choose to use the intersections of piecewise linear interpolations of our observed and expected 
rate limits in order to quote ranges of Higgs boson masses that are excluded and that are expected 
to be excluded.  The sensitivities of our searches to Higgs bosons are smooth functions of the Higgs 
boson mass and depend most strongly on the predicted cross sections and the decay branching ratios 
(the decay $H\rightarrow W^+W^-$ is the dominant decay for the region of highest sensitivity). The 
mass resolution of the channels is poor due to the presence of two highly energetic neutrinos in 
signal events.  We therefore use the linear interpolations to extend the results from the 5~GeV/$c^2$ 
mass grid investigated to points in between.  This procedure yields higher expected and observed 
interpolated limits than if the full dependence of the cross section and branching ratio were 
included as well, since the latter produces limit curves that are concave upwards.  The regions 
of Higgs boson masses excluded at the 95\% C.L. thus obtained are $158<m_{H}<175$~GeV/$c^{2}$ 
and $100<m_H<109$~GeV/$c^{2}$.  The expected exclusion region, given the current sensitivity, is 
$156<m_{H}<173$~GeV/$c^{2}$.  The excluded region obtained by finding the intersections of the 
linear interpolations of the observed $1-{\rm CL}_{\rm s}$ curve shown in Figure~\ref{fig:comboLLR-2} is 
slightly larger than that obtained with the Bayesian calculation.  As previously stated, we make 
the {\it a priori} choice to quote the exclusion region using the Bayesian calculation. 

The results presented in this paper significantly extend the individual limits of each
collaboration and those obtained in our previous combination.  The sensitivity of our combined search is 
sufficient to exclude a Higgs boson at high mass and is expected to grow substantially 
in the future as more data are added and further improvements are made to our analysis 
techniques.

 \begin{figure}[t]
 \begin{centering}
 \includegraphics[width=14.0cm]{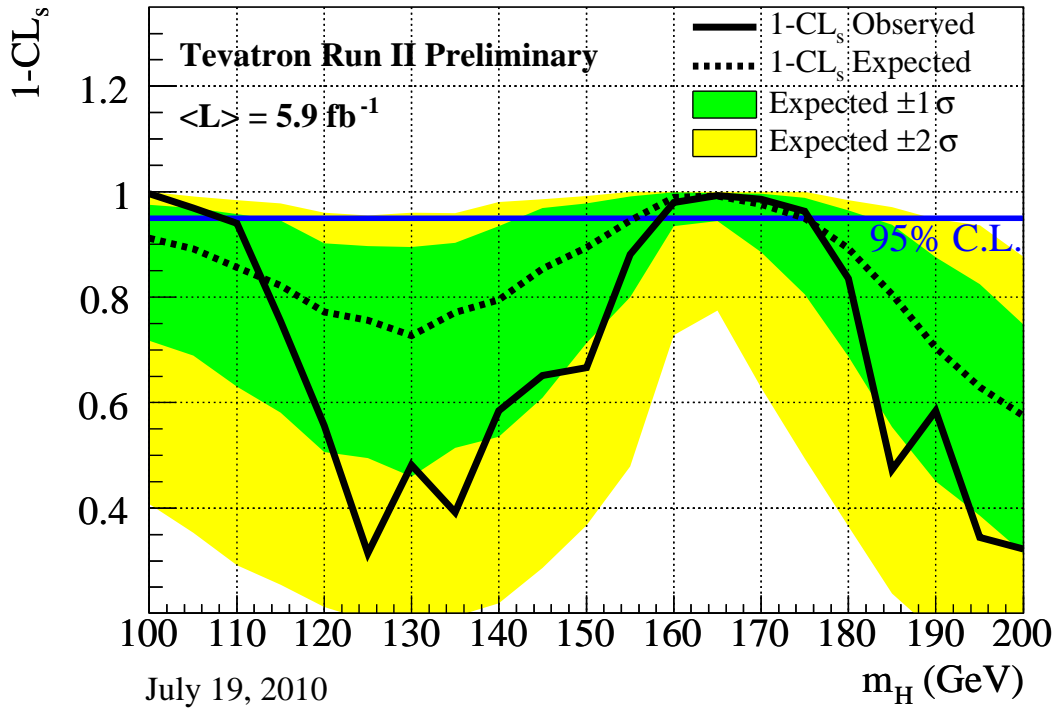}
 \caption{
 \label{fig:comboLLR-2}
 The exclusion strength 1-${\rm CL}_{\rm s}$ as a function of the Higgs boson mass
(in steps of 5 GeV/$c^2$), as obtained with ${\rm CL}_{\rm s}$ method
 for the combination of the
 CDF and D0 analyses. }
 \end{centering}
 \end{figure}


\begin{table}[htpb]
\caption{\label{tab:clsVals} The observed and expected 1-{\rm CL}$_{\rm s}$ values as functions of $m_H$, for the combined
CDF and \Dzero Higgs boson searches.}
\begin{ruledtabular}
\begin{tabular}{lcccccc}
$m_H$ (GeV/$c^2$) & 1-{\rm CL}$_{\rm s}^{\rm{obs}}$ &
1-{\rm CL}$_{\rm s}^{-2\sigma}$ &
1-{\rm CL}$_{\rm s}^{-1\sigma}$ &
1-{\rm CL}$_{\rm s}^{\rm{median}}$ &
1-{\rm CL}$_{\rm s}^{+1\sigma}$ &
1-{\rm CL}$_{\rm s}^{+2\sigma}$ \\ \hline
100 & 0.996 & 1.000 & 0.975 & 0.912 & 0.717 & 0.408 \\ 
105 & 0.969 & 0.991 & 0.969 & 0.892 & 0.689 & 0.353 \\ 
110 & 0.940 & 0.984 & 0.958 & 0.856 & 0.630 & 0.292 \\ 
115 & 0.754 & 0.978 & 0.945 & 0.822 & 0.581 & 0.254 \\ 
120 & 0.556 & 0.960 & 0.902 & 0.772 & 0.506 & 0.213 \\ 
125 & 0.315 & 0.955 & 0.896 & 0.756 & 0.495 & 0.187 \\ 
130 & 0.482 & 0.960 & 0.895 & 0.727 & 0.460 & 0.183 \\ 
135 & 0.392 & 0.959 & 0.903 & 0.771 & 0.514 & 0.195 \\ 
140 & 0.585 & 0.981 & 0.934 & 0.795 & 0.535 & 0.219 \\ 
145 & 0.652 & 0.986 & 0.968 & 0.853 & 0.609 & 0.286 \\ 
150 & 0.667 & 0.992 & 0.977 & 0.894 & 0.713 & 0.367 \\ 
155 & 0.881 & 1.000 & 0.991 & 0.945 & 0.800 & 0.479 \\ 
160 & 0.979 & 1.000 & 0.998 & 0.989 & 0.935 & 0.728 \\ 
165 & 0.993 & 1.000 & 0.997 & 0.991 & 0.944 & 0.774 \\ 
170 & 0.986 & 1.000 & 0.996 & 0.976 & 0.885 & 0.629 \\ 
175 & 0.963 & 1.000 & 0.988 & 0.950 & 0.805 & 0.494 \\ 
180 & 0.835 & 0.984 & 0.965 & 0.892 & 0.688 & 0.365 \\ 
185 & 0.473 & 0.971 & 0.937 & 0.805 & 0.554 & 0.238 \\ 
190 & 0.585 & 0.948 & 0.876 & 0.705 & 0.451 & 0.163 \\ 
195 & 0.345 & 0.936 & 0.825 & 0.630 & 0.386 & 0.130 \\ 
200 & 0.323 & 0.876 & 0.748 & 0.575 & 0.315 & 0.100 \\ 
\end{tabular}
\end{ruledtabular}
\end{table}

\clearpage
\newpage


\appendix
\appendixpage
\addappheadtotoc
\section{Systematic Uncertainties}


\begin{table}[h]
\begin{scriptsize}
\begin{center}
\caption{\label{tab:cdfsystwh2jet} Systematic uncertainties on the signal and 
background contributions for CDF's $WH\rightarrow\ell\nu b{\bar{b}}$ tight 
double tag (TDT), loose double tag (LDT), looser double tag (LDTX), and single 
tag (ST) 2 jet channels.  Systematic uncertainties are listed by name; see the 
original references for a detailed explanation of their meaning and on how 
they are derived.  Systematic uncertainties for $WH$ shown in this table are 
obtained for $m_H=115$ GeV/$c^2$.  Uncertainties are relative, in percent, and 
are symmetric unless otherwise indicated.}
\vskip 0.1cm
{\centerline{CDF: tight and loose double-tag (TDT and LDT) $WH\rightarrow\ell\nu b{\bar{b}}$ channel relative uncertainties (\%)}}
\vskip 0.099cm
\begin{ruledtabular}
\begin{tabular}{lcccccc}\\
Contribution              & $W$+HF & Mistags & Top & Diboson & Non-$W$ & $WH$  \\ \hline
Luminosity ($\sigma_{\mathrm{inel}}(p{\bar{p}})$)
                          & 0      & 0       & 3.8 & 3.8     & 0       &    3.8   \\
Luminosity Monitor        & 0      & 0       & 4.4 & 4.4     & 0       &    4.4   \\
Lepton ID                 & 0      & 0       & 2   & 2       & 0       &    2   \\
Jet Energy Scale          & 0      & 0       & 0   & 0       & 0       &    2   \\
Mistag Rate               & 0      & 35     & 0   & 0       & 0       &    0   \\
$B$-Tag Efficiency          & 0      & 0       & 8.6 & 8.6     & 0       &    8.6   \\
$t{\bar{t}}$ Cross Section  & 0    & 0       & 10  & 0       & 0       &    0   \\
Diboson Rate              & 0      & 0       & 0   & 11.5    & 0       &    0   \\
Signal Cross Section      & 0      & 0       & 0   & 0       & 0       &    5 \\
HF Fraction in W+jets     &    45  & 0       & 0   & 0       & 0       &    0   \\
ISR+FSR+PDF               & 0      & 0       & 0   & 0       & 0       &    5 \\
QCD Rate                  & 0      & 0       & 0   & 0       & 40      &    0   \\
\end{tabular}
\end{ruledtabular}

\vskip 0.3cm
{\centerline{CDF: looser double-tag (LDTX) $WH\rightarrow\ell\nu b{\bar{b}}$ channel relative uncertainties (\%)}}
\vskip 0.099cm
\begin{ruledtabular}
\begin{tabular}{lcccccc}\\
Contribution              & $W$+HF & Mistags & Top & Diboson & Non-$W$ & $WH$  \\ \hline
Luminosity ($\sigma_{\mathrm{inel}}(p{\bar{p}})$)
                          & 0      & 0       & 3.8 & 3.8     & 0       &    3.8   \\
Luminosity Monitor        & 0      & 0       & 4.4 & 4.4     & 0       &    4.4   \\
Lepton ID                 & 0      & 0       & 2   & 2       & 0       &    2   \\
Jet Energy Scale          & 0      & 0       & 0   & 0       & 0       &    2.2   \\
Mistag Rate               & 0      & 36     & 0   & 0       & 0       &    0   \\
$B$-Tag Efficiency          & 0      & 0       & 13.6 & 13.6     & 0       &    13.6   \\
$t{\bar{t}}$ Cross Section  & 0    & 0       & 10  & 0       & 0       &    0   \\
Diboson Rate              & 0      & 0       & 0   & 11.5    & 0       &    0   \\
Signal Cross Section      & 0      & 0       & 0   & 0       & 0       &    5 \\
HF Fraction in W+jets     &    45  & 0       & 0   & 0       & 0       &    0   \\
ISR+FSR+PDF               & 0      & 0       & 0   & 0       & 0       &    7.7 \\
QCD Rate                  & 0      & 0       & 0   & 0       & 40      &    0   \\
\end{tabular}
\end{ruledtabular}

\vskip 0.3cm
{\centerline{CDF: single tag (ST) $WH\rightarrow\ell\nu b{\bar{b}}$ channel relative uncertainties (\%)}}
\vskip 0.099cm
\begin{ruledtabular}
\begin{tabular}{lcccccc}\\
Contribution              & $W$+HF & Mistags & Top & Diboson & Non-$W$ & $WH$  \\ \hline
Luminosity ($\sigma_{\mathrm{inel}}(p{\bar{p}})$)
                          & 0      & 0       & 3.8 & 3.8     & 0       &    3.8   \\
Luminosity Monitor        & 0      & 0       & 4.4 & 4.4     & 0       &    4.4   \\
Lepton ID                 & 0      & 0       & 2   & 2       & 0       &    2   \\
Jet Energy Scale          & 0      & 0       & 0   & 0       & 0       &    2   \\
Mistag Rate               & 0      & 35    & 0   & 0       & 0       &    0   \\
$B$-Tag Efficiency          & 0      & 0       & 4.3 & 4.3     & 0       &    4.3   \\
$t{\bar{t}}$ Cross Section  & 0    & 0       & 10  & 0       & 0       &    0   \\
Diboson Rate                & 0      & 0       & 0   & 11.5    & 0       &    0   \\
Signal Cross Section        & 0      & 0       & 0   & 0       & 0       &    5 \\
HF Fraction in W+jets       &    42  & 0       & 0   & 0       & 0       &    0   \\
ISR+FSR+PDF                 & 0      & 0       & 0   & 0       & 0       &    3.0 \\
QCD Rate                    & 0      & 0       & 0   & 0       & 40      &    0   \\
\end{tabular}
\end{ruledtabular}

\end{center}
\end{scriptsize}
\end{table}


\begin{table}[t]
\begin{center}
\caption{\label{tab:cdfsystwh3jet} Systematic uncertainties on the signal and background 
contributions for CDF's $WH\rightarrow\ell\nu b{\bar{b}}$ tight double tag (TDT), loose 
double tag (LDT), and single tag (ST) 3 jet channels.  Systematic uncertainties are listed 
by name; see the original references for a detailed explanation of their meaning and on how 
they are derived.  Systematic uncertainties for $WH$ shown in this table are obtained for 
$m_H=115$ GeV/$c^2$.  Uncertainties are relative, in percent, and are symmetric unless 
otherwise indicated.}

\vskip 0.1cm
{\centerline{CDF: tight and loose double-tag (TDT and LDT) $WH\rightarrow\ell\nu b{\bar{b}}$ channel relative uncertainties (\%)}}
\vskip 0.099cm
\begin{ruledtabular}
\begin{tabular}{lcccccc}\\
Contribution              & $W$+HF & Mistags & Top & Diboson & Non-$W$ & $WH$  \\ \hline
Luminosity ($\sigma_{\mathrm{inel}}(p{\bar{p}})$)
                          & 0      & 0       & 3.8 & 3.8     & 0       &    3.8   \\
Luminosity Monitor        & 0      & 0       & 4.4 & 4.4     & 0       &    4.4   \\
Lepton ID                 & 0      & 0       & 2   & 2       & 0       &    2   \\
Jet Energy Scale          & 0      & 0       & 0   & 0       & 0       &    13.5   \\
Mistag Rate               & 0      & 9     & 0   & 0       & 0       &    0   \\
$B$-Tag Efficiency          & 0      & 0       & 8.4 & 8.4     & 0       &    8.4   \\
$t{\bar{t}}$ Cross Section  & 0    & 0       & 10  & 0       & 0       &    0   \\
Diboson Rate              & 0      & 0       & 0   & 10    & 0       &    0   \\
Signal Cross Section      & 0      & 0       & 0   & 0       & 0       &    10 \\
HF Fraction in W+jets     &    30  & 0       & 0   & 0       & 0       &    0   \\
ISR+FSR+PDF               & 0      & 0       & 0   & 0       & 0       &    21.4 \\
QCD Rate                  & 0      & 0       & 0   & 0       & 40      &    0   \\
\end{tabular}
\end{ruledtabular}

\vskip 0.3cm
{\centerline{CDF: single tag (ST) $WH\rightarrow\ell\nu b{\bar{b}}$ channel relative uncertainties (\%)}}
\vskip 0.099cm
\begin{ruledtabular}
\begin{tabular}{lcccccc}\\
Contribution              & $W$+HF & Mistags & Top & Diboson & Non-$W$ & $WH$  \\ \hline
Luminosity ($\sigma_{\mathrm{inel}}(p{\bar{p}})$)
                          & 0      & 0       & 3.8 & 3.8     & 0       &    3.8   \\
Luminosity Monitor        & 0      & 0       & 4.4 & 4.4     & 0       &    4.4   \\
Lepton ID                 & 0      & 0       & 2   & 2       & 0       &    2   \\
Jet Energy Scale          & 0      & 0       & 0   & 0       & 0       &    15.8   \\
Mistag Rate               & 0      & 13.3    & 0   & 0       & 0       &    0   \\
$B$-Tag Efficiency          & 0      & 0       & 3.5 & 3.5     & 0       &    3.5   \\
$t{\bar{t}}$ Cross Section  & 0    & 0       & 10  & 0       & 0       &    0   \\
Diboson Rate              & 0      & 0       & 0   & 10    & 0       &    0   \\
Signal Cross Section      & 0      & 0       & 0   & 0       & 0       &    10 \\
HF Fraction in W+jets     &    30  & 0       & 0   & 0       & 0       &    0   \\
ISR+FSR+PDF               & 0      & 0       & 0   & 0       & 0       &    13.1 \\
QCD Rate                  & 0      & 0       & 0   & 0       & 40      &    0   \\
\end{tabular}
\end{ruledtabular}

\end{center}
\end{table}


\begin{table}[h]
\begin{center}
\caption{\label{tab:d0systwh1} Systematic uncertainties on the signal and background 
contributions for D0's $WH\rightarrow\ell\nu b{\bar{b}}$ single (ST) and double tag 
(DT) channels.
Systematic uncertainties are listed by name, see the original
references for a detailed explanation of their meaning and on how they are derived.
Systematic uncertainties for $WH$ shown in this table are obtained for $m_H=115$ GeV/$c^2$.
  Uncertainties are
relative, in percent, and are symmetric unless otherwise indicated.  }
\vskip 0.2cm
{\centerline{D0: single tag (ST) $WH \rightarrow\ell\nu b\bar{b}$ channel relative uncertainties (\%)}}
\vskip 0.099cm
\begin{ruledtabular}
\begin{tabular}{l c c c c c c c }\\
Contribution  &~WZ/WW~&Wbb/Wcc&Wjj/Wcj&$~~~t\bar{t}~~~$&single top&Multijet& ~~~WH~~~\\
\hline
Luminosity                &  6    &  6    &  6    &  6    &  6    &  0    &  6    \\ 
Trigger eff.              &  2--3 &  2--3 &  2--3 &  2--3 &  2--3 &  0    &  2--3 \\       
EM ID/Reco eff./resol.    &     3 &     3 &     3 &     3 &     3 &  0    &     3 \\       
Muon ID/Reco eff./resol.  &   4.1 &   4.1 &   4.1 &   4.1 &   4.1 &  0    &   4.1 \\        
Jet ID/Reco eff.          &     1 &     1 &     2 &     1 &     1 &  0    &     1 \\ 
Jet Energy Scale          &  2--5 &   2-5 &  2--5 &  2--4 &  2--5 &  0    &  2--5 \\       
$b$-tagging/taggability   &   5-6 &   3-4 &   8-9 &   2-4 &   2-4 &  0    &   2-4 \\ 
Cross Section             &     6 &     9 &     9 &    10 &    10 &  0    &     6 \\       
Heavy-Flavor K-factor     &  0    &    20 &     0 &  0    &  0    &  0    &  0    \\       
Instrumental-WH           &  0    &     0 &     0 &  0    &  0    &  1    &  0    \\ 
PDF, reweighting          &  0--1 &  0--2 & 2--3  & 2--3  & 0--4  &  0    &  0--1 \\
\end{tabular}
\end{ruledtabular}
\vskip 0.5cm
{\centerline{D0: double tag (DT) $WH \rightarrow\ell\nu b\bar{b}$ channel relative uncertainties (\%)}}
\vskip 0.099cm
\begin{ruledtabular}
\begin{tabular}{ l c c c c c c c }   \\
Contribution  &~WZ/WW~&Wbb/Wcc&Wjj/Wcj&$~~~t\bar{t}~~~$&single top&Multijet& ~~~WH~~~\\
\hline
Luminosity                &  6    &  6    &  6    &  6    &  6    &  0    &  6    \\ 
Trigger eff.              &  2--3 &  2--3 &  2--3 &  2--3 &  2--3 &  0    &  2--3 \\       
EM ID/Reco eff./resol.    &     3 &     3 &     3 &     3 &     3 &  0    &     3 \\       
Muon ID/Reco eff./resol.  &   4.1 &   4.1 &   4.1 &   4.1 &   4.1 &  0    &   4.1 \\        
Jet ID/Reco eff.          &     1 &     1 &     2 &     2 &     1 &  0    &  1--2 \\ 
Jet Energy Scale          &  2--5 &  2--5 &  2--5 &  2--3 &  1--2 &  0    &  2--5 \\       
$b$-tagging/taggability   & 9--11 & 9--11 &     7 &11--14 &11--14 &  0    &11--14 \\       
Cross Section             &     6 &     9 &     9 &    10 &    10 &  0    &     6 \\       
Heavy-Flavor K-factor     &  0    &    20 &     0 &  0    &  0    &  0    &  0    \\       
Instrumental-WH           &  0    &     0 &     0 &  0    &  0    &  1    &  0    \\
PDF, reweighting          &  0--1 &  0--1 & 1--2  & 2--3  & 0--1  &  0    &  0--1 \\ 
\end{tabular}
\end{ruledtabular}
\end{center}
\end{table}


\begin{table}
\begin{center}
\caption{\label{tab:cdfvvbb1} Systematic uncertainties on the signal and background contributions for CDF's 
$WH,ZH\rightarrow\MET b{\bar{b}}$ tight double tag (TDT), loose double tag (LDT), and single tag (ST) channels.  
Systematic uncertainties are listed by name; see the original references for a detailed explanation of their 
meaning and on how they are derived.  Systematic uncertainties for $ZH$ and $WH$ shown in this table are 
obtained for $m_H=120$~GeV/$c^2$.  Uncertainties are relative, in percent, and are symmetric unless otherwise 
indicated.}
\vskip 0.1cm
{\centerline{CDF: tight double-tag (TDT) $WH,ZH\rightarrow\MET b{\bar{b}}$ channel relative uncertainties (\%)}}
\vskip 0.099cm
\begin{ruledtabular}
      \begin{tabular}{lcccccccc}\\
        Contribution     & ZH & WH &Multijet& Top Pair & S. Top & Di-boson  & W + h.f.  & Z + h.f. \\\hline
        Luminosity       & 3.8 & 3.8 &       & 3.8 & 3.8 & 3.8     & 3.8     & 3.8     \\
        Lumi Monitor      & 4.4 & 4.4 &       & 4.4 & 4.4 & 4.4     & 4.4     & 4.4     \\
        Tagging SF        & 10.4& 10.4&       & 10.4& 10.4& 10.4    & 10.4    & 10.4    \\
      Trigger Eff. (shape)& 1.0 & 1.2 & 1.1 & 0.7 & 1.1 & 1.6     & 1.7     & 1.3     \\
        Lepton Veto       & 2.0 & 2.0 &       & 2.0 & 2.0 &2.0      & 2.0     & 2.0     \\
        PDF Acceptance    & 2.0 & 2.0 &       & 2.0 & 2.0 &2.0      & 2.0     & 2.0     \\
        JES (shape)       & $^{+3.0}_{-3.0}$
                                  & $^{+3.5}_{-4.7}$
                                          &  $^{-4.0}_{+3.8}$
                                                  & $^{+1.1}_{-1.1}$
                                                          & $^{+2.4}_{-4.7}$
                                                                  & $^{+8.2}_{-6.1}$
                                                                             & $^{+7.3}_{-11.8}$
                                                                                          & $^{+6.5}_{-8.3}$    \\
        ISR               & \multicolumn{2}{c}{$^{+4.4}_{+3.7}$} &       &       &       &           &           &      \\
        FSR               & \multicolumn{2}{c}{$^{+1.8}_{+4.4}$} &       &       &       &           &           &      \\
        Cross-Section     &  5  & 5 &       & 10 & 10 & 6    & 30      & 30      \\
        Multijet Norm.  (shape)   &       & & 22 &       & &          &           &           \\
      \end{tabular}
\end{ruledtabular}

\vskip 0.3cm
{\centerline{CDF: loose double-tag (LDT) $WH,ZH\rightarrow\MET b{\bar{b}}$ channel relative uncertainties (\%)}}
\vskip 0.099cm
 \begin{ruledtabular}
     \begin{tabular}{lcccccccc}\\
        Contribution & ZH & WH & Multijet & Top Pair & S. Top  & Di-boson  & W + h.f.  & Z + h.f. \\\hline
        Luminosity       & 3.8  & 3.8  &     & 3.8  & 3.8  & 3.8      & 3.8      & 3.8     \\
        Lumi Monitor      & 4.4  & 4.4  &     & 4.4  & 4.4  & 4.4      & 4.4      & 4.4     \\
        Tagging SF        & 11.6 & 11.6 &     & 11.6 & 11.6 & 11.6     & 11.6     & 11.6     \\
     Trigger Eff. (shape) & 1.2  & 1.3  &1.1& 0.7  & 1.2  & 1.2      & 1.8      & 1.3     \\
        Lepton Veto       & 2.0  & 2.0  &     & 2.0  & 2.0  &2.0       & 2.0      & 2.0     \\
        PDF Acceptance    & 2.0  & 2.0  &     & 2.0  & 2.0  &2.0       & 2.0      & 2.0     \\
        JES (shape)       & $^{+3.7}_{-3.7}$
                                   & $^{+4.0}_{-4.0}$
                                            & $^{-5.4}_{+5.2}$
                                                  & $^{+1.1}_{-0.7}$
                                                           & $^{+4.2}_{-4.2}$
                                                                    & $^{+7.0}_{-7.0}$
                                                                                 & $^{+1.3}_{-7.6}$
                                                                                              & $^{+6.2}_{-7.1}$    \\
        ISR               & \multicolumn{2}{c}{$^{+1.4}_{-2.9}$} &       &       &       &           &           &      \\
        FSR               & \multicolumn{2}{c}{$^{+5.3}_{+2.5}$} &       &       &       &           &           &      \\
        Cross-Section     &  5.0   & 5.0 &      & 10 & 10 & 6    & 30      & 30      \\
        Multijet Norm.  (shape)   &       & & 11 &       & &          &           &           \\
      \end{tabular}
\end{ruledtabular}

\vskip 0.3cm
{\centerline{CDF: single-tag (ST) $WH,ZH\rightarrow\MET b{\bar{b}}$ channel relative uncertainties (\%)}}
\vskip 0.099cm
\begin{ruledtabular}
      \begin{tabular}{lcccccccc}\\
        Contribution & ZH & WH & Multijet & Top Pair & S. Top  & Di-boson  & W + h.f.  & Z + h.f. \\\hline
        Luminosity       & 3.8  & 3.8  &     & 3.8  & 3.8  & 3.8      & 3.8      & 3.8     \\
        Lumi Monitor      & 4.4  & 4.4  &     & 4.4  & 4.4  & 4.4      & 4.4      & 4.4     \\
        Tagging SF        & 5.2  & 5.2  &     & 5.2  & 5.2  & 5.2      & 5.2      & 5.2     \\
     Trigger Eff. (shape) & 0.9  & 1.1  &1.1& 0.7  & 1.1  & 1.3      & 2.0      & 1.4     \\
        Lepton Veto       & 2.0  & 2.0  &     & 2.0  & 2.0  &2.0       & 2.0      & 2.0     \\
        PDF Acceptance    & 2.0  & 2.0  &     & 2.0  & 2.0  &2.0       & 2.0      & 2.0     \\
        JES (shape)       & $^{+3.8}_{-3.8}$
                                   & $^{+3.8}_{-3.8}$
                                            & $^{-5.2}_{+5.6}$
                                                  & $^{+0.7}_{-0.8}$
                                                           & $^{+4.6}_{-4.6}$
                                                                    & $^{+7.0}_{-5.6}$
                                                                                 & $^{+12.4}_{-12.7}$
                                                                                              & $^{+8.3}_{-8.1}$    \\
        ISR               & \multicolumn{2}{c}{$^{-1.0}_{-1.5}$} &       &       &       &           &           &      \\
        FSR               & \multicolumn{2}{c}{$^{+2.0}_{-0.1}$} &       &       &       &           &           &      \\
        Cross-Section     &  5.0   & 5.0 &      & 10 & 10 & 6    & 30      & 30      \\
        Multijet Norm.  (shape)   &       & & 10 &       & &          &           &           \\
      \end{tabular}
\end{ruledtabular}

\end{center}
\end{table}


\begin{table}
\begin{center}
\caption{\label{tab:d0vvbb} Systematic uncertainties on the signal and background contributions for 
D0's $ZH\rightarrow \nu \nu b{\bar{b}}$ single tag (ST) and tight-loose double tag (TLDT) channels.
Systematic uncertainties are listed by name; see the original references for a detailed explanation 
of their meaning and on how they are derived.  Systematic uncertainties for $ZH$, $WH$ shown in this 
table are obtained for $m_H=115$ GeV/$c^2$.  Uncertainties are relative, in percent, and are symmetric 
unless otherwise indicated. Shape uncertainties are labeled with an ``s''. }

\vskip 0.1cm
{\centerline{D0: single tag (ST)~ $ZH \rightarrow \nu\nu b \bar{b}$ channel relative uncertainties (\%)}}
\vskip 0.099cm
\begin{ruledtabular}
\begin{tabular}{ l c c c c c c } \\
Contribution                          &~WZ/ZZ~        &~Z+jets~      &~W+jets~      &~~~$t\bar{t}$    &~~ZH,WH~~\\ \hline
Jet Energy Scale pos/neg (S)          &  $\pm$5.5     &  $\pm$5.5    &$\pm$7.0      &  $\mp$1.5       & $\pm$ 1.9      \\
Jet ID (S)                            &  1.1          &  0.7         &  1.0         &  0.8            &  0.7      \\
Jet Resolution pos/neg (S)            &  $\pm$0.5     &  $\pm$2.7    &  $\pm$3.8    &  $\mp$0.6       &  $\pm$0.7      \\
MC Heavy flavor $b$-tagging pos/neg (S) & $\pm$4.5    &  $\pm$2.9    &  $\pm$2.8    &  $\pm$3.5       &  $\pm$3.5      \\
MC light flavor $b$-tagging pos/neg (S) &  $\pm$3.1   &  $\pm$4.9    &  $\pm$6.5    &  $\pm$0.6       &  $\pm$0.1      \\
Direct taggability (S)                &  1.6          &  1.3         &  1.7         &  0.5            &  1.9      \\
Trigger efficiency (S)                &  3.5          &  3.5         &  3.5         &  3.5            &  3.5      \\
ALPGEN MLM pos/neg(S)                 &  -            &  Shape Only  &  Shape only  &  -              &  -      \\
ALPGEN Scale (S)                      &  -            &  Shape Only  &  Shape only  &  -              &  -      \\
Underlying Event (S)                  &  -            &  Shape Only  &  Shape only  &  -              &  -      \\
Parton Distribution Function (S)      &  0.0          &  0.4         &  0.2         &  2.0            &  0.0      \\
EM ID                                 &  0.3          &  0           &  0.3         &  0.6            &  1.0      \\
Muon ID                               &  1.1          &  0.5         &  1.4         &  1.9            &  0.2      \\
Cross Section                         &  7            &  6.0         &  6.0         &  10             &  6.0      \\
Heavy Flavor Ratio                    &  -            &  20          &  20          &  -              &  -      \\
Luminosity                            &  6.1          &  6.1         &  6.1         &  6.1            &  6.1  \\
\end{tabular}
\end{ruledtabular}

\vskip 0.3cm
{\centerline{D0: double tag (TLDT)~ $ZH \rightarrow \nu\nu b \bar{b}$ channel relative uncertainties (\%)}}
\vskip 0.099cm
\begin{ruledtabular}
\begin{tabular}{ l  c  c  c  c  c  c } \\
Contribution                          &~WZ/ZZ~        &~Z+jets~      &~W+jets~      &~~~$t\bar{t}$    &~~ZH,WH~~\\ \hline
Jet Energy Scale pos/neg (S)          &  $\pm$ 5.1    & $\pm$ 7.1    &  $\pm$6.6    &  $\mp$0.5       &  $\pm$1.6      \\
Jet ID (S)                            &  1.1          &  $\pm$1.2    &  0.8         &  0.1            &  1.1      \\
Jet Resolution pos/neg  (S)           &  $\mp$1.6     &  $\pm$2.0    &  $\pm$1.9    &  $\mp$2.0       & $\mp$ 1.6      \\
MC Heavy flavor $b$-tagging pos/neg (S) & $\pm$ 8.0   &  $\pm$0.6    &  $\pm$8.5    &  $\pm$10.2      &  $\pm$9.9      \\
MC light flavor $b$-tagging pos/neg (S) &  1.5        &  $\pm$12.6   &  $\pm$1.2    &  $\pm$0.1       &  0.0      \\
Direct taggability \& Vertex Confirmation(S) pos/neg&  7.4/1.5&  $\pm$9.0 &  $\pm$6.8    &  5.2/0.1        &  8.3/0.0      \\
Trigger efficiency (S)                &  3.5          &  3.5         &  3.5         &  3.5            &  3.5      \\
ALPGEN MLM pos/neg (S)                &  -            &  Shape only  &  Shape only  &  -              &  -      \\
ALPGEN Scale (S)                      &  -            &  Shape only  &  Shape only  &  -              &  -      \\
Underlying Event (S)                  &  -            &  Shape only  &  Shape only  &  -              &  -      \\
Parton Distribution Function (S)      &  $\pm$0.1     &  0.0        &  $\pm$0.4    &  0.6/-0.5       &  0.6/0.9      \\
EM ID                                 &  0.3          &  -           &  0.6         &  0.8            &  0.3      \\
Muon ID                               &  1.1          &  0.5         &  1.0         &  1.8            &  1.0      \\
Cross Section                         &  7.0          &  6.0         &  6.0         & 10              &  6.0      \\
Heavy Flavor Ratio                    &  -            &  20          &  20          &  -              &  -      \\
Luminosity                            &  6.1          &  6.1         &  6.1         &  6.1            &  6.1  \\
\end{tabular}
\end{ruledtabular}
\end{center}
\end{table}

%

\begin{table}[t] 
\begin{center}
\caption{\label{tab:cdfllbb1} Systematic uncertainties on the signal and background contributions for CDF's 
$ZH\rightarrow \ell^+\ell^-b{\bar{b}}$ single tag (ST), tight double tag (TDT), and loose double tag (LDT) 
channels.  The channels are further divided into low and high $s/b$ categories.  Systematic uncertainties 
are listed by name; see the original references for a detailed explanation of their meaning and on how they 
are derived.  Systematic uncertainties for $ZH$  shown in this table are obtained for $m_H=115$ GeV/$c^2$.  
Uncertainties are relative, in percent, and are symmetric unless otherwise indicated.}
\vskip 0.1cm
{\centerline{CDF: single tag (ST) high $s/b$~ $ZH \rightarrow \ell\ell b \bar{b}$ channel relative uncertainties (\%)}}
\vskip 0.099cm
\begin{ruledtabular}
\begin{tabular}{lcccccccc} \\ 
Contribution   & ~Fakes~ & ~~~Top~~~  & ~~$WZ$~~ & ~~$ZZ$~~ & ~$Z+b{\bar{b}}$~ & ~$Z+c{\bar{c}}$~& ~$Z+$mistag~ & ~~~$ZH$~~~ \\ \hline
Luminosity ($\sigma_{\mathrm{inel}}(p{\bar{p}})$)          & 0     &    3.8 &    3.8 &    3.8 &    3.8           &    3.8          & 0        &    3.8  \\
Luminosity Monitor        & 0     &    4.4 &    4.4 &    4.4 &    4.4           &    4.4          & 0        &    4.4  \\
Lepton ID    & 0     &    1 &    1 &    1 &    1           &    1          & 0        &    1  \\
Lepton Energy Scale    & 0     &    1.5 &    1.5 &    1.5 &    1.5           &    1.5          & 0        &    1.5  \\
$ZH$ Cross Section    & 0     &    0 &    0 &    0 &    0           &    0          & 0        &    5 \\
Fake Leptons       & 50    & 0    & 0    & 0    & 0              & 0             & 0        & 0     \\
Jet Energy Scale  (shape dep.)       & 0     &
  $^{+2.0}_{-2.2}$   & 
  $^{+3.1}_{-4.7}$   & 
  $^{+3.5}_{-5.1}$   & 
  $^{+10.6}_{-9.6}$   & 
  $^{+9.5}_{-9.4}$   & 
  0   & 
  $^{+2.2}_{-2.6}$   \\ 
Mistag Rate (shape dep.)      & 0     & 0    & 0    & 0    & 0              & 0             &   $^{+14.7}_{-14.8}$     & 0     \\
B-Tag Efficiency      & 0     &    4 &    4 &    4 &    4           &   4          & 0        &    4  \\
$t{\bar{t}}$ Cross Section         & 0     &   10 & 0    & 0    & 0              & 0             & 0        & 0     \\
Diboson Cross Section        & 0     & 0    & 6   & 6    & 0              & 0             & 0        & 0     \\
$\sigma(p{\bar{p}}\rightarrow Z+HF)$      & 0     & 0    & 0    & 0    &  40            & 40           & 0        & 0     \\
ISR (shape dep.)           & 0     & 0    & 0    & 0    & 0              & 0             & 0        &   $^{-4.1}_{-4.8}$     \\
FSR (shape dep.)           & 0     & 0    & 0    & 0    & 0              & 0             & 0        &   $^{-1.2}_{-2.4}$     \\
\end{tabular}
\end{ruledtabular}

\vskip 0.3cm
{\centerline{CDF: single tag (ST) low $s/b$~ $ZH \rightarrow \ell\ell b \bar{b}$ channel relative uncertainties (\%)}}
\vskip 0.099cm
\begin{ruledtabular}
\begin{tabular}{lcccccccc} \\
Contribution   & ~Fakes~ & ~~~Top~~~  & ~~$WZ$~~ & ~~$ZZ$~~ & ~$Z+b{\bar{b}}$~ & ~$Z+c{\bar{c}}$~& ~$Z+$mistag~ & ~~~$ZH$~~~ \\ \hline
Luminosity ($\sigma_{\mathrm{inel}}(p{\bar{p}})$)          & 0     &    3.8 &    3.8 &    3.8 &    3.8           &    3.8          & 0        &    3.8  \\
Luminosity Monitor        & 0     &    4.4 &    4.4 &    4.4 &    4.4           &    4.4          & 0        &    4.4  \\
Lepton ID    & 0     &    1 &    1 &    1 &    1           &    1          & 0        &    1  \\
Lepton Energy Scale    & 0     &    1.5 &    1.5 &    1.5 &    1.5           &    1.5          & 0        &    1.5  \\
$ZH$ Cross Section    & 0     &    0 &    0 &    0 &    0           &    0          & 0        &    5 \\
Fake Leptons       & 50    & 0    & 0    & 0    & 0              & 0             & 0        & 0     \\
Jet Energy Scale  (shape dep.)       & 0     &
  $^{+1.8}_{-1.6}$   & 
  $^{+6.8}_{-4.7}$   & 
  $^{+2.9}_{-6.2}$   & 
  $^{+11.6}_{-10.2}$   & 
  $^{+10.0}_{-10.3}$   & 
  0   & 
  $^{+3.9}_{-1.4}$   \\ 
Mistag Rate (shape dep.)      & 0     & 0    & 0    & 0    & 0              & 0             &   $^{+14.8}_{-14.9}$     & 0     \\
B-Tag Efficiency      & 0     &    4 &    4 &    4 &    4           &   4          & 0        &    4  \\
$t{\bar{t}}$ Cross Section         & 0     &   10 & 0    & 0    & 0              & 0             & 0        & 0     \\
Diboson Cross Section        & 0     & 0    & 6   & 6    & 0              & 0             & 0        & 0     \\
$\sigma(p{\bar{p}}\rightarrow Z+HF)$      & 0     & 0    & 0    & 0    &  40            & 40           & 0        & 0     \\
ISR (shape dep.)           & 0     & 0    & 0    & 0    & 0              & 0             & 0        &   $^{+7.4}_{-2.5}$     \\
FSR (shape dep.)           & 0     & 0    & 0    & 0    & 0              & 0             & 0        &   $^{+6.9}_{+1.9}$     \\
\end{tabular}
\end{ruledtabular}

\vskip 0.3cm
{\centerline{CDF: tight double tag (TDT) high $s/b$~ $ZH \rightarrow \ell\ell b \bar{b}$ channel relative uncertainties (\%)}}
\vskip 0.099cm
\begin{ruledtabular}
\begin{tabular}{lcccccccc} \\
Contribution   & ~Fakes~ & ~~~Top~~~  & ~~$WZ$~~ & ~~$ZZ$~~ & ~$Z+b{\bar{b}}$~ & ~$Z+c{\bar{c}}$~& ~$Z+$mistag~ & ~~~$ZH$~~~ \\ \hline
Luminosity ($\sigma_{\mathrm{inel}}(p{\bar{p}})$)          & 0     &    3.8 &    3.8 &    3.8 &    3.8           &    3.8          & 0        &    3.8  \\
Luminosity Monitor        & 0     &    4.4 &    4.4 &    4.4 &    4.4           &    4.4          & 0        &    4.4  \\
Lepton ID    & 0     &    1 &    1 &    1 &    1           &    1          & 0        &    1  \\
Lepton Energy Scale    & 0     &    1.5 &    1.5 &    1.5 &    1.5           &    1.5          & 0        &    1.5  \\
$ZH$ Cross Section    & 0     &    0 &    0 &    0 &    0           &    0          & 0        &    5 \\
Fake Leptons       & 50    & 0    & 0    & 0    & 0              & 0             & 0        & 0     \\
Jet Energy Scale  (shape dep.)       & 0     &
  $^{+1.5}_{-1.1}$   & 
  $^{+0.0}_{-0.0}$   & 
  $^{+1.8}_{-2.7}$   & 
  $^{+5.9}_{-6.9}$   & 
  $^{+6.0}_{-6.0}$   & 
  0   & 
  $^{+1.6}_{-0.3}$   \\ 
Mistag Rate (shape dep.)      & 0     & 0    & 0    & 0    & 0              & 0             &   $^{+30.9}_{-26.8}$     & 0     \\
B-Tag Efficiency      & 0     &    8 &    8 &    8 &    8           &   8          & 0        &    8  \\
$t{\bar{t}}$ Cross Section         & 0     &   10 & 0    & 0    & 0              & 0             & 0        & 0     \\
Diboson Cross Section        & 0     & 0    & 6   & 6    & 0              & 0             & 0        & 0     \\
$\sigma(p{\bar{p}}\rightarrow Z+HF)$      & 0     & 0    & 0    & 0    &  40            & 40           & 0        & 0     \\
ISR (shape dep.)           & 0     & 0    & 0    & 0    & 0              & 0             & 0        &   $^{-2.1}_{+0.4}$     \\
FSR (shape dep.)           & 0     & 0    & 0    & 0    & 0              & 0             & 0        &   $^{-1.7}_{-0.7}$     \\
\end{tabular}
\end{ruledtabular}

\end{center}
\end{table}

\begin{table}[t]
\begin{center}

\vskip 0.3cm
{\centerline{CDF: tight double tag (TDT) low $s/b$~ $ZH \rightarrow \ell\ell b \bar{b}$ channel relative uncertainties (\%)}}
\vskip 0.099cm
\begin{ruledtabular}
\begin{tabular}{lcccccccc} \\
Contribution   & ~Fakes~ & ~~~Top~~~  & ~~$WZ$~~ & ~~$ZZ$~~ & ~$Z+b{\bar{b}}$~ & ~$Z+c{\bar{c}}$~& ~$Z+$mistag~ & ~~~$ZH$~~~ \\ \hline
Luminosity ($\sigma_{\mathrm{inel}}(p{\bar{p}})$)          & 0     &    3.8 &    3.8 &    3.8 &    3.8           &    3.8          & 0        &    3.8  \\
Luminosity Monitor        & 0     &    4.4 &    4.4 &    4.4 &    4.4           &    4.4          & 0        &    4.4  \\
Lepton ID    & 0     &    1 &    1 &    1 &    1           &    1          & 0        &    1  \\
Lepton Energy Scale    & 0     &    1.5 &    1.5 &    1.5 &    1.5           &    1.5          & 0        &    1.5  \\
$ZH$ Cross Section    & 0     &    0 &    0 &    0 &    0           &    0          & 0        &    5 \\
Fake Leptons       & 50    & 0    & 0    & 0    & 0              & 0             & 0        & 0     \\
Jet Energy Scale  (shape dep.)       & 0     &
  $^{+0.5}_{-0.9}$   & 
  $^{+0.0}_{-0.0}$   & 
  $^{+0.0}_{-3.3}$   & 
  $^{+5.7}_{-6.2}$   & 
  $^{+7.2}_{-5.6}$   & 
  0   & 
  $^{+1.5}_{-0.6}$   \\ 
Mistag Rate (shape dep.)      & 0     & 0    & 0    & 0    & 0              & 0             &   $^{+31.5}_{-27.2}$     & 0     \\
B-Tag Efficiency      & 0     &    8 &    8 &    8 &    8           &   8          & 0        &    8  \\
$t{\bar{t}}$ Cross Section         & 0     &   10 & 0    & 0    & 0              & 0             & 0        & 0     \\
Diboson Cross Section        & 0     & 0    & 6   & 6    & 0              & 0             & 0        & 0     \\
$\sigma(p{\bar{p}}\rightarrow Z+HF)$      & 0     & 0    & 0    & 0    &  40            & 40           & 0        & 0     \\
ISR (shape dep.)           & 0     & 0    & 0    & 0    & 0              & 0             & 0        &   $^{-1.0}_{-2.7}$     \\
FSR (shape dep.)           & 0     & 0    & 0    & 0    & 0              & 0             & 0        &   $^{-5.3}_{-2.8}$     \\
\end{tabular}
\end{ruledtabular}

\vskip 0.3cm
{\centerline{CDF: loose double tag (LDT) high S/B~ $ZH \rightarrow \ell\ell b \bar{b}$ channel relative uncertainties (\%)}}
\vskip 0.099cm                                                                                                          
\begin{ruledtabular}
\begin{tabular}{lcccccccc} \\
Contribution   & ~Fakes~ & ~~~Top~~~  & ~~$WZ$~~ & ~~$ZZ$~~ & ~$Z+b{\bar{b}}$~ & ~$Z+c{\bar{c}}$~& ~$Z+$mistag~ & ~~~$ZH$~~~ \\ \hline
Luminosity ($\sigma_{\mathrm{inel}}(p{\bar{p}})$)          & 0     &    3.8 &    3.8 &    3.8 &    3.8           &    3.8          & 0        &    3.8  \\
Luminosity Monitor        & 0     &    4.4 &    4.4 &    4.4 &    4.4           &    4.4          & 0        &    4.4  \\
Lepton ID    & 0     &     &     &    1 &    1          &    1          & 0        &    1  \\
Lepton Energy Scale    & 0     &    1.5 &    1.5 &    1.5 &    1.5           &    1.5          & 0        &    1.5  \\
$ZH$ Cross Section    & 0     &    0 &    0 &    0 &    0           &    0          & 0        &    5 \\
Fake Leptons       & 50    & 0    & 0    & 0    & 0              & 0             & 0        & 0     \\
Jet Energy Scale  (shape dep.)       & 0     & 
  $^{+1.3}_{-0.6}$   & 
  $^{+3.2}_{-4.3}$   & 
  $^{+3.2}_{-3.0}$   & 
  $^{+7.4}_{-7.3}$   & 
  $^{+6.3}_{-6.0}$   & 
  0   & 
  $^{+1.04}_{-0.6}$   \\ 
Mistag Rate (shape dep.)      & 0     & 0    & 0    & 0    & 0              & 0             &   $^{+32.1}_{-25.7}$     & 0     \\
B-Tag Efficiency      & 0     &    11 &    11 &    11 &    11           &   11          & 0        &    11  \\
$t{\bar{t}}$ Cross Section         & 0     &   10 & 0    & 0    & 0              & 0             & 0        & 0     \\
Diboson Cross Section        & 0     & 0    & 6   & 6    & 0              & 0             & 0        & 0     \\
$\sigma(p{\bar{p}}\rightarrow Z+HF)$      & 0     & 0    & 0    & 0    &  40            & 40           & 0        & 0     \\
ISR (shape dep.)           & 0     & 0    & 0    & 0    & 0              & 0             & 0        &   $^{+1.4}_{-0.6}$     \\
FSR (shape dep.)           & 0     & 0    & 0    & 0    & 0              & 0             & 0        &   $^{+0.4}_{-2.0}$     \\
\end{tabular}
\end{ruledtabular}

\vskip 0.3cm
{\centerline{CDF: loose double tag (LDT) low S/B~ $ZH \rightarrow \ell\ell b \bar{b}$ channel relative uncertainties (\%)}}
\vskip 0.099cm                                                                                                          
\begin{ruledtabular}
\begin{tabular}{lcccccccc} \\
Contribution   & ~Fakes~ & ~~~Top~~~  & ~~$WZ$~~ & ~~$ZZ$~~ & ~$Z+b{\bar{b}}$~ & ~$Z+c{\bar{c}}$~& ~$Z+$mistag~ & ~~~$ZH$~~~ \\ \hline
Luminosity ($\sigma_{\mathrm{inel}}(p{\bar{p}})$)          & 0     &    3.8 &    3.8 &    3.8 &    3.8           &    3.8          & 0        &    3.8  \\
Luminosity Monitor        & 0     &    4.4 &    4.4 &    4.4 &    4.4           &    4.4          & 0        &    4.4  \\
Lepton ID    & 0     &     &     &    1 &    1          &    1          & 0        &    1  \\
Lepton Energy Scale    & 0     &    1.5 &    1.5 &    1.5 &    1.5           &    1.5          & 0        &    1.5  \\
$ZH$ Cross Section    & 0     &    0 &    0 &    0 &    0           &    0          & 0        &    5 \\
Fake Leptons       & 50    & 0    & 0    & 0    & 0              & 0             & 0        & 0     \\
Jet Energy Scale  (shape dep.)       & 0     & 
  $^{+1.7}_{-0.2}$   & 
  $^{-0.0}_{-3.4}$   & 
  $^{+3.1}_{-1.0}$   & 
  $^{+8.2}_{-8.6}$   & 
   $^{+8.0}_{-8.8}$   & 
  0   & 
  $^{+0.3}_{-1.8}$   \\ 
Mistag Rate (shape dep.)      & 0     & 0    & 0    & 0    & 0              & 0             &   $^{+31.7}_{-26.0}$     & 0     \\
B-Tag Efficiency      & 0     &    11 &    11 &    11 &    11           &   11          & 0        &    11  \\
$t{\bar{t}}$ Cross Section         & 0     &   10 & 0    & 0    & 0              & 0             & 0        & 0     \\
Diboson Cross Section        & 0     & 0    & 6   & 6    & 0              & 0             & 0        & 0     \\
$\sigma(p{\bar{p}}\rightarrow Z+HF)$      & 0     & 0    & 0    & 0    &  40            & 40           & 0        & 0     \\
ISR (shape dep.)           & 0     & 0    & 0    & 0    & 0              & 0             & 0        &   $^{+1.8}_{+5.3}$     \\
FSR (shape dep.)           & 0     & 0    & 0    & 0    & 0              & 0             & 0        &   $^{+23.0}_{+7.9}$     \\
\end{tabular}
\end{ruledtabular}

\end{center}
\end{table}

\begin{table}[t]
\begin{center}
\caption{\label{tab:cdfllbb2} Systematic uncertainties on the signal and background contributions for CDF's 
$ZH\rightarrow \ell^+\ell^-b{\bar{b}}$ single tag (ST), loose double tag (LDT), and tight double tag (TDT)
loose muon channels. The channels are further divided to separate events collected from either the muon or 
missing $E_T$ trigger path.  Systematic uncertainties are listed by name; see the original references for a 
detailed explanation of their meaning and on how they are derived.  Systematic uncertainties for $ZH$ shown 
in this table are obtained for $m_H=115$ GeV/$c^2$.  Uncertainties are relative, in percent, and are symmetric 
unless otherwise indicated.}
\vskip 0.1cm
{\centerline{CDF: single tag (ST) loose muons (muon trigger) ~ $ZH \rightarrow \ell\ell b \bar{b}$ channel relative uncertainties (\%)}}
\vskip 0.099cm
\begin{ruledtabular}
\begin{tabular}{lcccccccc} \\
Contribution   & ~Fakes~ & ~~~Top~~~  & ~~$WZ$~~ & ~~$ZZ$~~ & ~$Z+b{\bar{b}}$~ & ~$Z+c{\bar{c}}$~& ~$Z+$mistag~ & ~~~$ZH$~~~ \\ \hline
Luminosity ($\sigma_{\mathrm{inel}}(p{\bar{p}})$)          & 0     &    3.8 &    3.8 &    3.8 &    3.8           &    3.8          & 0        &    3.8  \\
Luminosity Monitor        & 0     &    4.4 &    4.4 &    4.4 &    4.4           &    4.4          & 0        &    4.4  \\
Lepton ID    & 0     &    1 &    1 &    1 &    1           &    1          & 0        &    1  \\
Lepton Energy Scale    & 0     &    1.5 &    1.5 &    1.5 &    1.5           &    1.5          & 0        &    1.5  \\
$ZH$ Cross Section    & 0     &    0 &    0 &    0 &    0           &    0          & 0        &    5 \\
Fake Leptons       & 50    & 0    & 0    & 0    & 0              & 0             & 0        & 0     \\
Jet Energy Scale  (shape dep.)       & 0     &
  $^{+0.01}_{-0.01}$   & 
  $^{+0.0}_{-1.3}$   & 
  $^{+1.3}_{-2.1}$   & 
  $^{+2.9}_{-2.8}$   & 
  $^{+3.2}_{-2.3}$   & 
  0   & 
  $^{+0.2}_{-0.3}$   \\ 
Mistag Rate (shape dep.)      & 0     & 0    & 0    & 0    & 0              & 0             &   $^{+14.3}_{-14.4}$     & 0     \\
B-Tag Efficiency      & 0     &    4 &    4 &    4 &    4           &   4          & 0        &    4  \\
$t{\bar{t}}$ Cross Section         & 0     &   10 & 0    & 0    & 0              & 0             & 0        & 0     \\
Diboson Cross Section        & 0     & 0    & 6   & 6    & 0              & 0             & 0        & 0     \\
$\sigma(p{\bar{p}}\rightarrow Z+HF)$      & 0     & 0    & 0    & 0    &  40            & 40           & 0        & 0     \\
ISR/FSR           & 0     & 0    & 0    & 0    & 0              & 0             & 0        &   5     \\
NN Trigger Model           & 0     & 5    & 5    & 5   & 5              & 5             & 0        &   5     \\
\end{tabular}
\end{ruledtabular}

\vskip 0.3cm
{\centerline{CDF: loose double tag (LDT) loose muons (muon trigger) ~ $ZH \rightarrow \ell\ell b \bar{b}$ channel relative uncertainties (\%)}}
\vskip 0.099cm
\begin{ruledtabular}
\begin{tabular}{lcccccccc} \\
Contribution   & ~Fakes~ & ~~~Top~~~  & ~~$WZ$~~ & ~~$ZZ$~~ & ~$Z+b{\bar{b}}$~ & ~$Z+c{\bar{c}}$~& ~$Z+$mistag~ & ~~~$ZH$~~~ \\ \hline
Luminosity ($\sigma_{\mathrm{inel}}(p{\bar{p}})$)          & 0     &    3.8 &    3.8 &    3.8 &    3.8           &    3.8          & 0        &    3.8  \\
Luminosity Monitor        & 0     &    4.4 &    4.4 &    4.4 &    4.4           &    4.4          & 0        &    4.4  \\
Lepton ID    & 0     &    1 &    1 &    1 &    1           &    1          & 0        &    1  \\
Lepton Energy Scale    & 0     &    1.5 &    1.5 &    1.5 &    1.5           &    1.5          & 0        &    1.5  \\
$ZH$ Cross Section    & 0     &    0 &    0 &    0 &    0           &    0          & 0        &    5 \\
Fake Leptons       & 50    & 0    & 0    & 0    & 0              & 0             & 0        & 0     \\
Jet Energy Scale  (shape dep.)       & 0     &
  $^{+0.1}_{-0.9}$   & 
  $^{+0.0}_{-0.0}$   & 
  $^{+0.0}_{-0.0}$   & 
  $^{+3.7}_{-4.2}$   & 
  $^{+4.0}_{-1.6}$   & 
  0   & 
  $^{+0.1}_{-0.0}$   \\ 
Mistag Rate (shape dep.)      & 0     & 0    & 0    & 0    & 0              & 0             &   $^{+33.6}_{-26.2}$     & 0     \\
B-Tag Efficiency      & 0     &    11 &    11 &    11 &    11           &   11         & 0        &    11  \\
$t{\bar{t}}$ Cross Section         & 0     &   10 & 0    & 0    & 0              & 0             & 0        & 0     \\
Diboson Cross Section        & 0     & 0    & 6   & 6    & 0              & 0             & 0        & 0     \\
$\sigma(p{\bar{p}}\rightarrow Z+HF)$      & 0     & 0    & 0    & 0    &  40            & 40           & 0        & 0     \\
ISR/FSR           & 0     & 0    & 0    & 0    & 0              & 0             & 0        &   3     \\
NN Trigger Model           & 0     & 5    & 5    & 5   & 5              & 5             & 0        &   5     \\
\end{tabular}
\end{ruledtabular}

\vskip 0.3cm
{\centerline{CDF: tight double tag (TDT) loose muons (muon trigger) ~ $ZH \rightarrow \ell\ell b \bar{b}$ channel relative uncertainties (\%)}}
\vskip 0.099cm
\begin{ruledtabular}
\begin{tabular}{lcccccccc} \\
Contribution   & ~Fakes~ & ~~~Top~~~  & ~~$WZ$~~ & ~~$ZZ$~~ & ~$Z+b{\bar{b}}$~ & ~$Z+c{\bar{c}}$~& ~$Z+$mistag~ & ~~~$ZH$~~~ \\ \hline
Luminosity ($\sigma_{\mathrm{inel}}(p{\bar{p}})$)          & 0     &    3.8 &    3.8 &    3.8 &    3.8           &    3.8          & 0        &    3.8  \\
Luminosity Monitor        & 0     &    4.4 &    4.4 &    4.4 &    4.4           &    4.4          & 0        &    4.4  \\
Lepton ID    & 0     &    1 &    1 &    1 &    1           &    1          & 0        &    1  \\
Lepton Energy Scale    & 0     &    1.5 &    1.5 &    1.5 &    1.5           &    1.5          & 0        &    1.5  \\
$ZH$ Cross Section    & 0     &    0 &    0 &    0 &    0           &    0          & 0        &    5 \\
Fake Leptons       & 50    & 0    & 0    & 0    & 0              & 0             & 0        & 0     \\
Jet Energy Scale  (shape dep.)       & 0     &
  $^{+1.2}_{-0.0}$   & 
  $^{+0.0}_{-0.0}$   & 
  $^{+0.0}_{-0.0}$   & 
  $^{+2.1}_{-3.3}$   & 
  $^{+1.3}_{-0.0}$   & 
  0   & 
  $^{+0.0}_{-0.0}$   \\ 
Mistag Rate (shape dep.)      & 0     & 0    & 0    & 0    & 0              & 0             &   $^{+30.7}_{-26.6}$     & 0     \\
B-Tag Efficiency      & 0     &    8 &    8 &    8 &    8           &   8         & 0        &    8  \\
$t{\bar{t}}$ Cross Section         & 0     &   10 & 0    & 0    & 0              & 0             & 0        & 0     \\
Diboson Cross Section        & 0     & 0    & 6   & 6    & 0              & 0             & 0        & 0     \\
$\sigma(p{\bar{p}}\rightarrow Z+HF)$      & 0     & 0    & 0    & 0    &  40            & 40           & 0        & 0     \\
ISR/FSR           & 0     & 0    & 0    & 0    & 0              & 0             & 0        &   1     \\
NN Trigger Model           & 0     & 5    & 5    & 5   & 5              & 5             & 0        &   5     \\
\end{tabular}
\end{ruledtabular}

\end{center}
\end{table}

\begin{table}[t]
\begin{center}

\vskip 0.3cm
{\centerline{CDF: single tag (ST) loose muons (missing $E_T$ trigger) ~ $ZH \rightarrow \ell\ell b \bar{b}$ channel relative uncertainties (\%)}}
\vskip 0.099cm
\begin{ruledtabular}
\begin{tabular}{lcccccccc} \\ 
Contribution   & ~Fakes~ & ~~~Top~~~  & ~~$WZ$~~ & ~~$ZZ$~~ & ~$Z+b{\bar{b}}$~ & ~$Z+c{\bar{c}}$~& ~$Z+$mistag~ & ~~~$ZH$~~~ \\ \hline
Luminosity ($\sigma_{\mathrm{inel}}(p{\bar{p}})$)          & 0     &    3.8 &    3.8 &    3.8 &    3.8           &    3.8          & 0        &    3.8  \\
Luminosity Monitor        & 0     &    4.4 &    4.4 &    4.4 &    4.4           &    4.4          & 0        &    4.4  \\
Lepton ID    & 0     &    1 &    1 &    1 &    1           &    1          & 0        &    1  \\
Lepton Energy Scale    & 0     &    1.5 &    1.5 &    1.5 &    1.5           &    1.5          & 0        &    1.5  \\
$ZH$ Cross Section    & 0     &    0 &    0 &    0 &    0           &    0          & 0        &    5 \\
Fake Leptons       & 50    & 0    & 0    & 0    & 0              & 0             & 0        & 0     \\
Jet Energy Scale  (shape dep.)       & 0     &
  $^{+0.0}_{-0.1}$   & 
  $^{+0.0}_{-0.0}$   & 
  $^{+0.6}_{-0.4}$   & 
  $^{+0.6}_{-0.7}$   & 
  $^{+0.7}_{-1.0}$   & 
  0   & 
  $^{+0.0}_{-0.2}$   \\ 
Mistag Rate (shape dep.)      & 0     & 0    & 0    & 0    & 0              & 0             &   $^{+14.1}_{-14.1}$     & 0     \\
B-Tag Efficiency      & 0     &    4 &    4 &    4 &    4           &   4          & 0        &    4  \\
$t{\bar{t}}$ Cross Section         & 0     &   10 & 0    & 0    & 0              & 0             & 0        & 0     \\
Diboson Cross Section        & 0     & 0    & 6   & 6    & 0              & 0             & 0        & 0     \\
$\sigma(p{\bar{p}}\rightarrow Z+HF)$      & 0     & 0    & 0    & 0    &  40            & 40           & 0        & 0     \\
ISR/FSR           & 0     & 0    & 0    & 0    & 0              & 0             & 0        &   3    \\
NN Trigger Model           & 0     & 5    & 5    & 5   & 5              & 5             & 0        &   5     \\
\end{tabular}
\end{ruledtabular}

\vskip 0.3cm
{\centerline{CDF: loose double tag (LDT) loose muons (missing $E_T$ trigger) ~ $ZH \rightarrow \ell\ell b \bar{b}$ channel relative uncertainties (\%)}}
\vskip 0.099cm
\begin{ruledtabular}
\begin{tabular}{lcccccccc} \\
Contribution   & ~Fakes~ & ~~~Top~~~  & ~~$WZ$~~ & ~~$ZZ$~~ & ~$Z+b{\bar{b}}$~ & ~$Z+c{\bar{c}}$~& ~$Z+$mistag~ & ~~~$ZH$~~~ \\ \hline
Luminosity ($\sigma_{\mathrm{inel}}(p{\bar{p}})$)          & 0     &    3.8 &    3.8 &    3.8 &    3.8           &    3.8          & 0        &    3.8  \\
Luminosity Monitor        & 0     &    4.4 &    4.4 &    4.4 &    4.4           &    4.4          & 0        &    4.4  \\
Lepton ID    & 0     &    1 &    1 &    1 &    1           &    1          & 0        &    1  \\
Lepton Energy Scale    & 0     &    1.5 &    1.5 &    1.5 &    1.5           &    1.5          & 0        &    1.5  \\
$ZH$ Cross Section    & 0     &    0 &    0 &    0 &    0           &    0          & 0        &    5 \\
Fake Leptons       & 50    & 0    & 0    & 0    & 0              & 0             & 0        & 0     \\
Jet Energy Scale  (shape dep.)       & 0     &
  $^{+0.0}_{-0.4}$   & 
  $^{+0.0}_{-0.0}$   & 
  $^{+0.0}_{-0.0}$   & 
  $^{+0.7}_{-0.3}$   & 
  $^{+0.0}_{-1.3}$   & 
  0   & 
  $^{+0.2}_{-0.2}$   \\ 
Mistag Rate (shape dep.)      & 0     & 0    & 0    & 0    & 0              & 0             &   $^{+39.0}_{-29.5}$     & 0     \\
B-Tag Efficiency      & 0     &    11 &    11 &    11 &    11           &   11         & 0        &    11  \\
$t{\bar{t}}$ Cross Section         & 0     &   10 & 0    & 0    & 0              & 0             & 0        & 0     \\
Diboson Cross Section        & 0     & 0    & 6   & 6    & 0              & 0             & 0        & 0     \\
$\sigma(p{\bar{p}}\rightarrow Z+HF)$      & 0     & 0    & 0    & 0    &  40            & 40           & 0        & 0     \\
ISR/FSR           & 0     & 0    & 0    & 0    & 0              & 0             & 0        &   1     \\
NN Trigger Model           & 0     & 5    & 5    & 5   & 5              & 5             & 0        &   5     \\
\end{tabular}
\end{ruledtabular}

\vskip 0.3cm
{\centerline{CDF: tight double tag (TDT) loose muons (missing $E_T$ trigger) ~ $ZH \rightarrow \ell\ell b \bar{b}$ channel relative uncertainties (\%)}}
\vskip 0.099cm
\begin{ruledtabular}
\begin{tabular}{lcccccccc} \\
Contribution   & ~Fakes~ & ~~~Top~~~  & ~~$WZ$~~ & ~~$ZZ$~~ & ~$Z+b{\bar{b}}$~ & ~$Z+c{\bar{c}}$~& ~$Z+$mistag~ & ~~~$ZH$~~~ \\ \hline
Luminosity ($\sigma_{\mathrm{inel}}(p{\bar{p}})$)          & 0     &    3.8 &    3.8 &    3.8 &    3.8           &    3.8          & 0        &    3.8  \\
Luminosity Monitor        & 0     &    4.4 &    4.4 &    4.4 &    4.4           &    4.4          & 0        &    4.4  \\
Lepton ID    & 0     &    1 &    1 &    1 &    1           &    1          & 0        &    1  \\
Lepton Energy Scale    & 0     &    1.5 &    1.5 &    1.5 &    1.5           &    1.5          & 0        &    1.5  \\
$ZH$ Cross Section    & 0     &    0 &    0 &    0 &    0           &    0          & 0        &    5 \\
Fake Leptons       & 50    & 0    & 0    & 0    & 0              & 0             & 0        & 0     \\
Jet Energy Scale  (shape dep.)       & 0     &
  $^{+0.0}_{-0.0}$   & 
  $^{+0.0}_{-0.0}$   & 
  $^{+0.0}_{-0.0}$   & 
  $^{+0.4}_{-0.3}$   & 
  $^{+0.3}_{-0.1}$   & 
  0   & 
  $^{+0.5}_{-0.5}$   \\ 
Mistag Rate (shape dep.)      & 0     & 0    & 0    & 0    & 0              & 0             &   $^{+29.7}_{-25.8}$     & 0     \\
B-Tag Efficiency      & 0     &    8 &    8 &    8 &    8           &   8         & 0        &    8  \\
$t{\bar{t}}$ Cross Section         & 0     &   10 & 0    & 0    & 0              & 0             & 0        & 0     \\
Diboson Cross Section        & 0     & 0    & 6   & 6    & 0              & 0             & 0        & 0     \\
$\sigma(p{\bar{p}}\rightarrow Z+HF)$      & 0     & 0    & 0    & 0    &  40            & 40           & 0        & 0     \\
ISR/FSR           & 0     & 0    & 0    & 0    & 0              & 0             & 0        &   4     \\
NN Trigger Model           & 0     & 5    & 5    & 5   & 5              & 5             & 0        &   5     \\
\end{tabular}
\end{ruledtabular}

\end{center}
\end{table}

\begin{table}
\caption{\label{tab:d0llbb1}Systematic uncertainties on the contributions for D0's $ZH\rightarrow \ell^+\ell^-b{\bar{b}}$ channels.
Systematic uncertainties are listed by name; see the original references for a detailed explanation of their meaning and on how they
are derived.
Systematic uncertainties for $ZH$  shown in this table are obtained for $m_H=115$ GeV/$c^2$.
Uncertainties are relative, in percent, and are symmetric unless otherwise indicated. Shape uncertainties are 
labeled with an ``s''. }
\vskip 0.8cm
{\centerline{D0: ~ $ZH \rightarrow \ell\ell b \bar{b}$ analyses relative uncertainties (\%)}}
\vskip 0.099cm
\begin{ruledtabular}
\begin{tabular}{  l  c  c  c  c  c  c  c  c }   
Contribution              & Signal  & Multijet& $Z$+LF  &  $Z$\bb & $Z$\cc & Diboson & \ttbar\\ hline
Jet Energy Scale (S)      &  3.8  &         &  2.1   &  3.5   &  3.8   &  4.8   &  3.3  \\
Jet Energy Resolution (S) &  3.5  &         &  4.4   &   10   &  9.5   &  3.9   &  3.6  \\
Jet ID (S)                & 0.53  &         & 0.83   & 0.40   & 0.08   & 0.85   & 0.68  \\
Taggability (S)           & 3.5   &         &  2.6   &  1.9   &  2.6   & 4.7    & 3.5   \\
$Z p_T$ Model (S)         &       &         &  4.4   &  4.5   &  4.5   &        &        \\
HF Tagging Efficiency (S) &  1.6  &         &         &  3.7   &  6.4   &  6.9   &  1.3  \\
LF Tagging Efficiency (S) &        &   52   &   49   &         &         &  6.9  &        \\
$ee$ Multijet Shape (S)  &        &   13  &        &        &        &        &        \\
Multijet Normalization    &        &  20-50   &         &         &         &         &    \\
$Z$+jets Jet Angles (S)   &        &         & 0.87   & 0.52   & 0.49  &        &        \\
Alpgen MLM (S)            &        &         & 0.36  &        &        &        &        \\
Alpgen Scale (S)          &        &         & 0.23   & 0.16   & 0.15  &        &        \\
Underlying Event (S)      &        &         & 0.01   & 0.06   & 0.14  &        &        \\
Modeling (S)		  & 3      &         &  2     &  2     &  2     &  2      & 7    \\
Trigger (S)               & 0.52  &         & 0.64   & 0.41   & 0.38   & 0.49   & 0.69  \\
Cross Sections            & 6.0   &         &         & 20     & 20     & 7      & 10  \\
Normalization             & 11-14  &      &  2-10    & 2-10  & 2-10  & 10-15  & 10-15  \\
PDFs                      & 0.55  &         & 1      & 2.4    & 1.1    & 0.66   & 5.9 

\end{tabular}
\end{ruledtabular}
\vskip 0.8cm
{\centerline{D0: Double Tag (DT)~ $ZH \rightarrow \ell\ell b \bar{b}$ analysis relative uncertainties (\%)}}
\vskip 0.099cm
\begin{ruledtabular}
\begin{tabular}{  l  c c  c  c  c  c  c  c }  \\
Contribution             & Signal & Multijet& $Z$+LF  &  $Z$\bb & $Z$\cc & Diboson & \ttbar\\  \hline
Jet Energy Scale (S)     &  3.7  &         &  4.6   &  5.2   &  5.4   &  4.9   &  3.2  \\
Jet Energy Resolution(S) &  2.6  &         &   10   &   13   &   12   &  3.2   &  2.9  \\
JET ID (S)               &  1.4  &         &  3.0   &  1.0   &  1.4   &  2.2   & 0.96  \\
Taggability (S)          &  6.2  &         &  4.3   &  4.5   &  4.4   &  6.8   &  6.4  \\
$Z_{p_T}$ Model (S)       &        &         &  4.4   &  4.2   &  4.2   &         &        \\
HF Tagging Efficiency (S)&  8.3  &         &         &  7.7   &  9.9   &  8.4   &  8.9  \\
LF Tagging Efficiency (S)&        &   36   &   15   &         &         &  8.4   &        \\
$ee$ Multijet Shape (S) &        &  7.9   &         &         &         &         &        \\
Multijet Normalization   &        &  20-50   &         &         &         &         &    \\
$Z$+jets Jet Angles (S)  &        &         &  1.4   & 0.59   & 0.83   &         &        \\
Alpgen MLM (S)           &        &         & 0.33   &         &         &         &        \\
Alpgen Scale (S)         &        &         & 0.29   & 0.15   & 0.16   &         &        \\
Underlying Event(S)      &        &         & 0.05   & 0.11   & 0.16   &         &        \\
Modeling (S)              &2      &         & 0.6    & 2      &2       &2       &6      \\
Trigger (S)              & 0.62  &         &  1.3   & 0.55   & 0.74   & 0.70   & 0.86  \\
Cross Sections           & 6.0   &         &         & 20     & 20     & 7      & 10  \\
Normalization         & 11-14  &       &  2-10    & 2-10  & 2-10  & 10-15  & 10-15  \\
PDFs                     & 0.55  &         & 1      & 2.4    & 1.1    & 0.66   & 5.9 
\end{tabular}
\end{ruledtabular}
\end{table}

\clearpage


\begin{table}
\begin{center}
\caption{\label{tab:cdfsystww0} Systematic uncertainties on the signal and background contributions for CDF's 
$H\rightarrow W^+W^-\rightarrow\ell^{\pm}\ell^{\prime \mp}$ channels with zero, one, and two or more associated 
jets.  These channels are sensitive to gluon fusion production (all channels) and $WH, ZH$ and VBF production.  
Systematic uncertainties are listed by name (see the original references for a detailed explanation of their 
meaning and on how they are derived).  Systematic uncertainties for $H$ shown in this table are obtained for 
$m_H=160$ GeV/$c^2$.  Uncertainties are relative, in percent, and are symmetric unless otherwise indicated.  
The uncertainties associated with the different background and signal processed are correlated within individual 
jet categories unless otherwise noted.  Boldface and italics indicate groups of uncertainties which are correlated 
with each other but not the others on the line.}

\vskip 0.1cm
{\centerline{CDF: $H\rightarrow W^+W^-\rightarrow\ell^{\pm}\ell^{\prime \mp}$ with no associated jet channel relative uncertainties (\%)}}
\vskip 0.099cm
\begin{ruledtabular}
\begin{tabular}{lccccccccccc} \\
Contribution               &   $WW$     &  $WZ$         &  $ZZ$  &  $t\bar{t}$   &  DY    &  $W\gamma$   & $W$+jet &$gg\to H$&  $WH$ &  $ZH$  &  VBF  \\ \hline
{\bf Cross Section :}      &        	&        	&        	&        &        	&        &         &        &        &        & 	\\ 
Scale                      &        	&        	&        	&        &        	&        &         &  7.0  &        &        &       	\\ 
PDF Model                  &        	&        	&        	&        &        	&        &         &  7.6  &        &        &       	\\ 
Total                      & {\it 6.0 }& {\it 6.0 }   & {\it 6.0 }   & 10.0  &  		& 	 & 	   & 	    &  {\bf 5.0 } &  {\bf 5.0 } &       10.0           \\ 
{\bf Acceptance :}         &        	&        	&        	&        &        	&        &         &        &        &        & 	\\ 
Scale (leptons)            &        	&        	&        	&        &        	&        &         &  1.7  &        &        &       	\\ 
Scale (jets)               & {\it 0.3 }&        	&        	&        &        	&        &         &  1.5  &        &        &      	\\  
PDF Model (leptons)        &            &       	&        	&        &              &  	 &         &  2.7  &        &        &         \\  
PDF Model (jets)           & {\it 1.1 } &        	&        	&        &        	&        &         &  5.5  &        &        &      	\\  
Higher-order Diagrams      &            & {\it 10.0 }  & {\it 10.0 } 	& 10.0  &              & 10.0  &   	   &        & {\bf 10.0 } & {\bf 10.0 } & {\bf 10.0 }          \\ 
$\MET$ Modeling            &  	 	&      	        &     	        &        & 19.5  	&	 &         &        &        &        &       	\\ 
Conversion Modeling       &             &               &               &        &              & 10.0  &         &        &        &        &       	\\ 
Jet Fake Rates             &        	&        	&        	&        &        	&        &         &        &        &        &       	\\ 
(Low S/B)                  &        	&        	&        	&        &        	&        & 22.0   &        &        &        &       	\\
(High S/B)                 &        	&        	&        	&        &        	&        & 25.0   &        &        &        &       	\\ 
Jet Energy Scale          &  {\it 2.6 }  & {\it 6.1 }   &   {\it 3.4 }   &   {\it 26.0 }  &  {\it 17.5 } &   {\it 3.1 } &   & {\it 5.0 } & {\it 10.5 } & {\it 5.0 } & {\it 11.5 } \\ 
Lepton ID Efficiencies     & {\it 3.0 } & {\it 3.0 } 	& {\it 3.0 } 	&{\it 3.0 }&{\it 3.0 }&  	 &         & {\it 3.0 } &  {\it 3.0 } &  {\it 3.0 } &  {\it 3.0 } \\ 
Trigger Efficiencies       & {\it 2.0 } & {\it 2.0 } 	& {\it 2.0 } 	&{\it 2.0 }&{\it 2.0 }&  	 &         & {\it 2.0 } &  {\it 2.0 } &  {\it 2.0 } &  {\it 2.0 } \\ 
{\bf Luminosity}           & {\it 3.8 } & {\it 3.8 } 	& {\it 3.8 } 	&{\it 3.8 }&{\it 3.8 }&  	 &         & {\it 3.8 } &  {\it 3.8 } &  {\it 3.8 } &  {\it 3.8 } \\ 
{\bf Luminosity Monitor}   & {\it 4.4 } & {\it 4.4 } 	& {\it 4.4 } 	&{\it 4.4 }&{\it 4.4 }&  	 &         & {\it 4.4 } &  {\it 4.4 } &  {\it 4.4 } &  {\it 4.4 } \\ 
\end{tabular}
\end{ruledtabular}

\vskip 0.3cm
{\centerline{CDF: $H\rightarrow W^+W^-\rightarrow\ell^{\pm}\ell^{\prime \mp}$ with one associated jet channel relative uncertainties (\%)}}
\vskip 0.099cm
\begin{ruledtabular}
\begin{tabular}{lccccccccccc} \\ 
Contribution               &   $WW$     &  $WZ$         &  $ZZ$  &  $t\bar{t}$   &  DY     & $W\gamma$   & $W$+jet &$gg \to H$&  $WH$  &  $ZH$  &  VBF \\ \hline
{\bf Cross Section :}      &        	&        	&        	&        &        	&        &         &        &        &        &        \\ 
Scale                      &        	&        	&        	&        &        	&        &         & 23.5  &        &        &        \\ 
PDF Model                  &        	&        	&        	&        &        	&        &         & 17.3  &        &        &        \\ 
Total                      & {\it 6.0 }& {\it 6.0 }   & {\it 6.0 }   & 10.0  &  	 	& 	 & 	   & 	    &{\bf 5.0 }&{\bf 5.0 }& 10.0  \\ 
{\bf Acceptance :}         &        	&        	&        	&        &        	&        &         &        &        &        &        \\ 
Scale (leptons)            &        	&        	&        	&        &        	&        &         &  2.2  &        &        &        \\ 
Scale (jets)               &{\it -4.0 } &        	&        	&        &        	&        &         & -1.9  &        &        &        \\ 
PDF Model (leptons)        &            &               &               &        &              &  	 &	   &{\it 3.6 }&     &        &        \\ 
PDF Model (jets)           &{\it  4.7 }&        	&        	&        &        	&        &         & -6.3  &        &        &        \\ 
Higher-order Diagrams      &            & {\it 10.0 }  & {\it 10.0 }  & 10.0  &  	        & 10.0  &   	   &        &{\bf 10.0 }&{\bf 10.0 }&{\bf 10.0 }\\ 
$\MET$ Modeling            & 	 	&   	        &       	&        & 20.0  	& 	 &         &        &        &        &        \\ 
Conversion Modeling       &            &               &               &        &              & 10.0  &         &        &        &        &        \\ 
Jet Fake Rates             &        	&        	&        	&        &        	&        &         &        &        &        &        \\ 
(Low S/B)                  &        	&        	&        	&        &        	&        & 23.0   &        &        &        &        \\
(High S/B)                 &        	&        	&        	&        &        	&        & 28.0   &        &        &        &        \\ 
Jet Energy Scale &  {\it -5.5 }  & {\it -1.0 } &   {\it -4.3 }  &   {\it -13.0 } & {\it -6.5 } &   {\it -9.5 } &   & {\it -4.0 } & {\it -8.5 } & {\it -7.0 } & {\it -6.5 } \\ 
Lepton ID Efficiencies     & {\it 3.0 } & {\it 3.0 } 	& {\it 3.0 } 	&{\it 3.0 }&{\it 3.0 }& 	 &      &{\it 3.0 }&{\it 3.0 }&{\it 3.0 }&{\it 3.0 }\\ 
Trigger Efficiencies       & {\it 2.0 } & {\it 2.0 } 	& {\it 2.0 } 	&{\it 2.0 }&{\it 2.0 }& 	 &      &{\it 2.0 }&{\it 2.0 }&{\it 2.0 }&{\it 2.0 }\\ 
{\bf Luminosity}           & {\it 3.8 } & {\it 3.8 } 	& {\it 3.8 } 	&{\it 3.8 }&{\it 3.8 }& 	 &      &{\it 3.8 }&{\it 3.8 }&{\it 3.8 }&{\it 3.8 }\\ 
{\bf Luminosity Monitor}   & {\it 4.4 } & {\it 4.4 } 	& {\it 4.4 } 	&{\it 4.4 }&{\it 4.4 }& 	 &      &{\it 4.4 }&{\it 4.4 }&{\it 4.4 }&{\it 4.4 }\\ 
\end{tabular}
\end{ruledtabular}

\end{center}
\end{table}

\begin{table}
\begin{center}
\vskip 0.3cm
{\centerline{CDF: $H\rightarrow W^+W^-\rightarrow\ell^{\pm}\ell^{\prime \mp}$ with two or more associated jets channel relative uncertainties (\%)}}
\vskip 0.0999cm
\begin{ruledtabular}
\begin{tabular}{lccccccccccc} \\
Contribution               &  $WW$    &  $WZ$         &  $ZZ$   &  $t\bar{t}$  &  DY   &  $W\gamma$    & $W$+jet &$gg\to H$&  $WH$ &  $ZH$  &  VBF    \\ \hline
{\bf Cross Section :}      &        	&        	&        	&        &        	&        &         &        &        &        &        \\ 
Scale                      &        	&        	&        	&        &        	&        &         & 67.5  &        &        &        	\\ 
PDF Model                  &        	&        	&        	&        &        	&        &         & 29.7  &        &        &        	\\ 
Total                      & {\it 6.0 }& {\it 6.0 }   & {\it 6.0 }   & 10.0  &  	 	& 	 & 	   &        &{\bf 5.0 }&{\bf 5.0 }& 10.0  \\ 
{\bf Acceptance :}         &        	&        	&        	&        &        	&        &         &        &        &        &        \\ 
Scale (leptons)            &        	&        	&        	&        &        	&        &         &  3.1  &        &        &        	\\ 
Scale (jets)               & {\it -8.2 }&        	&        	&        &        	&        &         & -6.8  &        &        &        	\\ 
PDF Model (leptons)        & 		 &  		&  		&	 &	        &  	 &      &{\it 4.8 }&	     &        &         \\ 
PDF Model (jets)           & {\it 4.2 }&        	&        	&        &        	&        &         & -12.3  &        &        &        	\\ 
Higher-order Diagrams      &  		& {\it 10.0 }  & {\it 10.0 }  & 10.0  &              & 10.0  & 	   &        &{\bf 10.0 }&{\bf 10.0 }&{\bf 10.0 }\\ 
$\MET$ Modeling            &  		&   	        &     	        &        & 25.5  	& 	 &         &        &        &        &     	\\ 
Conversion Modeling      &            &               &               &        &              & 10.0  &         &        &        &        &     	\\ 
Jet Fake Rates             &        	&        	&        	&        &        	&        & 28.0   &        &        &        &        	\\ 
Jet Energy Scale& {\it -14.8 } & {\it -12.9 } & {\it -12.1 }  & {\it -1.7 }  & {\it -29.2 } & {\it -22.0 } &  & {\it -17.0 } & {\it -4.0 } & {\it -2.3 } & {\it -4.0 } \\ 
$b$-tag Veto               &        	&        	&        	&  3.8  &        	&        &         &        &        &        &        	\\ 
Lepton ID Efficiencies     & {\it 3.0 } &  {\it 3.0 } &  {\it 3.0 }  &{\it 3.0 }&{\it 3.0 }&   	 &      &{\it 3.0 }&{\it 3.0 }&{\it 3.0 }&{\it 3.0 }\\ 
Trigger Efficiencies       & {\it 2.0 } &  {\it 2.0 } &  {\it 2.0 }  &{\it 2.0 }&{\it 2.0 }&  	 &      &{\it 2.0 }&{\it 2.0 }&{\it 2.0 }&{\it 2.0 }\\ 
{\bf Luminosity}           & {\it 3.8 } &  {\it 3.8 } &  {\it 3.8 }  &{\it 3.8 }&{\it 3.8 }&  	 &      &{\it 3.8 }&{\it 3.8 }&{\it 3.8 }&{\it 3.8 }\\ 
{\bf Luminosity Monitor}   & {\it 4.4 } &  {\it 4.4 } &  {\it 4.4 }  &{\it 4.4 }&{\it 4.4 }&  	 &      &{\it 4.4 }&{\it 4.4 }&{\it 4.4 }&{\it 4.4 }\\ 
\end{tabular}
\end{ruledtabular}
\end{center}
\end{table}

%
%

\begin{table}
\begin{center}
\caption{\label{tab:cdfsystww4} Systematic uncertainties on the signal and background contributions for CDF's low-$M_{\ell\ell}$ 
$H\rightarrow W^+W^-\rightarrow\ell^{\pm}\ell^{\prime \mp}$ channel with zero or one associated jets.  This channel is sensitive 
to only gluon fusion production.  Systematic uncertainties are listed by name (see the original references for a detailed 
explanation of their meaning and on how they are derived).  Systematic uncertainties for $H$ shown in this table are obtained 
for $m_H=160$ GeV/$c^2$.  Uncertainties are relative, in percent, and are symmetric unless otherwise indicated.  The uncertainties 
associated with the different background and signal processed are correlated within individual categories unless otherwise noted.  
In these special cases, the correlated uncertainties are shown in either italics or bold face text.}
\vskip 0.1cm
{\centerline{CDF: low $M_{\ell\ell}$ $H\rightarrow W^+W^-\rightarrow\ell^{\pm}\ell^{\prime \mp}$ with zero or one associated jets channel relative uncertainties (\%)}}
\vskip 0.099cm
\begin{ruledtabular}
\begin{tabular}{lcccccccc} \\
Contribution            & $WW$       & $WZ$       & $ZZ$       & $t\bar{t}$ & DY      & $W\gamma$  & $W$+jet(s) & $gg\to H$ \\ \hline 
{\bf Cross Section :}   &            &            &            &            &         &            &            &           \\ 
Scale                   &            &            &            &            &         &            &            & 12.0       \\ 
PDF Model               &            &            &            &            &         &            &            & 10.7        \\ 
Total                   & {\it 6.0 }  & {\it 6.0 }  & {\it 6.0 }  & 10.0 &  5.0  &            &            & 16.1       \\ 
{\bf Acceptance :}      &            &            &            &            &         &            &            &           \\ 
Scale (leptons)         &            &            &            &            &         &            &            & 0.6        \\  
Scale (jets)            &            &            &            &            &         &            &            & 1.2        \\ 
PDF Model (leptons)     & 	     &            &            &            &         &            &         & {\it 1.0 }\\   
PDF Model (jets)        &{\it 1.6 } &            &            &            &         &            &            & 2.1        \\ 
Higher-order Diagrams   & 	     &  10.0     &   10.0    &   10.0    & 10.0   &            &            &           \\ 
Jet Energy Scale           & {\it 1.0 }  & {\it 2.3 } & {\it 2.0 } & {\it 12.9 } & {\it 6.4 } & {\it 1.3 } &            & {\it 2.4 } \\  
Conversion Modeling      &            &            &            &            &         & 10.0        &            &           \\ 
Jet Fake Rates          &            &            &            &            &         &            & 18.4        &           \\ 
Lepton ID Efficiencies     & {\it 3.0 } &  {\it 3.0 } &  {\it 3.0 } & {\it 3.0 } & {\it 3.0 } &  		  &         & {\it 3.0 } \\ 
Trigger Efficiencies       & {\it 2.0 } &  {\it 2.0 } &  {\it 2.0 } & {\it 2.0 } & {\it 2.0 } &  		  &         & {\it 2.0 } \\ 
{\bf Luminosity}           & {\it 3.8 } &  {\it 3.8 } &  {\it 3.8 } & {\it 3.8 } &   {\it 3.8 }    &          &        & {\it 3.8 }  \\ 
{\bf Luminosity Monitor}   & {\it 4.4 } &  {\it 4.4 } &  {\it 4.4 } & {\it 4.4 } &   {\it 4.4 }    &          &        & {\it 4.4 }  \\ 
\end{tabular}
\end{ruledtabular}
\end{center}
\end{table}


\begin{table}
\begin{center}
\caption{\label{tab:cdfsystww5} Systematic uncertainties on the signal and background contributions for CDF's 
$H\rightarrow W^+W^-\rightarrow e^{\pm} \tau^{\mp}$ and $H\rightarrow W^+W^-\rightarrow \mu^{\pm} \tau^{\mp}$ 
channels.  These channels are sensitive to gluon fusion production, $WH, ZH$ and VBF production.  Systematic 
uncertainties are listed by name (see the original references for a detailed explanation of their meaning 
and on how they are derived).  Systematic uncertainties for $H$ shown in this table are obtained for 
$m_H=160$ GeV/$c^2$.  Uncertainties are relative, in percent, and are symmetric unless otherwise indicated.  
The uncertainties associated with the different background and signal processed are correlated within individual 
categories unless otherwise noted.  In these special cases, the correlated uncertainties are shown in either 
italics or bold face text.}
\vskip 0.1cm
{\centerline{CDF: $H\rightarrow W^+W^-\rightarrow e^{\pm} \tau^{\mp}$ channel relative uncertainties ( )}}
\vskip 0.099cm
\begin{ruledtabular}
\begin{tabular}{lccccccccccccccc} \\
Contribution                 & $WW$  & $WZ$ & $ZZ$ & $t\bar{t}$  & $Z\rightarrow\tau\tau$  & $Z\rightarrow\ell\ell$  & $W$+jet  & $W\gamma$  & $gg\to H$  & $WH$ & $ZH$ & VBF  \\ \hline
Cross section                & 6.0   & 6.0  & 6.0  & 10.0 &  5.0 &  5.0 &       &      &  10.3  &   5   &   5   &  10  \\ 
Measured W cross-section     &       &      &      &      &      &      & 12    &      &        &       &       &      \\ 
PDF Model                    & 1.6   & 2.3  & 3.2  & 2.3  &  2.7 &  4.6 & 2.2   & 3.1  &   2.5  & 2.0   &  1.9  & 1.8  \\ 
Higher order diagrams        & 10    & 10   & 10   & 10   &  10  &  10  &       & 10   &        & 10    &  10   & 10   \\ 
Conversion modeling          &       &      &      &      &      &      &       & 10   &        &       &       &      \\ 
Trigger Efficiency           & 0.5   & 0.6  & 0.6  & 0.6  & 0.7  & 0.5  & 0.6   & 0.6  &   0.5  & 0.5   &  0.6  & 0.5  \\ 
Lepton ID Efficiency         & 0.4   & 0.5  & 0.5  & 0.4  & 0.4  &  0.4 & 0.5   & 0.4  &   0.4  & 0.4   &  0.4  & 0.4  \\     
$\tau$ ID Efficiency         & 1.0   & 1.3  & 1.9  & 1.3  & 2.1  &      &       & 0.3  &   2.8  &  1.6  &  1.7  & 2.8  \\ 
Jet into $\tau$ Fake rate    & 5.8   & 4.8  & 2.0  & 5.1  &      &  0.1 & 8.8   &      &        &  4.2  &  4.0  & 0.4  \\ 
Lepton into $\tau$ Fake rate & 0.2   & 0.1  & 0.6  & 0.2  &      &  2.3 &       & 2.1  &  0.15  & 0.06  &  0.15 & 0.11 \\ 
W+jet scale                  &       &      &      &      &      &      & 1.6   &      &        &       &       &      \\ 
MC Run dependence            & 2.6   & 2.6  & 2.6  &      &      &      & 2.6   &      &        &       &       &      \\   
Luminosity                   & 3.8   & 3.8  & 3.8  & 3.8  & 3.8  & 3.8  & 3.8   & 3.8  &  3.8   & 3.8   &  3.8  & 3.8  \\ 
Luminosity Monitor           & 4.4   & 4.4  & 4.4  & 4.4  & 4.4  & 4.4  & 4.4   & 4.4  &  4.4   & 4.4   &  4.4  & 4.4  \\ 
\end{tabular}
\end{ruledtabular}

\vskip 0.3cm
{\centerline{CDF: $H\rightarrow W^+W^-\rightarrow \mu^{\pm} \tau^{\mp}$ channel relative uncertainties (\%)}}
\vskip 0.099cm
\begin{ruledtabular}
\begin{tabular}{lcccccccccccccc} \\
    Contribution                 & $WW$  & $WZ$ & $ZZ$ & $t\bar{t}$  & $Z\rightarrow\tau\tau$  & $Z\rightarrow\ell\ell$  & $W$+jet  & $W\gamma$  & $gg\to H$  & $WH$ & $ZH$ & VBF  \\ \hline
    Cross section                & 6.0  & 6.0  & 6.0  & 10.0 & 5.0 &  5.0 &     &      & 10.4  &   5   &   5   &  10  \\ 
    Measured W cross-section     &      &      &      &      &     &      & 12  &      &       &       &       &      \\ 
    PDF Model                    & 1.5  & 2.1  & 2.9  & 2.1  & 2.5 & 4.3  & 2.0 & 2.9  &  2.6  & 2.2   &  2.0  & 2.2  \\ 
    Higher order diagrams        & 10   & 10   & 10   & 10   &     &      &     & 11   &       & 10    &  10   & 10   \\ 
    Trigger Efficiency           & 1.3  & 0.7  & 0.7  & 1.1  & 0.9 & 1.3  & 1.0 & 1.0  &  1.3  & 1.3   &  1.2  & 1.3  \\ 
    Lepton ID Efficiency         & 1.1  & 1.4  & 1.4  & 1.1  & 1.2 & 1.1  & 1.4 & 1.3  &  1.0  & 1.0   &  1.0  & 1.0  \\     
    $\tau$ ID Efficiency         & 1.0  & 1.2  & 1.4  & 1.6  & 1.9 &      &     &      &  2.9  &  1.6  &  1.7  & 2.8  \\ 
    Jet into $\tau$ Fake rate    & 5.8  & 5.0  & 4.4  & 4.4  &     & 0.2  & 8.8 &      &       &  4.5  &  4.2  & 0.4  \\ 
    Lepton into $\tau$ Fake rate & 0.06 & 0.05 & 0.09 & 0.04 &     & 1.9  &     & 1.2  & 0.04  & 0.02  &  0.02 & 0.04 \\ 
    W+jet scale                  &      &      &      &      &     &      & 1.4 &      &       &       &       &      \\ 
    MC Run dependence            & 3.0  & 3.0  & 3.0  &      &     &      & 3.0 &      &       &       &       &      \\   
    Luminosity                   & 3.8  & 3.8  & 3.8  & 3.8  & 3.8 & 3.8  & 3.8 & 3.8  & 3.8   & 3.8   &  3.8  & 3.8  \\ 
    Luminosity Monitor           & 4.4  & 4.4  & 4.4  & 4.4  & 4.4 & 4.4  & 4.4 & 4.4  & 4.4   & 4.4   &  4.4  & 4.4  \\ 
\end{tabular}
\end{ruledtabular}
\end{center}
\end{table}


\begin{table}
\begin{center}
\caption{\label{tab:cdfsystwww} Systematic uncertainties on the signal and background contributions for 
CDF's $WH\rightarrow WWW \rightarrow\ell^{\pm}\ell^{\prime \pm}$ channel with one or more associated 
jets and $WH\rightarrow WWW \rightarrow \ell^{\pm}\ell^{\prime \pm} \ell^{\prime \prime \mp}$ channel.  
These channels are sensitive to only $WH$ and $ZH$ production.  Systematic uncertainties are listed 
by name (see the original references for a detailed explanation of their meaning and on how they are 
derived).  Systematic uncertainties for $H$ shown in this table are obtained for $m_H=160$ GeV/$c^2$.  
Uncertainties are relative, in percent, and are symmetric unless otherwise indicated.  The uncertainties 
associated with the different background and signal processed are correlated within individual categories 
unless otherwise noted.  In these special cases, the correlated uncertainties are shown in either italics 
or bold face text.}
\vskip 0.1cm
{\centerline{CDF: $WH \rightarrow WWW \rightarrow\ell^{\pm}\ell^{\prime\pm}$ channel relative uncertainties (\%)}}
\vskip 0.099cm
\begin{ruledtabular}
\begin{tabular}{lccccccccc} \\
Contribution               & $WW$          &   $WZ$       &  $ZZ$        & $t\bar{t}$   &  DY          & $W\gamma$    & $W$+jet &  $WH$        &  $ZH$        \\ \hline
{\bf Cross Section}        &   {\it 6.0 } &  {\it 6.0 } &  {\it 6.0 } &       10.0  &        5.0  &              &     	&  {\bf 5.0 } &  {\bf 5.0 } \\ 
Scale (Acceptance)         &  {\it -6.1 } &              &              &              &              &              &         &              &              \\ 
PDF Model (Acceptance)     &   {\it 5.7 } &  		  &  		 &  		&  	       &  	      &         &  	       & 	      \\  
Higher-order Diagrams      &   		   & {\it 10.0 } & {\it 10.0 } &       10.0  &       10.0  &       10.0  &         & {\bf 10.0 } & {\bf 10.0 } \\ 
Conversion Modeling        &               &              &              &              &              &       10.0  &         &              &              \\ 
Jet Fake Rates             &               &              &              &              &              &              &  39.1  &              &              \\ 
Jet Energy Scale           & {\it -14.0 } & {\it -3.9 } & {\it -2.8 } & {\it -0.6 } & {\it -9.3 } & {\it -7.6 } &         & {\it -1.0 } & {\it -0.7 } \\ 
Charge Mismeasurement Rate &  {\it 19.0 } &              &              & {\it 19.0 } & {\it 19.0 } &              &         &              &              \\ 
Lepton ID Efficiencies     &   {\it 3.0 } &  {\it 3.0 } &  {\it 3.0 } & {\it 3.0 }  &  {\it 3.0 } &  	      &         &  {\it 3.0 } &  {\it 3.0 } \\ 
Trigger Efficiencies       &   {\it 2.0 } &  {\it 2.0 } &  {\it 2.0 } & {\it 2.0 }  &  {\it 2.0 } &  	      &         &  {\it 2.0 } &  {\it 2.0 } \\ 
{\bf Luminosity}           &   {\it 3.8 } &  {\it 3.8 } &  {\it 3.8 } & {\it 3.8 }  &  {\it 3.8 } &   	      &         &  {\it 3.8 } &  {\it 3.8 } \\ 
{\bf Luminosity Monitor}   &   {\it 4.4 } &  {\it 4.4 } &  {\it 4.4 } & {\it 4.4 }  &  {\it 4.4 } &   	      &         &  {\it 4.4 } &  {\it 4.4 } \\
\end{tabular}
\end{ruledtabular}

\vskip 0.3cm
{\centerline{CDF: $WH\rightarrow WWW \rightarrow \ell^{\pm}\ell^{\prime \pm} \ell^{\prime \prime \mp}$ channel relative uncertainties (\%)}}
\vskip 0.0999cm
\begin{ruledtabular}
\begin{tabular}{lccccccc} \\
Contribution                & $WZ$        & $ZZ$        & $Z\gamma$     & $t\bar{t}$  & Fakes       & $WH$         & $ZH$           \\ \hline
{\bf Cross Section}         & {\it 6.0 } &  {\it 6.0 } &    10.0     &      10.0  &             & {\bf 5.0 }  &   {\bf 5.0 }  \\ 
Higher-order Diagrams       & {\it 10.0 }& {\it 10.0 } &    15.0     &      10.0  &             & {\bf 10.0 } & {\bf 10.0 }    \\ 
Jet Energy Scale            &             &              &  {\it -2.7 }&             &             &              &                  \\ 
Jet Fake Rates              &             &              &              &             & 24.8       &              &                 \\ 
$b$-Jet Fake Rates          &             &              &              &  27.3      &             &              &                 \\ 
MC Run Dependence           &             &              &   5.0       &	      &             &  		   &                 \\  
Lepton ID Efficiencies      & {\it 3.0 } &  {\it 3.0 } &              & {\it 3.0 } &             & {\it 3.0 }  &   {\it 3.0 }    \\ 
Trigger Efficiencies        & {\it 2.0 } &  {\it 2.0 } &              & {\it 2.0 } &             & {\it 2.0 }  &   {\it 2.0 }    \\ 
{\bf Luminosity}            & {\it 3.8 } &  {\it 3.8 } &              & {\it 3.8 } &             &  	{\it 3.8 }&   {\it 3.8 }    \\ 
{\bf Luminosity Monitor}    & {\it 4.4 } &  {\it 4.4 } &              & {\it 4.4 } &             &  	{\it 4.4 }&   {\it 4.4 }    \\
\end{tabular}
\end{ruledtabular}

\end{center}
\end{table}

\begin{table}
\begin{center}
\caption{\label{tab:cdfsystzww} Systematic uncertainties on the signal and background contributions for 
CDF's $ZH\rightarrow ZWW \rightarrow \ell^{\pm}\ell^{\mp} \ell^{\prime \pm}$ channels with 1 jet and 2 
or more jets.  These channels are sensitive to only $WH$ and $ZH$ production.  Systematic uncertainties 
are listed by name (see the original references for a detailed explanation of their meaning and on how 
they are derived).  Systematic uncertainties for $H$ shown in this table are obtained for $m_H=160$ 
GeV/$c^2$.  Uncertainties are relative, in percent, and are symmetric unless otherwise indicated.  The 
uncertainties associated with the different background and signal processed are correlated within 
individual categories unless otherwise noted.  In these special cases, the correlated uncertainties are 
shown in either italics or bold face text.}
\vskip 0.1cm
{\centerline{CDF: $ZH\rightarrow ZWW \rightarrow \ell^{\pm}\ell^{\mp} \ell^{\prime \pm}$ with one associated jet channel relative uncertainties (\%)}}
\vskip 0.0999cm
\begin{ruledtabular}
\begin{tabular}{lccccccc} \\
Contribution                & $WZ$        & $ZZ$         & $Z\gamma$    & $t\bar{t}$  & Fakes       & $WH$         & $ZH$           \\ \hline
{\bf Cross Section}         & {\it 6.0 } &  {\it 6.0 } &   10.0      &      10.0  &             & {\bf 5.0 }  &   {\bf 5.0 }  \\ 
Higher-order Diagrams       & {\it 10.0 }& {\it 10.0 } &    15.0     &      10.0  &             & {\bf 10.0 } & {\bf 10.0 }    \\ 
Jet Energy Scale            & {\it -7.6 }& {\it -2.3 } &  {\it -5.3 }& {\it  9.4 }&             &  {\it -9.0 }& {\it  8.1 }    \\ 
Jet Fake Rates              &             &              &              &             & 25.8       &              &                 \\ 
$b$-Jet Fake Rates          &             &              &              &  42.0      &             &              &                 \\ 
MC Run Dependence           &             &              &   5.0       &	      &             &  		   &                 \\  
Lepton ID Efficiencies      & {\it 3.0 } &  {\it 3.0 } &              & {\it 3.0 } &             & {\it 3.0 }  &   {\it 3.0 }   \\ 
Trigger Efficiencies        & {\it 2.0 } &  {\it 2.0 } &              & {\it 2.0 } &             &   {\it 2.0 }&   {\it 2.0 }   \\ 
{\bf Luminosity}            & {\it 3.8 } &  {\it 3.8 } &              & {\it 3.8 } &             &  	{\it 3.8 }&   {\it 3.8 }    \\ 
{\bf Luminosity Monitor}    & {\it 4.4 } &  {\it 4.4 } &              & {\it 4.4 } &             &  	{\it 4.4 }&   {\it 4.4\%}    \\
\end{tabular}
\end{ruledtabular}

\vskip 0.3cm
{\centerline{CDF: $ZH\rightarrow ZWW \rightarrow \ell^{\pm}\ell^{\mp} \ell^{\prime \pm}$ with two or more associated jets channel relative uncertainties (\%)}}
\vskip 0.0999cm
\begin{ruledtabular}
\begin{tabular}{lccccccc} \\
Contribution                & $WZ$        & $ZZ$         & $Z\gamma$    & $t\bar{t}$  & Fakes       & $WH$         & $ZH$           \\ \hline
{\bf Cross Section}         & {\it 6.0 } &  {\it 6.0 } &   10.0      &      10.0  &             & {\bf 5.0 }  &   {\bf 5.0 }  \\ 
Higher-order Diagrams       & {\it 10.0 }& {\it 10.0 } &    15.0     &      10.0  &             & {\bf 10.0 } & {\bf 10.0 }    \\ 
Jet Energy Scale            &{\it -17.8 }& {\it -13.1 }& {\it -18.2 }& {\it -3.6 }&             & {\it -15.4 }& {\it -4.9 }    \\ 
Jet Fake Rates              &             &              &              &             & 25.4       &              &                 \\ 
$b$-Jet Fake Rates          &             &              &              &  22.2      &             &              &                 \\ 
MC Run Dependence           &             &              &   5.0       &	      &             &  		   &                 \\  
Lepton ID Efficiencies      & {\it 3.0 } &  {\it 3.0 } &              & {\it 3.0 } &             & {\it 3.0 }  &   {\it 3.0 }   \\ 
Trigger Efficiencies        & {\it 2.0 } &  {\it 2.0 } &              & {\it 2.0 } &             &   {\it 2.0 }&   {\it 2.0 }   \\ 
{\bf Luminosity}            & {\it 3.8 } &  {\it 3.8 } &              & {\it 3.8 } &             &  	{\it 3.8 }&   {\it 3.8 }    \\ 
{\bf Luminosity Monitor}    & {\it 4.4 } &  {\it 4.4 } &              & {\it 4.4 } &             &  	{\it 4.4 }&   {\it 4.4 }    \\ 
\end{tabular}
\end{ruledtabular}

\end{center}
\end{table}


\begin{table}
\begin{center}
\caption{\label{tab:d0systww} Systematic uncertainties on the signal and background contributions for D0's
$H\rightarrow WW \rightarrow\ell^{\pm}\ell^{\prime \mp}$ channels.  Systematic uncertainties are listed by 
name; see the original references for a detailed explanation of their meaning and on how they are derived.  
Systematic uncertainties shown in this table are obtained for the $m_H=165$ GeV/c$^2$ Higgs selection.
Uncertainties are relative, in percent, and are symmetric unless otherwise indicated.}
\vskip 0.1cm
{\centerline{D0: $H\rightarrow WW \rightarrow e^{\pm} e^{ \mp}$ channel relative uncertainties (\%)}}
\vskip 0.099cm
\begin{ruledtabular}
\begin{tabular}{ l  c  c  c  c  c  c  c}  \\
Contribution & Diboson & ~~$Z/\gamma^* \rightarrow \ell\ell$~~&$~~W+jet/\gamma$~~ &~~~~$t\bar{t}~~~~$    & ~~Multijet~~  & ~~~~$H$~~~~      \\
\hline
Lepton ID                        &  6           &   6           & 6             & 6            & --   &   6        \\
Charge mis-ID                    &  1           &   1           & 1             & 1            & --   &   1        \\
Jet Energy Scale (s)               &  1           &   1        & 1          & 1          & --   &   1      \\
Jet identification (s)              &  1           &   1           & 1             & 1            & --   &   1        \\
Cross Section                     &  7           &   7           & 7            & 10           & 2  &   11         \\
Luminosity                       &  6           &   6           & 6             & 6            & --   &   6      \\
Modeling (s)~~~~~                  &  0           &   1           & 1             & 0            &  --   &   1     \\
\end{tabular}
\end{ruledtabular}

\vskip 0.3cm
{\centerline{D0: $H\rightarrow WW \rightarrow e^{\pm} \mu^{\mp}$ channel relative uncertainties (\%)}}
\vskip 0.099cm
\begin{ruledtabular}
\begin{tabular}{ l  c  c  c  c  c  c  c} \\
Contribution & Diboson & ~~$Z/\gamma^* \rightarrow \ell\ell$~~&$~~W+jet/\gamma$~~ &~~~~$t\bar{t}~~~~$    & ~~Multijet~~  & ~~~~$H$~~~~      \\
\hline
Trigger                          &  2           &   2           & 2             & 2            & --   &   2         \\
Lepton ID                        &  3           &   3           & 3             & 3            & --   &   3        \\
Momentum resolution (s)          &  0           &   3           & 1             & 0            & --   &   0       \\
Jet Energy Scale (s)              &  1           &   5           & 1             & 1            & --   &   1       \\
Jet identification (s)              &  1            &   3           & 1             & 1            & --   &   1        \\
Cross Section                     &  7           &   7           & 7            & 10           & 10   &   11       \\
Luminosity                       &  6           &   6           & 6             & 6            & --   &   6      \\
Modeling (s)~~~~~                  &  1           &   1           & 3             & 0            &  0   &   1      \\
\end{tabular}
\end{ruledtabular}

\vskip 0.3cm
{\centerline{D0: $H\rightarrow WW \rightarrow \mu^{\pm} \mu^{\mp}$ channel relative uncertainties (\%)}}
\vskip 0.099cm
\begin{ruledtabular}
\begin{tabular}{ l  c  c  c  c  c  c  c} \\
Contribution & Diboson & ~~$Z/\gamma^* \rightarrow \ell\ell$~~&$~~W+jet/\gamma$~~ &~~~~$t\bar{t}~~~~$    & ~~Multijet~~  & ~~~~$H$~~~~      \\
\hline
Lepton ID                        &  4           &   4           & 4             & 4            & --   &   4      \\
Momentum resolution (s)          &  1           &   1           & 2             & 1            & --   &   1      \\
Charge mis-ID                    &  1           &   1           & 1             & 1            & --   &   1      \\
Jet Energy Scale (s)             &  1           &   1           & 1             & 1            & --   &   1     \\
Jet identification               &  1           &   1           & 3             & 1            & --   &   1      \\
Cross Section                    &  7           &   7           & 7            & 10           &  15   &   11     \\
Luminosity                       &  6           &   6           & 6             & 6            & --   &   6     \\
Modeling~~~~~                    &  0           &   0           & 1             & 0            &  0   &   1      \\
\end{tabular}
\end{ruledtabular}

\end{center}
\end{table}


\begin{table}
\begin{center}
\caption{\label{tab:d0systwww} Systematic uncertainties on the signal and background contributions for D0's
$WH \rightarrow WWW \rightarrow\ell^{\prime \pm}\ell^{\prime \pm}$ channel.  Systematic uncertainties are 
listed by name; see the original references for a detailed explanation of their meaning and on how they are 
derived. Shape uncertainties are labeled with the ``shape'' designation.  Systematic uncertainties for signal 
shown in this table are obtained for $m_H=165$ GeV/$c^2$.  Uncertainties are relative, in percent, and are 
symmetric unless otherwise indicated.}
\vskip 0.1cm
{\centerline{D0: $VH \rightarrow\ell^{\pm}\ell^{\prime\pm} + X $ run IIa channel relative uncertainties (\%)}}
\vskip 0.099cm
\begin{ruledtabular}
\begin{tabular}{ l  c  c  c  c c } \\
Contribution		&WZ/ZZ		&W+jet		&ChargeFlip	&Multijet 		& $VH \rightarrow llX$	\\
\hline
Cross section		& 7		& 6		& 0		& 0			& 0		\\
Normalization		& 4		& 4		& 0		& 0			& 0		\\
Trigger (mumu)		& 0		& 0		& 0		& 0			& 2		\\
LeptonID (ee)		& 8.6		& 8.6 		& 0		& 0			& 8.6		\\
LeptonID (mumu)	 	& 4		& 4		& 0		& 0			& 4		\\
LeptonID (emu)		& 6.3		& 6.3		& 0		& 0			& 6.3		\\
JetID/JES		& 2		& 2		& 0		& 0			& 2	\\
Jet-Lepton Fake		& 0		& 20		& 0		& 0			& 0		\\
Instrumental ($ee$) 	& 0		& 0		& 0		& 52			& 44	\\
Instrumental ($e\mu$ 	& 0		& 0		& 0		& 0			& 29		\\
Instrumental ($\mu\mu$) & 0		& 0		& 0		& 155			& 42		\\
Instrumental Model	& -		& -		& shape		& shape			& -		\\ 
\end{tabular}
\end{ruledtabular}

\vskip 0.3cm
{\centerline{D0: $VH \rightarrow\ell^{\pm}\ell^{\prime\pm} + X $ run IIb channel relative uncertainties (\%)}}
\vskip 0.099cm
\begin{ruledtabular}
\begin{tabular}{ l  c  c  c  c c } \\
Contribution 	&WZ/ZZ		&W+jet		&ChargeFlip	&Multijet 		& $VH \rightarrow llX$	\\               
\hline
Cross section		& 7		& 6		& 0			& 0			& 0		\\
Normalization		& 4		& 4		& 0			& 0			& 0		\\
Trigger (mumu)		& 0		& 0		& 0			& 0			& 5		\\
LeptonID (ee)		& 8.6		& 8.6 		& 0			& 0			& 8.6		\\
LeptonID (mumu)	 	& 4		& 4		& 0			& 0			& 4		\\
LeptonID (emu)		& 6.3		& 6.3		& 0			& 0			& 6.3		\\
JetID/JES		& 2		& 2		& 0			& 0			& 2	\\
Jet-Lepton Fake		& 0		& 20		& 0			& 0			& 0		\\
Instrumental ($ee$)	& 0		& 0		& 0			& 23			& 31		\\
Instrumental ($e\mu$)	& 0		& 0		& 0			& 0			& 19		\\
Instrumental ($\mu\mu$)	& 0		& 0		& 0			& 43			& 28		\\
Instrumental Model	& -		& -		& shape			& shape			& -		\\
\end{tabular}
\end{ruledtabular}
\end{center}
\end{table}


\begin{table}
\begin{center}
\caption{\label{tab:d0lvjj} Systematic uncertainties on the signal and background contributions for D0's 
$H\rightarrow W W^{*} \rightarrow lvjj$ electron and muon channels.  Systematic uncertainties are listed 
by name; see the original references for a detailed explanation of their meaning and on how they are 
derived.  Systematic uncertainties for $gg \rightarrow H\rightarrow W W^{*} \rightarrow lvjj$ shown in 
this table are obtained for $m_H=165$ GeV/$c^2$.  Uncertainties are relative, in percent, and are symmetric 
unless otherwise indicated. Shape uncertainties are labeled with an ``s''.}
\vskip 0.1cm
{\centerline{D0: $H\rightarrow W W^{*} \rightarrow lvjj$ run IIa channel relative uncertainties (\%)}}
\vskip 0.099cm
\begin{ruledtabular}
\begin{tabular}{ l c c c c c c } \\
Contribution                     & ~W+jets~     & ~Z+jets~      & ~~~Top~    &	~Diboson~ & ~ggHWWlnujj~~\\ \hline
Jet Energy Scale pos/neg (S)     & Shape Only   & Shape Only    & 7.5/21.0   & 10.0/8.75  & 5.25/5.5      \\
Wbb Jet Energy Scale	         & Shape Only	& -	        & -	     & -	  & - \\
Top Jet Energy Scale	         & -            & 1.8	        & -          & -	  & - \\
Jet ID (S)                       & Shape Only	& Shape Only	& 5.0	     &	3.0	  & 1.0 \\
Jet Resolution pos/neg (S)       & Shape Only	& Shape Only    & 1.75/0.25  & 2.3/1.25	  & 1.5/0.5      \\
SingleMuOR Trigger (S)           & $\pm$0.1	& $\pm$0.1	& $\pm$0.1   & $\pm$0.1	  & $\pm$0.1     \\
ALPGEN MLM pos/neg(S)            & Shape Only	& Shape Only    & -          & -          & -      \\
ALPGEN Scale (S)                 & Shape Only	& Shape Only	& -          & -	  & -      \\
Underlying Event (S)             & Shape Only	& Shape Only	& -          & -	  & -      \\
Parton Distribution Function (S) & 1.6/1.9	& 0.6/1.25	& 2.0/0.9    & $\pm$0.05  & 1.0/1.0      \\
EM ID                            & $\pm$4.0	& $\pm$4.0	& $\pm$4.0   & $\pm$4.0	  & $\pm$4.0      \\
Muon ID                          & $\pm$4.0	& $\pm$4.0	& $\pm$4.0   & $\pm$4.0	  & $\pm$4.0      \\
Cross Section                    & 6.0	  	& 6.0		& 10.0	     & 7.0        & 10.0      \\
Luminosity                       & 6.1		& 6.1		& 6.1	     & 6.1	  & 6.1  \\
\end{tabular}
\end{ruledtabular}

\vskip 0.3cm
{\centerline{D0: $H\rightarrow W W^{*} \rightarrow lvjj$ run IIb channel relative uncertainties (\%)}}
\vskip 0.099cm
\begin{ruledtabular}
\begin{tabular}{ l c c c c c c } \\
Contribution                     & ~W+jets~    	& ~Z+jets~     	& ~~~Top~    & ~Diboson~  & ~ggHWWlnujj~~\\ \hline
Jet Energy Scale pos/neg (S)     & Shape Only	& Shape Only	& $\pm$6.0   & 3.25/3.5	  & 3.25/2.0      \\
Wbb Jet Energy Scale		 & Shape Only	& -		& -	     & -	  & - \\
Top Jet Energy Scale		 & -		& 1.8		& -	     & -	  & - \\
Jet ID (S)                     	 & Shape Only	& Shape Only	& 3.25	     & 1.25	  & 3.5 \\
Jet Resolution pos/neg (S)       & Shape Only	& Shape Only	& 0.5/0.3    & 1.0/0.5	  & 2.0/1.75 \\
Vertex Confirmation (S)          & Shape Only	& Shape Only	& 3.75	     & 3.75	  & 4.75      \\
SingleMuOR Trigger (S)           & $\pm$0.5	& $\pm$0.5	& $\pm$0.5   & $\pm$0.25  & $\pm$0.25     \\
ALPGEN MLM pos/neg(S)            & Shape Only	& Shape Only	& -          & -	  & -      \\
ALPGEN Scale (S)                 & Shape Only	& Shape Only	& -          & -	  & -      \\
Underlying Event (S)             & Shape Only	& Shape Only	& -          & -          & -      \\
Parton Distribution Function (S) & 3.5/2.5	& 8.0/1.5	& 2.25/3.6   & $\pm$0.25  & 1.75/3.75      \\
EM ID                            & $\pm$4.0	& $\pm$4.0	& $\pm$4.0   & $\pm$4.0	  & $\pm$4.0      \\
Muon ID                          & $\pm$4.0	& $\pm$4.0	& $\pm$4.0   & $\pm$4.0	  & $\pm$4.0      \\
Cross Section                    & 6.0		& 6.0		& 10.0	     & 7.0	  & 10.0      \\
Luminosity                       & 6.1		& 6.1		& 6.1	     & 6.1	  & 6.1  \\
\end{tabular}
\end{ruledtabular}
\end{center}
\end{table}


\begin{table}
\begin{center}
\caption{\label{tab:d0systtth} Systematic uncertainties on the signal and background contributions for D0's
$t \bar{t} H\rightarrow t \bar{t} b \bar{b}$ channel.  The systematic uncertainties for $ZH$, $WH$ shown 
in this table are obtained for $m_H=115$ GeV/$c^2$.  Systematic uncertainties are listed by name; see the 
original references for a detailed explanation of their meaning and on how they are derived.  Uncertainties 
are relative, in percent, and are symmetric unless otherwise indicated.}
\vskip 0.1cm
{\centerline{D0: $t \bar{t} H\rightarrow t \bar{t} b \bar{b}$ channel relative uncertainties (\%)}}
\vskip 0.099cm
\begin{ruledtabular}
\begin{tabular}{lcc} \\
Contribution                             &  ~~~background~~~ & ~~~$t \bar{t} H$~~~    \\
\hline
Luminosity~~~~                           &  6          &  6    \\
lepton ID efficiency                     &  2--3       &  2--3    \\
Event preselection                       &  1          &  1    \\
$W$ +jet modeling                        &   15        & -     \\
Cross Section                            &  10--50     &  10    \\
\end{tabular}
\end{ruledtabular}
\end{center}
\end{table}


\begin{table}
\begin{center}
\caption{\label{tab:cdfsysttautau} Systematic uncertainties on the signal and background contributions for CDF's
$H\rightarrow \tau^+\tau^-$ channels.  Systematic uncertainties are listed by name; see the original references 
for a detailed explanation of their meaning and on how they are derived. Systematic uncertainties for the Higgs 
signal shown in these tables are obtained for $m_H=120$ GeV/$c^2$.  Uncertainties are relative, in percent, and 
are symmetric unless otherwise indicated.}
\vskip 0.1cm
{\centerline{CDF: $H \rightarrow \tau^+ \tau^-$ channel relative uncertainties (\%)}}
\vskip 0.099cm
\begin{ruledtabular}
\begin{tabular}{lccccccccc}\\
Contribution & $Z/\gamma^* \rightarrow ll$  & $t\bar{t}$ & diboson  & fakes from SS
&W+jets & $WH$      & $ZH$  & VBF      & $gg\rightarrow H$ \\
\hline
PDF Uncertainty                                      &  1    &  1   &  1  &  -   &    -     &  1.2  &  0.9  &  2.2  &  4.9   \\
ISR 1 JET                                                 &  -     &  -   &  -  &  -   &    -     & -6.1 & -1.7 & -2.9 & 13.0 \\
ISR $\ge$ 2 JETS                                     &  -     &  -   &  -  &  -   &    -     & -1.5 & 0.1 & -2.7 & 15.5 \\
FSR 1 JET                                                 &  -     &  -   &  -  &  -   &    -     &  4.3 & 1.0 & 1.7 & -5.0 \\
FSR $\ge$ 2 JETS                                     &  -     &  -   &  -  &  -   &    -     & -2.1 & 0.4 & -1.1 & -5.2 \\
JES (shape) 1 JET                                    & 6.2    &  -7.7   & 7.1  &  -   &    -     &  -4.8  &  -5.3  &  -3.7  & 5.1   \\
JES (shape) $\ge$ 2 JETS                        & 14.2    & 3.2   & 11.7 &  -   &    -     &  5.4  &  4.8  &  -5.2  & 13.2   \\
Cross Section or Norm. 1 JET                   &  2.2  &   10   & 6  &  10  &   18     &  5  &  5  & 10  & 23.5   \\
Cross Section or Norm. $\ge$2 JETS        &  2.2  &   10   & 6  &  10  &   30     &  5  &  5  & 10  & 67.5   \\
MC Acceptance                                         & 2.3   &  -   &  -  &  -   &    -     &  -  &  -  &  -  &  -   \\
tau ID scale factor:                                   &         &       &     &      &           &     &      &     &      \\
N$_{obs}$                                                &2.8    & 2.8  & 2.8  & - & - & 2.8 & 2.8 & 2.8 & 2.8 \\
N$_{SSdata}$                                           &-3.3  & -3.3  & -3.3  & - & - & -3.3 & -3.3 & -3.3 & -3.3 \\
N$_{W+jets}$                                          &-0.3    & -0.3  & -0.3  & - & - & -0.3 & -0.3 & -0.3 & -0.3 \\
Cross section (DY)                                    &-2.1    & -2.1  & -2.1  & - & - & -2.1 & -2.1 & -2.1 & -2.1 \\
MC Acceptance (DY)                                 &-2.2    & -2.2  & -2.2  & - & - & -2.2 & -2.2 & -2.2 & -2.2 \\
\end{tabular}
\end{ruledtabular}
\end{center}
\end{table}


\begin{table}
\begin{center}
\caption{\label{tab:cdfallhadsyst} Systematic uncertainties on the signal and background contributions for CDF's 
$WH+ZH \rightarrow jjbb$ and $VBF \rightarrow jjbb$ channels.  Systematic uncertainties are listed by name; see the original references for a 
detailed explanation of their meaning and on how they are derived.  Uncertainties with provided shape systematics 
are labeled with ``s''.  Systematic uncertainties for $H$ shown in this table are obtained for $m_H=115$ GeV/$c^2$.
Uncertainties are relative, in percent, and are symmetric unless otherwise indicated.  The cross section uncertainties 
are uncorrelated with each other (except for single top and $t{\bar{t}}$, which are treated as correlated).  The QCD 
uncertainty is also uncorrelated with other channels' QCD rate uncertainties.}
\vskip 0.1cm
{\centerline{CDF: $WH+ZH\rightarrow jjbb$ and $VBF \rightarrow jjbb$ channel relative uncertainties (\%)}}
\vskip 0.099cm
\begin{ruledtabular}
\begin{tabular}{lccccc} \\
Contribution              & $t\bar{t}$ & diboson & $W/Z$+Jets & VH  & VBF \\ \hline
Jet Energy Correction     &            &         &            & 7 s & 7 s \\
PDF Modeling              &            &         &            & 2   & 2   \\ 
SecVtx+SecVtx             & 7.6        & 7.6     & 7.6        & 7.6 & 7.6 \\ 
SecVtx+JetProb            & 9.7        & 9.7     & 9.7        & 9.7 & 9.7 \\
Luminosity                & 6          & 6       & 6          & 6   & 6   \\ 
ISR/FSR modeling          &            &         &            & 2 s & 3 s \\
Jet Moment                &            &         &            & s   & s   \\
Trigger                   & 4          & 4       & 4          & 4   & 4   \\  
QCD Interpolation         &            &         &            & s   & s   \\ 
QCD MJJ Tuning            &            &         &            & s   & s   \\ 
QCD Jet Moment Tuning     &            &         &            & s   & s   \\ 
cross section             & 10         & 6       & 50         &     &     \\
\end{tabular}
\end{ruledtabular}
\end{center}
\end{table}


\begin{table}
\begin{center}
\caption{\label{tab:cdfsystgg} Systematic uncertainties on the signal and background contributions for CDF's 
$H\rightarrow \gamma \gamma$ channel.  Systematic uncertainties for the Higgs signal shown in this table are 
obtained for $m_H=120$ GeV/$c^2$.  Systematic uncertainties are listed by name; see the original references 
for a detailed explanation of their meaning and on how they are derived.  Uncertainties are relative, in 
percent, and are symmetric unless otherwise indicated.}
\vskip 0.1cm
{\centerline{CDF: $H \rightarrow \gamma \gamma$ channel relative uncertainties (\%)}}
\vskip 0.099cm
\begin{ruledtabular}
\begin{tabular}{lcc}\\
Contribution & background & signal \\\hline
Luminosity                                      &    & 6 \\
$\sigma_{ggH}$ / $\sigma_{VH}$ / $\sigma_{VBF}$ &    & 12 / 5 / 10 \\ 
PDF                                             &    & 1 \\
ISR                                             &    & 2 \\
FSR                                             &    & 2 \\
Energy Scale                                    &    & 0.1 \\
Vertex                                          &    & 0.2 \\
Conversions                                     &    & 0.2 \\
Photon/Electron ID                              &    & 1.0 \\
Run Dependence                                  &    & 1.5 \\
Data/MC fits                                    &    & 0.2 \\
Background Shape                                & 4  &     \\
\end{tabular}
\end{ruledtabular}
\end{center}
\end{table}

\begin{table}
\begin{center}
\caption{\label{tab:d0systgg} Systematic uncertainties on the signal and background contributions for D0's
$H\rightarrow \gamma \gamma$ channel. Systematic uncertainties for the Higgs signal shown in this table are 
obtained for $m_H=115$ GeV/$c^2$.  Systematic uncertainties are listed by name; see the original references 
for a detailed explanation of their meaning and on how they are derived.  Uncertainties are relative, in 
percent, and are symmetric unless otherwise indicated.}
\vskip 0.1cm
{\centerline{D0: $H \rightarrow \gamma \gamma$ channel relative uncertainties (\%)}}
\vskip 0.099cm
\begin{ruledtabular}
\begin{tabular}{lcc}\\
Contribution &  ~~~background~~~  & ~~~signal~~~    \\
\hline
Luminosity~~~~                            &  6     &  6    \\
Acceptance                                &  --    &   2   \\
electron ID efficiency                    &  2     &  2    \\
electron track-match inefficiency         & 10--20 & -     \\
Photon ID efficiency                      &  7     &   7     \\
Photon energy scale                       &  --    &   2     \\
Acceptance                                &  --    &  2    \\
$\gamma$-jet and jet-jet fakes          &  26    &  --    \\
Cross Section ($Z$)                       &  4     &  6    \\
Background subtraction                    &  8--14 &  -    \\
\end{tabular}
\end{ruledtabular}
\end{center}
\end{table}

\end{document}